\RequirePackage{lineno}
\documentclass[
 preprint,
showpacs,
preprintnumbers,
nofootinbib,
 amsmath,
 amssymb,
 aps,
 onecolumn,
floatfix,
]{revtex4-1}

\usepackage{graphicx}
\usepackage{dcolumn}
\usepackage{bm}
\usepackage{color}
\usepackage{verbatim}
\usepackage{hyperref}
\usepackage{amsfonts}
\usepackage{geometry}
\usepackage{longtable}
\usepackage{natbib}

\graphicspath{{figs/}}

\newcommand{\pbp}{\bar{\psi}\psi}
\newcommand{\pbgp}{\bar{\psi}\gamma_5\psi}
\newcommand{\lpbpr}{\langle\bar{\psi}\psi\rangle}

\newcommand{\density}{\rho(\lambda,m)}
\newcommand{\dL}{\df\lambda}

\newcommand{\ub}{\bar{u}}
\newcommand{\db}{\bar{d}}
\newcommand{\ubl}{\bar{u}_L}
\newcommand{\dbl}{\bar{d}_L}
\newcommand{\ubr}{\bar{u}_R}
\newcommand{\dbr}{\bar{d}_R}
\newcommand{\ul}{u_L}
\newcommand{\dl}{d_L}
\newcommand{\ur}{u_R}
\newcommand{\dr}{d_R}

\newcommand{\df}{\mathrm{d}}

\newcommand{\mres}{m_\text{res}}
\newcommand{\qtop}{Q_\text{top}}
\newcommand{\sua}{SU(2)_L\times SU(2)_R}
\newcommand{\ua}{U(1)_A}
\newcommand{\Nt}{N_\tau}
\newcommand{\Ns}{N_\sigma}
\newcommand{\pmd}{\chi_\pi-\chi_\delta}

\def\slashed#1{\kern+0.1em /\kern-0.65em #1}

\newcommand{\Tspace}{\rule{0pt}{2.6ex}}
\newcommand{\Bspace}{\rule[-1.2ex]{0pt}{0pt}}

\newcounter{run}
\newenvironment{run}{\refstepcounter{run}\therun}{}
\newcommand{\runlabel}[1]{\begin{run}\label{run:#1}\end{run}}

\newcommand{\runn}[1]{\ref{run:#1}}

\begin{document}
\title{The chiral transition and \boldmath $U(1)_A$ symmetry restoration from lattice QCD using Domain Wall Fermions}

\author{A. Bazavov$^{\rm a}$, Tanmoy Bhattacharya$^{\rm b}$, Michael I. Buchoff$^{\rm c}$, Michael Cheng$^{\dagger \rm c}$, \\
N.H. Christ$^{\rm d}$, H.-T. Ding$^{\rm a}$, Rajan Gupta$^{\rm b}$,
Prasad Hegde$^{\rm a}$, Chulwoo Jung$^{\rm a}$,\\
 F. Karsch$^{\rm a,e}$,  Zhongjie Lin$^{\rm d}$, R.D. Mawhinney$^{\rm d}$, Swagato Mukherjee$^{\rm a}$,\\
  P. Petreczky$^{\rm a}$,  R.A. Soltz$^{\rm b}$, P.M. Vranas$^{\rm c}$, and Hantao Yin$^{\rm d}$
\\[1mm]
{\textbf{(HotQCD Collaboration)}}
}
\affiliation{
$^{\rm a}$ Physics Department, Brookhaven National Laboratory,Upton, NY 11973, USA\\
$^{\rm b}$ Theoretical Division, Los Alamos National Laboratory, Los Alamos, NM 87545, USA\\
$^{\rm c}$ Physics Division, Lawrence Livermore National Laboratory, Livermore CA 94550, USA\\
$^{\rm d}$ Physics Department, Columbia University, New York, NY 10027, USA\\
$^{\rm e}$ Fakult\"at f\"ur Physik, Universit\"at Bielefeld, D-33615 Bielefeld, Germany\\
$^\dagger$: Current Address:\\Center for Computational Science, Boston University, Boston, MA 02215, USA\\
}

\preprint{CU-TP-1200}
\preprint{BNL-97301-2012-JA}
\preprint{LA-UR-12-21329}
\preprint{LLNL-JRNL-557574}

\date{May 11, 2012}

\begin{abstract}

We present results on both the restoration of the spontaneously broken chiral symmetry
and the effective restoration of the anomalously broken $U(1)_A$ symmetry 
in finite temperature QCD at zero chemical potential using lattice QCD. 
We employ domain wall fermions on lattices with fixed temporal extent $\Nt = 8$ 
and spatial extent $\Ns = 16$ in a temperature range of $T = 139 - 195 ~\textrm{MeV}$, 
corresponding to lattice spacings of $a \approx 0.12 - 0.18~\textrm{fm}$.  In these 
calculations, we include two degenerate light quarks and a strange quark 
at fixed pion mass $m_\pi = 200~\textrm{MeV}$.   The strange quark mass is set near
its physical value.  We also present results from a second set of finite
temperature gauge configurations at the same volume and temporal
extent with slightly heavier pion mass.
To study chiral symmetry restoration, we 
calculate the chiral condensate, the disconnected chiral susceptibility, and 
susceptibilities in several meson channels of different quantum numbers.
To study $U(1)_A$ restoration, we calculate spatial correlators in the scalar 
and pseudo-scalar channels, as well as the corresponding susceptibilities. 
Furthermore, we also show results for the eigenvalue spectrum of the Dirac 
operator as a function of temperature, which can be connected to both $U(1)_A$  
and chiral symmetry restoration via Banks-Casher relations.
\end{abstract}

\pacs{11.15.Ha, 12.38.Gc}

\maketitle

\section{Introduction}
\label{sec:intro}

In the limit of vanishing up and down quark masses, Quantum Chromodynamics (QCD) 
posseses a chiral $\sua$ symmetry.   However, the QCD vacuum does not respect this
symmetry.  Instead the non-vanishing vacuum expectation value of the $\sua$ non-invariant
operators $\overline{\psi}_l\psi_l$, for $l=u$, $d$ reflect a smaller, $SU(2)_V$ vacuum 
symmetry.  This symmetry-breaking vacuum order is expected to disappear at high
temperature implying a phase transition separating a low temperature chirally asymmetric 
phase from a high-temperature phase with restored chiral symmetry.  The chirally 
symmetric, high temperature phase of QCD was present during the evolution of the 
early universe and is also expected to be created in heavy-ion collision experiments.  
Thus, studies of chiral symmetry restoration at high temperatures are of great physical 
importance.

At the classical level QCD posseses an additional $\ua$ symmetry which is broken by 
the axial anomaly.  This results in both the anomalous term in the conservation law 
for the $\ua$ axial current of  Adler~\cite{Adler:1969gk} and Bell and Jackiw \cite{Bell:1969ts}
as well as `t Hooft's explicit violation of the global symmetry~\cite{'tHooft:1976up} arising 
from fermion zero modes associated with topologically non-trivial gauge field configurations.
At low temperatures this anomalous $\ua$ symmetry is also broken by the QCD vacuum.
However, above the QCD phase transition vacuum symmetry breaking has disappeared 
and the effects of  the axial anomaly can be studied directly.

Lattice QCD is ideally suited to study these symmetries and their degree of restoration 
with increasing temperature.  However, such studies are complicated by the fermion doubling 
problem.  This fundamental difficulty, present in any discrete theory of fermions, sharply 
reduces the chiral symmetry that is present in a lattice fermion formulation.   The Wilson 
formulation shows chiral symmetry only in the continuum limit.  Staggered fermions are 
more successful and preserve a single, non-anomalous $U(1)$ axial symmetry at finite lattice 
spacing.  

In this paper, we employ the domain wall fermion (DWF) formulation of
Kaplan~\cite{Kaplan:1992bt} and Shamir~\cite{Furman:1994ky} which, at the classical level, 
shows the full  $\sua \times \ua$ symmetry, with lattice symmetry breaking controlled by the 
size, $L_s$, of an additional fifth dimension.  For the results reported here $L_s$ varies 
between 32 and 96 and is sufficiently large that the residual quark mass induced by lattice 
effects is on the order of 10 MeV or smaller -- sufficiently small that its effects can be easily 
incorporated as an additive shift in the quark mass.   Most previous lattice studies of the chiral 
transition in QCD use staggered fermions, for which the issue of anomalous symmetry is
somewhat subtle, involving possible non-commutativity of the continuum and chiral limits
and the non-unitarity of the rooted theory at finite lattice spacing \cite{Sharpe:2006re,Donald:2011if, 
Adams:2009eb}.  In contrast, the DWF 
formulation posseses an easily understood anomalous $\ua$ symmetry~\cite{Furman:1994ky},
broken by the same topological effects which produce anomalous symmetry breaking in the 
continuum, with explicit  lattice artifacts appearing at order $m_{\rm res} a^2$.   Thus, the 
degree of anomalous symmetry restoration with increasing temperature is a natural focus 
of this paper.

At sufficiently high temperatures anomalous $\ua$ symmetry breaking can be 
studied using the dilute instanton gas approximation \cite{Gross:1980br}.  In this
approximation one finds exponential suppression of the instanton density as the
gauge coupling decreases so that the $\ua$ symmetry becomes exact in the limit 
$T\to\infty$.
When the dilute instanton gas approximation is justified, the $\ua$ symmetry breaking
effects it predicts are very small.   With decreasing temperature, the semi-classical 
approximation underlying the dilute instanton gas picture becomes unreliable and the 
degree of anomalous symmetry breaking becomes a non-perturbative question well
suited to a DWF lattice study.  While one might imagine that anomalous $\ua$ 
breaking remains small as the temperature decreases from asymptotically
large values, even down to the critical temperature, $T_c$, it is also possible that 
new, non-perturbative phenomena emerge at lower temperatures leading to a 
significant topological charge density and to large $\ua$ symmetry breaking.

The degree of $\ua$ symmetry breaking may have interesting physical
consequences. For example, if the $\ua$ breaking is sufficiently large near the 
phase transition for QCD with two massless flavors then this transition can be second 
order, belonging to the three-dimensional $O(4)$ universality 
class \cite{Pisarski:1983ms,Butti:2003nu}.  On the other hand, if the axial symmetry 
breaking is negligible then this  $O(4)$ universality class is no longer appropriate
for the larger symmetry of the long-distance variables and the chiral transition is expected to 
be first order \cite{Pisarski:1983ms,Butti:2003nu}, although in this case a second-order
transition is also allowed with a 
different symmetry breaking pattern, $U(2)_L \times U(2)_R/U(2)_V$~\cite{Basile:2005hw}. 
Hence, the nature of the chiral phase transition itself may depend critically on the 
strength of the $\ua$ symmetry breaking. 

In heavy-ion collision experiments, it may also be possible to observe signatures of 
$\ua$ symmetry restoration through measurements of low-mass dileptons~\cite{Rapp:1999ej}. 
Moreover, an effective restoration of the axial $\ua$ symmetry above $T_c$ may lead to
softening of the $\eta'$ mass resulting in interesting
experimental signatures \cite{Shuryak:1993ee,Huang:1995fc,Kapusta:1995ww}. 
In fact, recently it has been
claimed that the results from the Relativistic Heavy-Ion Collider (RHIC)
suggest softening of the $\eta^\prime$ mass indicating partial restoration of
the $\ua$ symmetry in hot and dense matter \cite{Csorgo:2009pa}. Hence, studies
related to $\ua$ symmetry restoration with increasing temperature have  important 
theoretical and phenomenological consequences.

As discussed above, chiral symmetry restoration, as well as the degree of $\ua$ 
symmetry breaking above $T_c$, are essentially non-perturbative in nature.  At 
present, lattice QCD, as the most reliable non-perturbative technique, is ideally 
suited for such studies.  In fact, extensive lattice QCD studies of chiral symmetry 
restoration have been carried out.  For a review and summary of recent lattice QCD 
results see Refs.~\cite{Petreczky:2012rq,Mukherjee:2011td}. Most of these lattice studies 
have been performed using staggered fermion discretization schemes. Staggered 
fermions have also been used to study the degree of axial symmetry restoration in 
high temperature QCD \cite{Bernard:1996iz, Chandrasekharan:1998yx, Kogut:1998rh, 
Cheng:2010fe,Ohno:2011yr}.  However, as described earlier, for staggered fermions 
at non-zero lattice spacing, chiral symmetry, the axial anomaly and its relation to the 
index theorem suffer from significant complications.  Thus, a study using the DWF 
discretization scheme, which preserves the full $\sua$ symmetry and reproduces 
the correct anomaly even for non-zero values of lattice spacing, is well motivated.
To date, there have been only a few fully dynamical calculations using chiral
fermion formulations -- domain wall fermions~\cite{Chen:2000zu,Cheng:2009be} and
overlap fermions~\cite{Borsanyi:2012xf}.

In this paper we study the chiral transition and degree of restoration of $\ua$
symmetry for $T \ge T_c$ by performing lattice QCD simulations using the
DWF action with two degenerate light (up and down) and one heavier (strange)
quarks. We employ lattices with spatial size $\Ns = 16$ and temporal extent
$\Nt = 8$, with lattice spacings in the range $a \approx 0.12 - 0.18
~\textrm{fm}$, corresponding to a temperature range of $T = 137 - 198~ \textrm{MeV}$.
We work on a line of constant physics, {\it i.e.}, the strange quark mass is fixed to
near its physical value, while for most of the results presented here the two light 
quark masses have been chosen so that $m_\pi \approx 200 ~\textrm{MeV}$. 
This extends earlier studies of the QCD transition with domain wall fermions 
\cite{Chen:2000zu,Cheng:2009be} by going to a lighter quark mass, using
a gauge action optimized for the relatively large lattice spacing needed for such an 
$\Nt=8$ study, and exploring in more detail the chiral aspects of the QCD transition.
We also present a thorough study of the eigenvalue spectrum of the Dirac operator
employing a variant of the method of Giusti and L\"uscher~\cite{Giusti:2008vb} to 
convert the spectrum of the hermitian DWF Dirac operator to a spectrum evaluated 
in the $\overline{{\rm MS}}$ scheme which has a well-defined continuum limit.  This 
allows us to examine the density of eigenvalues near zero as a function of temperature.  
This density can be directly related to both $\sua$ and $\ua$ breaking and restoration 
through Banks-Casher type formulae.

This paper is organized as follows. We start in Sec.~\ref{sec:details} with a
discussion of the setup of our lattice calculation, including the choice of lattice
action and the determination of the line of constant physics.  
In Sec. \ref{sec:dirac} we present details of our eigenvalue calculations with DWF,
including the methods used to convert the  low-lying eigenvalue spectrum of the
hermitian DWF Dirac operator to a spectrum meaningful in the continuum limit. 
In Sec.~\ref{sec:basics} we introduce the basic observables which we will use to 
explore the chiral aspects of the QCD transition, emphasizing the role of the $\ua$ 
symmetry for the transition.  
Sec. \ref{sec:chiral} examines the restoration of $\sua$ chiral symmetry through the
subtracted chiral condensate, disconnected chiral susceptibility, and vector and
axial vector screening masses.  Sec.~\ref{sec:ua1} deals with the restoration of
$\ua$ symmetry by examining the scalar and pseudo-scalar screening
correlators, their respective susceptibilities, and their relation to the topological
charge.  We discuss our results and give conclusions in Sec. \ref{sec:conclusions}.  
 Appendix~\ref{sec:spectral_normalization} gives further details on the normalization 
of the eigenvalue spectrum,  Appendix~\ref{sec:renormalization} discusses the 
renormalization of the disconnected, staggered chiral susceptibility while Appendix~\ref{sec:RHMC_evolution} gives the details of the evolution algorithms 
used to generate our gauge field emsembles.  Finally Appendix~\ref{sec:g5-top_suscept}
examines a discrepancy between the topological and disconnected 
$\overline{\psi}\gamma^5\psi$ susceptibilities and concludes that the combination of 
APE smearing and improved gauge field operator~\cite{deForcrand:1997sq} used 
here to determine the topological charge contains large lattice artifacts when applied 
at non-zero temperatures on the coarse ensembles studied in this paper.

\section{Calculation Details}
\label{sec:details}
\subsection{Fermion and Gauge Action}
For this calculation, we use the domain wall fermion action.  
At the lattice spacings at which we work, \textit{i.e.}, those appropriate to study
the finite temperature transition region with temporal extent $\Nt = 8$, the residual chiral symmetry
breaking, parameterized by the residual mass $\mres$, becomes quite large because
of the proliferation of localized topology-changing dislocations in the gauge field.  This
leads to eigenstates of the five-dimensional transfer matrix with unit eigenvalue, mixing the 
left- and right-handed chiral modes~\cite{Furman:1994ky,Antonio:2008zz}.  Because
$\mres$ acts as an additive renormalization to the quark mass, a large $\mres$ 
makes it difficult to explore the transition region with a reasonably small pion mass.

In this work, we have used two different approaches to reduce the residual chiral symmetry
breaking.  The first is to choose a large value for the size of the fifth dimension, $L_s = 96$.
This is coupled with judicious choices for the input quark masses, $m_l$ and $m_s$ so that
the \textit{total} quark masses, \textit{i.e.}, $(m_l + \mres)$ and $(m_s + \mres)$
are fixed in lattice units.  (Throughout this paper we will express dimensional quantities in
lattice units unless physical units are explicitly specified.)  This results in pion masses of 
$m_\pi \approx 225-275~\textrm{MeV}$ in
the transition region.  However, because $\mres$ only falls linearly with $L_s$ in this 
regime $(m_\pi \sim 1/\sqrt{L_s})$, it is computationally very costly to perform calculations at small 
$m_\pi$ by simply increasing $L_s$ \cite{Antonio:2008zz}.

An alternative to increasing $L_s$ is to directly suppress the localized modes which are the
primary contribution to $\mres$ at coarse lattice spacings.  This is done by
augmenting our action with a ratio of determinants of the twisted-mass Wilson Dirac
operator.  This determinant ratio, which we call the ``Dislocation Suppressing Determinant Ratio'' (DSDR),
 suppresses those gauge field configurations which contribute most to the mixing between
left and right-handed walls.  This method is a further development of earlier applications of
the 4-d Wilson fermion determinant for this purpose with both domain wall and overlap fermions 
\cite{Vranas:1999rz,Vranas:2006zk,Fukaya:2006vs}. 

For both approaches with and without the DSDR method, we employ the Iwasaki gauge action \cite{Iwasaki:1983ck}
for the gauge links.  The Iwasaki gauge action has been used extensively in zero temperature calculations coupled 
with domain wall fermions \cite{Antonio:2006px,Allton:2007hx,Allton:2008pn,Aoki:2010dy}.  The RBC-UKQCD 
collaboration has also begun a large-scale study of zero temperature physics using the Iwasaki gauge action and the DSDR method.  Zero temperature results with the DSDR method have been presented in \cite{Renfrew:2009wu, Blum:2011ng, Kelly:2012uy}.

\subsection{Dislocation Suppressing Determinant Ratio}

To lowest order in $a^2$, the residual chiral symmetry breaking caused by the finite extent in the
fifth dimension acts as an additive renormalization to the bare quark mass. This additive renormalization
is known as the residual mass $\mres$.  At fixed bare coupling, the dependence of
$\mres$ on the extent of the fifth direction $L_s$ can be parameterized as \cite{Antonio:2008zz}:
\begin{equation}
\mres = c_1 \rho_H(\lambda_c)\frac{e^{-\lambda_c L_s}}{L_s} + c_2 \rho_H(0)\frac{1}{L_s},
\label{eq:mres}
\end{equation}
where $\rho_H(\lambda)$ represents the density of eigenmodes of the effective 4-d Hamiltonian
$\mathcal{H} = -\log(\mathcal{T})$, where $\mathcal{T}$ is the transfer matrix in the fifth 
direction that controls the mixing of chiral modes between the 4-d boundaries.  The 4-d Hamiltonian,
$\mathcal{H}$ is closely related to the hermitian Wilson operator, $H_W = \gamma^5 D_W(-M_5)$, via
$\mathcal{H} = 2 \tanh^{-1}\left(H_W/(2+D_W)\right)$, and it has been shown that the zero
modes of $\mathcal{H}$ and $H_W$ coincide \cite{Furman:1994ky}.

The first term in Eq. \eqref{eq:mres} represents contributions from eigenmodes with 
eigenvalues $\lambda$ greater than the mobility edge, $\lambda_c$.  These modes
have extended 4-d support and their contributions to $\mres$ are
exponentially suppressed with $L_s$.
The second term corresponds to contributions from near zero eigenmodes of the 4-d Hamiltonian,
or equivalently eigenmodes where the 5-d transfer matrix $\mathcal{T}$ is near unity, thus
allowing nearly unsuppressed mixing of the domain walls in the fifth direction.  These near-zero
eigenmodes come largely from localized dislocations in the gauge field corresponding to 
topology change~\cite{Golterman:2003qe, Golterman:2004cy, Golterman:2005fe}.   At strong
coupling, gauge field dislocations rapidly become more common, so that the
dominant contribution to $\mres$ comes from the near-zero eigenmodes of 
$\mathcal{H}$ and the second, power-suppressed term in Eq.~\eqref{eq:mres}. 

 One method to reduce the large residual chiral symmetry breaking 
is to augment the gauge action with the determinant of the 4-d hermitian
Wilson Dirac operator, $H_W(-M_5) = \gamma^5 D_W(-M_5)$ 
\cite{Vranas:1999rz, Vranas:2006zk,Fukaya:2006vs}, where
$M_5$ is the domain wall height ($M_5 = 1.8$ in our calculation).  
Including this determinant as a factor in the path integral 
explicitly suppresses those configurations which have a small eigenvalue of
$H_W$, and thus also those configurations with near-zero modes of $\mathcal{H}$.

Unfortunately, the suppression of the zero modes of $H_W$ also 
suppresses exactly those configurations that change topology during a molecular dynamics
evolution.  Therefore, in order to allow for the correct sampling of all
topological sectors, we augment the Wilson Dirac operator with a chirally 
twisted mass,

\begin{equation}
D_W(-M_5) \rightarrow D_W(-M_5 + i \epsilon \gamma^5) \; .
\end{equation}

We then employ the following weighting factor on the gauge fields:

\begin{eqnarray}
\label{eqn:DSDR}
\mathcal{W}(M_5, \epsilon_b, \epsilon_f) & = & \frac{\det\left[D^\dagger_W(-M_5 + i \epsilon_f \gamma^5)D_W(-M_5 + i \epsilon_f \gamma^5)\right]}{\det\left[D^\dagger_W(-M_5 + i \epsilon_b \gamma^5)D_W(-M_5 + i \epsilon_b \gamma^5)\right]}\\
& = & \frac{\det\left[D^\dagger_W(-M_5)D_W(-M_5) + \epsilon^2_f\right]}{\det\left[D^\dagger_W(-M_5)D_W(-M_5) + \epsilon^2_b\right]}\nonumber.
\end{eqnarray}

The bosonic and fermionic ``twisted-mass'' parameters $\epsilon_b, \epsilon_f$ can be tuned so that 
gauge field topology changes during HMC evolution, but the localized dislocations which 
contribute to the residual mass are suppressed.  We call the weighting factor $\mathcal{W}(M_5, \epsilon_b, \epsilon_f)$
the Dislocation Suppressing Determinant Ratio (DSDR).
Employing this ratio of determinants ensures that the ultraviolet modes of the theory
are minimally affected so that bare parameters such as
$\beta$ and the quark masses do not shift significantly compared to those used with the standard 
domain wall fermion action.

\subsection{Lattice Ensembles}

\subsubsection{$L_s = 96$ ensembles}

The finite temperature ensembles that we generated with $L_s = 96$ all have spatial 
volume of $16^3$ and temporal extent $N_t = 8$.  We generated nine different lattice
ensembles for temperatures in the range $T \in [137,198]$ MeV.  
The bare couplings $\beta \in [1.965,2.10]$ span approximately the same range used in a
previous study of the transition region with domain wall fermions with $L_s = 32$ by the 
RBC-Bielefeld Collaboration \cite{Cheng:2009be}.  Since the only change in the lattice action
on these ensembles is the choice of the size of the fifth dimension, to leading order this mainly
affects residual chiral symmetry breaking and has a minimal affect on the 
bare coupling and the lattice cut-off.
We therefore use the same interpolation as in \cite{Cheng:2009be} to determine the temperatures of
each of our lattice ensembles.  

The input light and strange quark masses, $m_l$ and $m_s$ are chosen so that
the total quark masses, including the contributions from the residual mass, are given
by $m_l + \mres = 0.00675$ and $m_s + \mres = 0.045$.  However,
these quark masses are not along a line of constant physics.  At $\beta = 2.025$, we can
directly compare our quark masses with the determination of $m_\pi$ in \cite{Cheng:2009be}. 
Our choice gives $m_\pi \approx 250~\textrm{MeV}$.  The choice of a fixed bare light
quark mass implies that $m_\pi$ in physical units will vary across the set of bare couplings
that we use.  The change in temperature from $\beta = 2.025$ to the extremal points
in our range suggests a $10\%$ variation for $m_\pi$ in either direction.  This
gives a range of $m_\pi \in [225, 275]$ MeV, with $m_\pi$ being heavier at higher temperatures. 

Table \ref{tab:Ls96} gives the details for these ensembles.

\begin{table}[hbt]
\begin{center}
\begin{tabular}{cccccc}
$T$(MeV) & $\beta$  & $m_l$ & $m_s$ & $\mres$ & Traj. \\
\hline
137 & 1.965  & 0.00045 & 0.0387 & 0.0063 & 1720\\
146 & 1.9875 & 0.00245 & 0.0407 & 0.0043 & 1640\\
151 & 2.00 & 0.00325 & 0.0415 & 0.0035 & 1540\\
156 & 2.0125 & 0.00395 & 0.0422 & 0.0028 & 1465\\
162 & 2.025 & 0.00435 & 0.0426 & 0.0024 & 1835\\
167 & 2.0375 & 0.00485 & 0.0431 & 0.0019 & 1690\\
173 & 2.05 & 0.00525 & 0.0435 & 0.0015 & 1570\\
188 & 2.08 & 0.00585 & 0.0441 & 0.0009 & 2006\\
198 & 2.10 & 0.00585  & 0.0441 & 0.0006 & 1648\\
\hline
\end{tabular}
\caption{Summary of the $16^3 \times 8$, $L_s=96$ finite temperature ensembles without DSDR.  The total molecular dynamics time per trajectory is $\tau = 0.5$. Quark masses were chosen so that the $m_l + \mres \approx 0.00675$ and $m_s + \mres \approx 0.045$.  Residual masses are estimated from those reported in Ref.~\cite{Cheng:2009be} assuming $\mres \sim 1/L_s$ scaling.  Note here and in the following all dimensional quantities are expressed in lattice units unless other physical units are specified.}
\label{tab:Ls96}
\end{center}
\end{table}

\begin{table}[hbt]
\begin{center}
\begin{tabular}{cccccccccccc}
\multicolumn{12}{c}{Finite Temperature Ensembles}\\
\hline
Label &$T$ (MeV)   &$\beta$&$\Ns$&$\Nt$&$L_s$&$m_l$&$m_s$& $\mres$ &$m_\pi$ (MeV) &Traj. & $\left<U_{\scriptscriptstyle\Box}\right>$\\
\hline
\runlabel{140} & 139(6)&1.633&16&8&48&-0.00136&0.0519& 0.00588(39) &191(7) &2996 & 0.46913(8)\\
\runlabel{150_32} & 149(5)&1.671&16&8&32&-0.00189&0.0464& 0.00643(9) & 199(5) & 6000 & 0.48491(3)\\
\runlabel{150_48} & 149(5)&1.671&16&8&48&0.00173 &0.0500& 0.00295(3) &202(5) &7000 & 0.48407(2)\\
\runlabel{160} & 159(4)&1.707&16&8&32&0.000551&0.0449& 0.00377(11) &202(3) &3659 & 0.49777(4)\\
\runlabel{170} & 168(4)&1.740&16&8&32&0.00175 &0.0427& 0.00209(9) &197(2) & 3343 & 0.50912(4)\\
\runlabel{180} & 177(4)&1.771&16&8&32&0.00232 &0.0403& 0.00132(6) &198(2) & 3540 & 0.51916(4)\\
\runlabel{190} & 186(5)&1.801&16&8&32&0.00258 &0.0379& 0.00076(3) &195(3) & 4715 & 0.52845(3)\\
\runlabel{200} & 195(6)&1.829&16&8&32&0.00265 &0.0357& 0.00047(1) &194(4) & 6991 & 0.53672(3)\\
\hline
\multicolumn{12}{c}{Zero Temperature Ensembles}\\
\hline
\runlabel{1.70ml0.013} & - & 1.70 & 16 & 32 & 32 & 0.013 & 0.047 & 0.00420(2) &394(9) & 1360 & 0.49510(3)\\
\runlabel{1.70ml0.006} & - & 1.70 & 16 & 32 & 32 & 0.006 & 0.047 & 0.00408(6) &303(7) & 1200 & 0.49509(3)\\
\runlabel{1.75ml0.006} & - & 1.75 & 16 & 16 & 32 & 0.006 & 0.037 & 0.00188 & - &1255 & 0.51222(3)\\
\runlabel{1.75ml0.0042} & - & 1.75$^*$ & 32 & 64 & 32 & 0.0042 & 0.045 & 0.00180(5) &246(5) & 1288 & 0.512203(7)\\
\runlabel{1.75ml0.001} & - & 1.75$^*$ & 32 & 64 & 32 & 0.001 & 0.045 & 0.00180(5) &172(4) & 1560 & 0.512235(7)\\
\runlabel{1.82ml0.013} & - & 1.82 & 16 & 32 & 32 & 0.013 & 0.040 & 0.00062(2) &398(9) & 2235 & 0.53384(1)\\
\runlabel{1.82ml0.007} & - & 1.82 & 16 & 32 & 32 & 0.007 & 0.040 & 0.00063(2) &304(7) & 2134 & 0.53386(2)\\
\hline
\end{tabular}
\end{center}
\caption{Summary of zero and finite temperature ensembles with DSDR. Each lattice ensemble is given a 
label for later reference. The total molecular dynamics time
per trajectory is $\tau = 1.0$.  The residual mass, $\mres$ and the average plaquette ($\left<U_{\scriptscriptstyle\Box}\right>$) are also tabulated.\\
$^*$The values given for $\beta = 1.75$ are zero temperature results from RBC-UKQCD~\protect{\cite{Blum:2011ng, Kelly:2012uy}}.}
\label{tab:para}
\end{table}

\subsubsection{DSDR ensembles}

For the gauge action augmented with DSDR, we generated several ensembles at
zero temperature ($\Nt = 32,~\Ns = 16$) in order to determine the bare couplings and
quark masses appropriate for exploring the transition region at $\Nt = 8$.  For
the twisted mass coefficients in the determinant ratio, we found that the choice
of $\epsilon_f = 0.02$ and $\epsilon_b = 0.5$ allows for a reasonable rate of 
tunneling between topological sectors while still suppressing residual chiral symmetry breaking \cite{Renfrew:2009wu}.
At two values of the coupling, $\beta = 1.70$ and $1.82$ we generated ensembles 
with two different quark masses, corresponding to $m_\pi \approx 300, 400$ MeV respectively.

We have also used preliminary results from the RBC-UKQCD calculation with $\Ns = 32,~\Nt=64$
at $\beta = 1.75$ to provide a better interpolation for the bare parameters of our
finite temperature ensembles.

At finite temperature, we produced ensembles at seven different temperatures in
the range $139~ \textrm{MeV} \le T \le 195 ~\textrm{MeV}$ with $\Nt = 8$ and spatial
extent $\Ns = 16$.  The quark masses are chosen so that
the physical pion masses are fixed, $m_\pi \approx 200~\textrm{MeV}$, while the strange
quark mass, $m_s$, is chosen so that $(m_l + \mres)/(m_s + \mres) = 0.088$, close to its
physical value.  Table \ref{tab:para} summarizes the parameters for both our finite and 
zero temperature ensembles.  Appendix~\ref{sec:RHMC_evolution} gives the details
of the various evolution algorithms used to generate these ensembles.

Except for the $T = 139, ~149~\textrm{MeV}$ ensembles, we use $L_s = 32$
for the extent of the fifth dimension.  Because of
the rapid growth of the residual mass as one moves to stronger coupling,
the use of a negative input light quark mass becomes necessary at the lowest 
temperatures so that the
total light quark mass $m_\textrm{tot} = m_l + \mres$ corresponds to a
fixed physical pion mass, $m_\pi \approx 200~\textrm{MeV}$.

In principle, the presence of a negative quark mass admits the possibility for a singular
fermion matrix, resulting in ``exceptional configurations'' that destroy the reliability of
the calculation.  However, the residual chiral symmetry breaking in our calculation
produces a dynamically generated mass, $\mres$ that additively renormalizes our
quark masses, theoretically moving one away from any singularities in the fermion 
matrix.  Of course, $\mres$ is only well-defined when one considers an ensemble average, so if one
uses a negative quark mass that is too large, \textit{i.e.}, $|m_l| \sim \mres$,
fluctuations in the gauge configurations may induce the unwanted singularities even if $m_\textrm{tot} > 0$.
  
For $T = 139~\textrm{MeV}$, we initially used a negative light quark mass
of $m_l = -0.00786$, with $\mres \approx 0.013$ at $L_s = 32$.  It was quickly discovered
that this resulted in a singular fermion matrix, signaled by the non-convergence of the conjugate
gradient inversion.  As a result, we switched to $L_s = 48$ at this temperature, where a smaller, but still
negative light quark $m_l  = -0.00136$ could be used to achieve the desired
total light quark mass.  At $L_s = 48$, we saw no exceptional configurations in our ensemble.

At $T = 149~\textrm{MeV}$ we produced configurations at both $L_s = 32$ and $L_s = 48$ in order
to verify that the use of a negative input quark mass had no effect on physical observables, beyond
small $O(a^2)$ effects.  With $L_s = 32$, 
a negative input quark mass, $m_l  = -0.00189$, is used, while at $L_s = 48$, we have
$m_l  = 0.00173$.   Both of these ensembles (ensembles 2 and 3 in Tab.~\ref{tab:para}) correspond
to approximately the same physical pion mass, $m_\pi \approx 200~\textrm{MeV}$.  We did not see any
large differences between these two ensembles in quantities such as the disconnected chiral
susceptibility, renormalization coefficients, or eigenvalue spectrum.  However, in the chiral condensate
we did see a significant difference in the two ensembles, presumably caused by the difference in
the leading-order ultraviolet divergent $m_l/a^2$ term that enters in the calculation of the chiral
condensate on the lattice.  Table~\ref{tab:para} also shows a 0.2\% difference in the average 
plaquette value, as we should expect from the small change in the fermion determinant caused by the 
increase in $L_s$ from 32 to 48.  (Recall that the ratio of  the physical fermion to Pauli-Villars 
DWF determinants should have an $L_s\to\infty$ limit.)

\subsection{Line of constant physics}

As discussed in the preceding subsection, the $L_s = 96$ ensembles do not lie on a line of
constant physics, but rather a line of constant bare quark mass.  This results in the
pion mass changing from $m_\pi \approx 225$ MeV at the lowest temperature in our ensemble
to $m_\pi \approx 275$ MeV at the highest temperature. 

For the DSDR ensembles, we have endeavored to move along
a line of fixed physical pion mass, $m_\pi = 200$ MeV.
Table~\ref{tab:zeroT} summarizes our results for $m_\pi , m_\rho,$ and $r_0$ on
the zero temperature ensembles. 

\begin{table}[hbt]
\begin{center}
\begin{tabular}{ccccccc}
Label & $\beta$ & $m_l$ & $r_0$ &$m_\rho$ & $m_\pi$ &$1/a^\dagger$ (GeV)\\
\hline
\runn{1.70ml0.013} & 1.70 & 0.013 & 2.895(11) & 0.68(2) & 0.310(1) & - \\
\runn{1.70ml0.006} & & 0.006 & 2.992(27) & 0.67(2) & 0.238(1) & -\\
\multicolumn{2}{c}{Extrapolated} & -0.0040 & 3.13(7) & 0.66(6) & - & 1.27(4)\\
     \hline
\runn{1.75ml0.0042} & 1.75 & 0.0042 & 3.349(20) & 0.57(2) & 0.1810(3) & -\\
\runn{1.75ml0.001}  & & 0.0010  & 3.356(22) & 0.56(2) & 0.1264(3) & -\\
\multicolumn{2}{c}{Extrapolated} & -0.0018 & 3.36(4) & 0.56(4)& - &1.36(3)\\
\hline
\runn{1.82ml0.013} & 1.82 & 0.013 & 3.743(28) & 0.56(2) & 0.255(2) & -\\
\runn{1.82ml0.007} &     & 0.007 & 3.779(37) & 0.53(2) & 0.195(2) & - \\
\multicolumn{2}{c}{Extrapolated}& -0.00064 & 3.83(9) & 0.49(5) & - & 1.55(5)\\
     \hline
\end{tabular}
\end{center}
\caption{\label{tab:zeroT}
Results for $r_0$, $m_\rho$, $m_\pi$, and the lattice scale, $a^{-1}$.
At each value of $\beta$, we perform simple linear extrapolations to $m_l = -\mres$, \textit{i.e.}, the
chiral limit, for $r_0$ and $m_\rho$.  The lattice scale is fixed using the extrapolated
value for $r_0$. $^\dagger$Lattice scale determined using $r_0 = 0.487(9)$ fm.}
\end{table}

In order to determine the lattice scale, we have used the Sommer parameter $r_0$,  
determined from the static quark potential.  The quantity $r_0$, extrapolated to the chiral limit, can
be related to the lattice scale using its physical value $r_0 = 0.487(9)$ fm, determined
using domain wall fermions \cite{Aoki:2010dy}.  The temperature is given by
$T = 1/\Nt a$.  The values for $r_0/a$ in Tab.~\ref{tab:zeroT} allow us to determine the bare couplings
needed for finite temperature lattice ensembles in the transition region.

\begin{figure}[hbt]
\begin{center}
\includegraphics[width=0.45\textwidth]{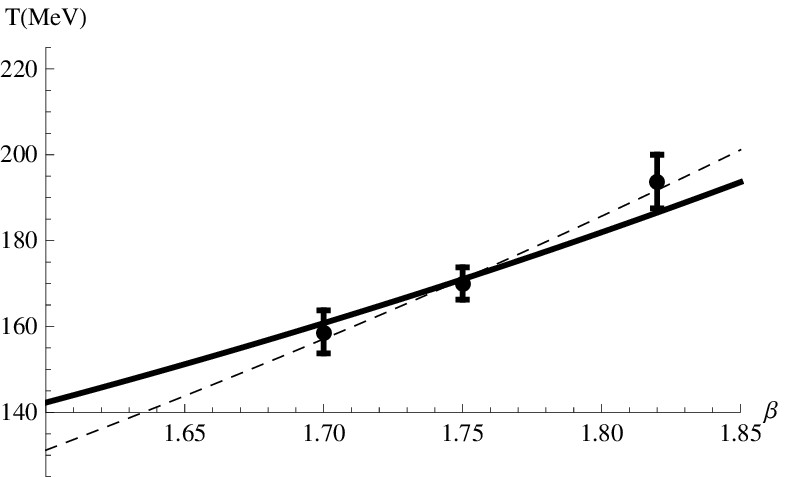}
\hspace{0.07\textwidth}
\includegraphics[width=0.45\textwidth]{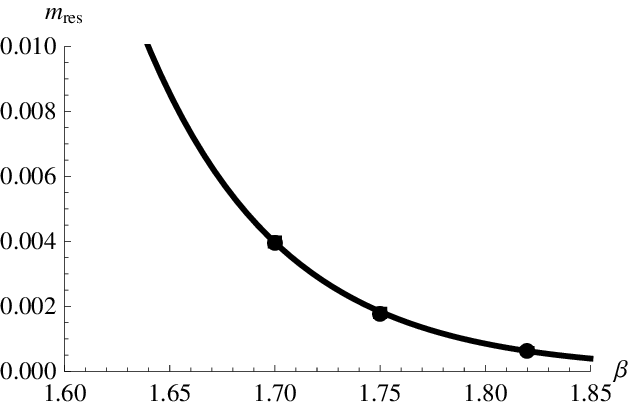}
\caption{Left panel: temperature for $\Nt=8$ is plotted versus $\beta$.  The solid curve is the fit
to the continuum RG running; $c_0 = 25.2(3)~\textrm{MeV}$.  The dashed curve is the result of the fit to 
Eq.~\eqref{eqn:RGfit}  which includes an added $a^2$ correction; $c_0 = 29.7(2.9)~\textrm{MeV},~c_1 
= -204(132)~\textrm{MeV}$.  Right panel: $\mres a$ is plotted versus $\beta$ with an exponential fit: 
$\mres(\beta) = A \exp\left( - B \beta\right)$; $A = 8.7(9.7) \times 10^8, B = 15.4(6)$.}
\label{fig:exponential}
\end{center}
\end{figure}

To describe $T(\beta)$ in physical units, we use a modified form of the two-loop renormalization 
group running, which includes an extra term for the $\mathcal{O}(a^2)$ lattice artifacts:
\begin{eqnarray}
\label{eqn:RGfit}
T(\beta) & = &\frac{1}{\Nt a(\beta)} = \left(c_0 + c_1 \hat{a}^2(\beta)\right)\frac{1}{\hat{a}(\beta)}\\ 
\hat{a}(\beta) & = & \exp\left(-\frac{\beta}{12 b_0}\right) \left(\frac{6 b_0}{\beta}\right)^{-b_1/(2 b_0^2)};~b_0 = \frac{9}{(4 \pi)^2};~b_1 = \frac{64}{(4 \pi)^4},
\end{eqnarray}
where $\hat{a}(\beta)$ is the continuum two-loop RG running for the lattice spacing.
The left panel of Fig.~\ref{fig:exponential} shows the result of the fit of the $\beta$-dependence
of the temperature to both the lattice-corrected RG fit of Eq. \eqref{eqn:RGfit}, and to the
continuum RG running, \textit{i.e.}, the case where $c_1 = 0$.  As can be seen, the lattice-corrected
fit provides a better description of the data.

The zero temperature ensembles show that the residual mass is strongly dependent on
the lattice spacing.  At coarser lattice spacings, the aforementioned dislocations are more
common and cause $\mres$ to increase rapidly as one moves from high to low
temperature.  The right panel of Fig.~\ref{fig:exponential} shows $\mres$ as a function of 
$\beta$.  We find that a simple exponential Ansatz describes the data well.

Finally, to ensure that we simulate along a line of fixed pion mass, we must account for the
running of the bare quark masses as the bare coupling is changed.  Since the residual
chiral symmetry breaking results in an additive shift in the quark mass, to leading order
in chiral perturbation theory, the pion mass depends on the total quark mass, 
$m_\textrm{tot} = m_l + \mres$, as:
\begin{displaymath}
m_\pi^2 \propto (m_l + \mres).
\end{displaymath}
This linear quark mass dependence is a surprisingly good description of earlier 
data~\cite{Aoki:2010dy} and sufficiently accurate for the present purpose.

\begin{figure}[hbt]
\begin{center}
\includegraphics[width=0.50\textwidth]{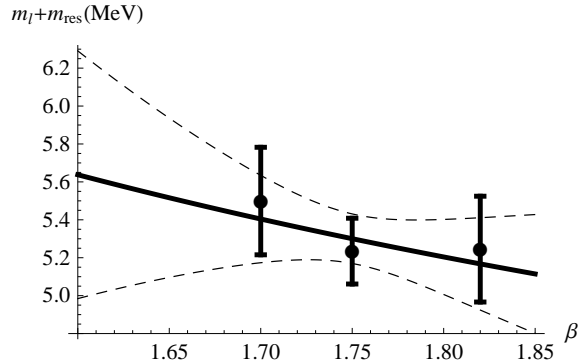}
\caption{Total light quark mass for $m_\pi = 200$ MeV line of constant physics, with a fit to the lattice-corrected mass anomalous dimension.  Dashed curves represent the 1-$\sigma$ error band.}
\label{fig:mRG}
\end{center}
\end{figure}

This allows us to determine the bare quark masses required for 
a specific line of constant physics on the zero temperature 
ensembles listed in Tab.~\ref{tab:zeroT}.  Figure~\ref{fig:mRG} 
shows the quark masses required for 
$m_\pi = 200$ MeV. We also fit these
results for $m_\textrm{tot}(\beta)$ to the lattice-corrected
two-loop running of the mass anomalous dimension:
\begin{equation}
m_\textrm{tot} \equiv (m_l + \mres) = \left(A + B\hat{a}^2(\beta)\right) \left(\frac{12 b_0}{\beta}\right)^{4/9} 
\label{eqn:mq}
\end{equation}
The lattice-corrected fit provides a good interpolation that
allows us to achieve a line of constant physics on the finite
temperature ensembles.
\section{Determining the Dirac Eigenvalue Spectrum}
\label{sec:dirac}

The spectrum of eigenvalues of the hermitian Dirac operator provides important
insight into the physics of QCD.  The Dirac spectrum depends dramatically
on the temperature and is fundamentally connected with both spontaneous and 
anomalous chiral symmetry breaking.  These topics will be explored in detail in 
later sections of this paper.

In this section we will explain how the continuum Dirac spectrum can be 
determined from the spectrum of the five-dimensional DWF Dirac operator,
including a method to determine its normalization.  The Ritz method used
to determine the lowest 100 eigenvalues for each of our finite temperature
ensembles will then be briefly described as well as the numerical details of 
our determination of the normalization of those eigenvalues.  A derivation for 
this normalization method, following the approach of Giusti and 
L\"uscher~\cite{Giusti:2008vb}, is given in Appendix~\ref{sec:spectral_normalization}.   
The resulting Dirac eigenvalue spectrum, computed and normalized following 
the methods described in this section, will be presented and analyzed in 
Sec.~\ref{sec:ua1}, in an effort to determine the temperature dependence and 
the origin of anomalous $U(1)_A$ symmetry breaking.

\subsection{Relating the continuum and DWF Dirac spectrum}

The domain wall fermion formulation can be viewed as a five-dimensional
theory whose low energy properties accurately reproduce four-dimensional 
QCD.  All low energy Green's functions and matrix elements are expected to
agree with those of a four-dimensional theory and it is only at high momenta 
or short distances that the five dimensional character of the theory 
becomes visible.  This perspective applies also to the five-dimensional 
DWF Dirac operator whose small eigenvalues and corresponding eigenstates
should closely approximate those of a continuum four-dimensional theory.
This can be shown explicitly for the free theory, order-by-order in
perturbation theory and by direct numerical evaluation in lattice QCD.
With the exception of gauge configurations which represent changing
topology, the modes with small eigenvalues are literally four-dimensional
with support concentrated on the four-dimensional left and right
walls of the original five-dimensional space.  

Thus, we can learn about the continuum Dirac eigenvalue spectrum by
directly studying that of the DWF Dirac operator, $ D^{\rm DWF}$, as
defined by Eqs. 1-3 in Ref.~\cite{Blum:2000kn}.  Of course, just 
as with other regulated versions of the continuum theory, explicit 
renormalization is needed to convert from a bare to a renormalized 
eigenvalue density.  Because the continuum Dirac operator, $\slashed{D}+m$, 
is linear in the quark mass, we should expect the Dirac eigenvalues to 
be related between different renormalization schemes by the same
factor $Z_m$ that connects the masses.  If we have two regularized
theories which describe the same long distance physics with bare
masses $m$ and $m'=Z_{m\to m'} m$, then we should expect that their 
eigenvalue densities would be related by:
\begin{equation}
  \rho'(\lambda') = \frac{1}{Z_{m\to m'}}
  \rho\left(\lambda'/Z_{m\to m'}\right).
\label{eq:spectral_renorm}
\end{equation}
Note this expectation is consistent with the form of the Banks-Casher
relation, $\langle \overline{\psi}\psi\rangle = \pi\rho(0)$, as the equality of 
the mass term in equivalent theories requires 
$\langle\overline{\psi}'\psi'\rangle = \langle\overline{\psi}\psi\rangle/Z_{m\to m'}$.

The renormalization of the bare input quark mass, $m_f$, for DWF has 
been extensively studied and the factor $Z_{m_f\to\overline{\rm MS}}(\mu^2)$ 
needed to convert this input bare mass to a continuum, $\overline{\rm MS}$ 
value at the scale $\mu$ is accurately known~\cite{Aoki:2010dy}.  However,
in contrast to the continuum theory or staggered or Wilson lattice fermions, 
the input quark mass for DWF does not enter as an additive constant but
instead appears as a coupling strength between the two four-dimensional
walls.  Thus, for DWF the Dirac spectrum and the quark mass will in general
be related to their continuum counterparts by {\it different} renormalization
factors.  To properly renormalize the DWF Dirac spectrum we should begin
with the hermitian operator $\gamma^5 R_5 D^{\rm DWF}$ and then add a 
multiple of the identity:
\begin{equation}
  \gamma^5 R_5 D^{\rm DWF} + m_{\rm tw}
  = \gamma^5 R_5 \left(D^{\rm DWF} +\gamma^5 R_5 m_{\rm tw}\right).
  \label{eq:tw_cont_norm}
\end{equation}
Here $R_5$ performs a simple reflection in the fifth dimension, taking the 
point $(x,s)$ to the point $(x,L_s-1-s)$ where $x$ is the space-time 
coordinate and $0\le s \le L_s-1$ the coordinate in the fifth dimension.  
The renormalization factor, $Z_{{\rm tw}\to\overline{\rm MS}}$, needed
to convert the DWF spectrum to the continuum, $\overline{\rm MS}$
spectrum then relates this new DWF pseudo-scalar operator to the 
corresponding $\overline{\rm MS}$ continuum operator:
\begin{equation}
  \left(\overline{\psi}(x) \gamma^5 \psi(x)\right)^{\overline{\rm MS}} 
  \approx \frac{1}{ Z_{{\rm tw}\to\overline{\rm MS}} }
  \sum_{s=0}^{L_s-1}\overline{\Psi}(x,s)\gamma^5\Psi(x,L_s-1-s),
  \label{eq:ps_equiv}
\end{equation}
where $\Psi(x,s)$ is the five-dimensional DWF field.  These two operators, 
which appear in different theories, are equated in Eq.~\eqref{eq:ps_equiv} in 
the sense that they give the same matrix elements when inserted in 
corresponding long-distance Green's functions.  

It is convenient to determine the renormalization constant 
$Z_{{\rm tw}\to\overline{\rm MS}}$ in two steps.  In the first we determine
the constant $Z_{{\rm tw} \to m_f}$ which relates this reflected 
pseudo-scalar term and the standard pseudo-scalar term belonging to the same
chiral representation as the usual DWF mass term $\overline{\psi}\psi$:
\begin{equation}
  \overline{\psi}(x)\gamma^5\psi(x) = \frac{1}{Z_{{\rm tw} \to m_f}}
  \overline{\Psi}(x)R_5\gamma^5\Psi(x),
\end{equation}
where the operator on the right-hand side is the same as that in the 
right-hand side of Eq.~\eqref{eq:ps_equiv} with the explicit sum over 
the $s$ coordinate suppressed.

Then in the second step we perform the well-understood conversion between
the standard DWF mass operator and a continuum, $\overline{\rm MS}$ 
normalized mass operator using $Z_{m_f\to\overline{\rm MS}}$:
\begin{equation}
 Z_{{\rm tw}\to\overline{\rm MS}} 
  = Z_{m_f\to\overline{\rm MS}} Z_{{\rm tw} \to m_f}.
\end{equation}

After the first step, we can compare the eigenvalue density $\rho(\lambda)$ 
for the lattice DWF operator with the usual lattice result for the chiral 
condensate using the Banks-Casher relation,
\begin{equation}
  \langle\overline{\psi}\psi\rangle = \frac{\pi}{Z_{{\rm tw} \to m_f}}\rho(0),
  \label{eq:Banks_Casher_lat}
\end{equation}
since both the left- and right-hand sides now use the same bare normalization
conventions.  In the second step we are simply dividing both sides of 
Eq.~\eqref{eq:Banks_Casher_lat} by the common factor $Z_{m_f\to\overline{\rm MS}}$ 
to convert from lattice to $\overline{\rm MS}$ normalization.

\subsection{Calculation of \boldmath{$Z_{{\rm tw} \to m_f}$}}

Because the operators $\overline{\psi}(x)\gamma^5\psi(x)$ and 
$\overline{\Psi}(x)R_5\gamma^5\Psi(x)/Z_{{\rm tw} \to m_f}$ are supposed to
be equivalent at long distances, we can determine the needed
factor $Z_{{\rm tw} \to m_f}$ by simply taking the ratio of
equivalent Green's functions, evaluated at distances greater than the lattice
spacing $a$, containing these two operators:
\begin{equation}
  Z_{{\rm tw} \to m_f}
  = \frac{\left\langle O_1\ldots O_n \overline{\Psi}(x)R_5\gamma^5\Psi(x)\right\rangle}
  {\left\langle O_1\ldots O_n \overline{\psi}(x)\gamma^5\psi(x)\right\rangle}\; ,
  \label{eq:univ_ratio}
\end{equation}
where the numerator and denominator in this expression are 
intended to represent identical Green's functions except for
the choice of pseudo-scalar vertex.  

We will now determine $Z_{{\rm tw} \to m_f}$ and test the 
accuracy to which the ratio given in Eq.~\eqref{eq:univ_ratio} 
defines a unique constant by studying the ratio of two type 
of matrix elements.  In the first we examine simple two-point 
correlators between each of the pseudo-scalar densities in 
Eq.~\eqref{eq:univ_ratio} and the operator $O_\pi(t)$ which 
creates a pion from a Coulomb gauge fixed wall source located 
at the time $t$:
\begin{equation}
  {\cal R}_{\pi}(t)
  = \frac{\left\langle \sum_{\vec x} 
  \overline{\Psi}(\vec x,t)R_5\gamma^5\Psi(\vec x,t )O_\pi(0)\right\rangle}
  {\left\langle \sum_{\vec x}
  \overline{\psi}(\vec x,t)\gamma^5\psi(\vec x,t)O_\pi(0)\right\rangle},
  \label{eq:univ_ratio_pi}
\end{equation}
which for large $t$ is the ratio of matrix elements of our two 
pseudo-scalar operators between a pion state and the vacuum.
Results are presented in Tab.~\ref{tab:ratio_pi}.

\begin{table}[t]
  \begin{center}
  \begin{tabular}{cccc}
    Label &$\beta$& T(MeV)&${\cal R}_{\pi}$\\
    \hline
    \runn{1.70ml0.006}& 1.70  & 0 &1.774(5) \\
    \runn{1.75ml0.006} &1.75  & 0 &1.570(4) \\
    \runn{1.82ml0.007} & 1.82 & 0 &1.397(2) \\
	\hline
    \runn{150_32} & 1.671 & 149 &1.905(6) \\
    \runn{150_48} & 1.671 & 149 &1.980(7) \\
    \runn{160}       & 1.707 & 159 &1.725(8) \\
    \runn{170}       & 1.740 & 168 &1.631(11)\\
    \runn{180}       & 1.771 & 177 &1.476(4) \\
    \runn{190}       & 1.801 & 186 &1.439(3) \\
    \runn{200}       & 1.829 & 195 &1.365(3) \\
\hline
\end{tabular}
\end{center}
 \caption{Values for the renormalization factor $Z_{{\rm tw}\to m_f}$ obtained
from the ratio of pseudo-scalar correlators ${\cal R}_{\pi}$ defined in 
Eq.~\eqref{eq:univ_ratio_pi}.}
\label{tab:ratio_pi}
\end{table}

Second we examine off-shell, three-point Green's functions evaluated in 
Landau gauge which again contain each of the pseudo-scalar densities being 
compared and a quark and an anti-quark field carrying momenta $p_1$ and 
$p_2$, allowing us to see the degree to which the ratio in Eq.~\eqref{eq:univ_ratio} 
does not depend on the small external momenta $p_1$ and $p_2$.
\begin{equation}
  {\cal R}_{\rm MOM}(p_1,p_2)
  = \frac{ {\rm Tr}\left\langle \sum_{x_2,x_1} e^{i(p_2x_2-p_1x_1)} 
  \psi(x_2)\overline{\Psi}(0)R_5\gamma^5\Psi(0)
  \overline{\psi}(x_1)\right\rangle}
  { {\rm Tr}\left\langle \sum_{x_1,x_2} e^{i(p_2x_2-p_1x_1)}
  \psi(x_2)\overline{\psi}(0)\gamma^5\psi(0),
  \overline{\psi}(x_1)\right\rangle}.
  \label{eq:univ_ratio_MOM}
\end{equation}
Here we are using the well-studied methods of Rome/Southampton 
non-perturbative renormalization~\cite{Martinelli:1995ty} to compare the normalizations 
of the operators $\overline{\Psi}R_5\gamma^5\Psi$ and 
$\overline{\psi}\gamma^5\psi$.  For a recent application of this method
to other operators in a DWF context see Ref.~\cite{Aoki:2007xm}.
For both Eqs.~\eqref{eq:univ_ratio_pi} and \eqref{eq:univ_ratio_MOM}, we
expect the ratio to be independent of $t$ and of $p_1$ and $p_2$ 
respectively and to yield the same value $Z_{{\rm tw} \to m_f}$.

\begin{figure}[htb]
  \begin{center}
   \input{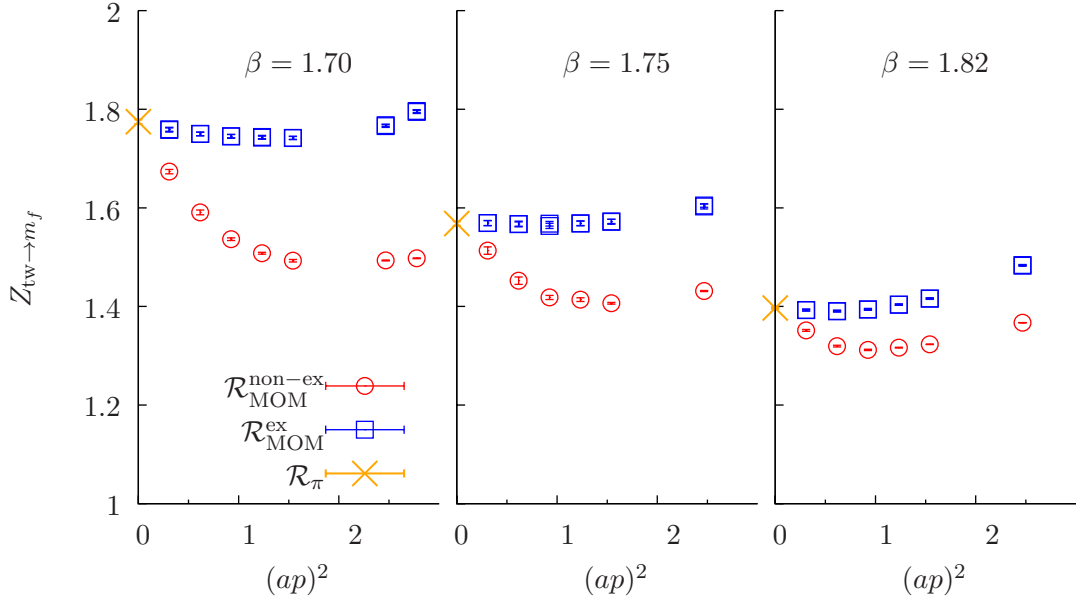}
  \end{center}
 \caption{Plots of the results for the quantity  $Z_{{\rm tw} \to m_f}$ given in 
Tabs.~\ref{tab:ratio_pi} and \ref{tab:npr} for each of the three values of $\beta$
that were studied at zero temperature.  The single value of  ${\cal R}_{\pi}$ is 
plotted as an`` $\times$'' in each panel and given the value $(pa)^2=0$.  (The
scale on the left-most $y$-axis applies to all three plots.)  As discussed in the text,
the discrepancies between ${\cal R^\textrm{non-ex}_\textrm{MOM}}$ and 
${\cal R^\textrm{non-ex}_\textrm{MOM}}$ are indicative of $O\left((pa)^2\right)$ errors, so we
use the value of ${\cal R_\pi}$ for $Z_{{\rm tw} \to m_f}$.}
\label{fig:npr}
\end{figure}

When evaluating the momentum space Green's functions in Eq.~\eqref{eq:univ_ratio_MOM}
we generate the needed quark propagators using a series of volume 
sources~\cite{Gockeler:1998ye}.  For each specific four-momentum $p$ we 
evaluate twelve propagators, one for each spin and color, using the sources
\begin{equation}
\eta(x,p)_{\alpha,a;\beta,b}=e^{ip\cdot x}\delta_{\alpha\beta}\delta_{ab},
\label{eq:vol_src}
\end{equation}
where $\alpha$ and $a$ are the spin and color indices of the source $\eta$
while $\beta$ and $b$ label the spins and colors of the twelve sources
evaluated for each four-momentum $p$.  We perform our calculation using both 
non-exceptional kinematics, $p_1^2=p_2^2=(p_1-p_2)^2$, and exceptional 
kinematics,  $p_1=p_2$.  Results for the ratios ${\cal R}_{\rm MOM}^{\rm non-ex}(p_1,p_2)$
and ${\cal R}_{\rm MOM}^{\rm ex}(p_1, p_2)$
for the three zero-temperature ensembles are presented in Tab.~\ref{tab:npr} and Fig.~\ref{fig:npr}.  
The specific momentum components used to construct $p_1$ and $p_2$ are listed in Tab.~\ref{tab:mom_comp}.

\begin{table}[htb]
\begin{center}
\begin{tabular}{ccc}
{$(pa)^2$} &$p_A L/2\pi$              &$p_B L/2\pi$ \\
\hline 
0.308        &(1,1,0,0)                     &(0,1,1,0)   \\
0.671        &(1,1,1,1)                     &(1,1,1,-1)  \\
0.925        &(2,1,1,0)                     &(2,0,-1,1)  \\
1.234        &(2,2,0,0)                     &(0,2,2,0)   \\
1.542        &(2,2,1,1)                     &(2,-1,2,1)  \\
2.467        &(2,2,2,2)                     &(2,2,2,-2)  \\
2.776        &(3,2,2,1)                     &(3,2,-1,-2) \\
\hline
\end{tabular}
\end{center}
  \caption{The components of the two momentum four-vectors $p_A$ and $p_B$ 
used to compute the quantities ${\cal R}_{\rm MOM}(p_1,p_2)$ given in 
Tab.~\ref{tab:npr}.  For non-exceptional momenta, we use $p_1 = p_A$ and
$p_2 = p_B$, while for exceptional momenta, only a single momentum,
either $p_1 = p_2 = p_A$ or $p_1 = p_2 = p_B$ is used. 
Here $L=16$ is the spatial size of the lattice.}
\label{tab:mom_comp}
\end{table}

\begin{table}[htb]
\begin{center}
  \begin{tabular}{c|cc|cc|cc}
~&\multicolumn{2}{c|}{$\beta=1.70$}&\multicolumn{2}{c|}{$\beta=1.75$}&\multicolumn{2}{c}{$\beta=1.82$}\\

        $(pa)^2$&${\cal R}_{\rm MOM}^{\rm non-ex}$&${\cal R}_{\rm MOM}^{\rm ex}$
        &${\cal R}_{\rm MOM}^{\rm non-ex}$&${\cal R}_{\rm MOM}^{\rm ex}$
        &${\cal R}_{\rm MOM}^{\rm non-ex}$&${\cal R}_{\rm MOM}^{\rm ex}$\\
    \hline
    0.308&1.673(5)  &1.759(4)&1.507(5) &1.566(4)&1.352(2) &1.393(2)\\
    0.617&1.591(5)  &1.745(4)&1.450(5) &1.562(4)&1.320(2) &1.390(2)\\
    0.925&1.536(3)  &1.745(3)&1.418(3) &1.562(4)&1.312(1) &1.394(2)\\
    1.234&1.508(2)  &1.744(3)&1.412(2) &1.564(4&1.3165(7)&1.404(1)\\
    1.542&1.493(2)  &1.742(3)&1.406(1) &1.570(4)&1.3233(6)&1.416(1)\\
    2.467&1.4933(10)&1.766(3)&1.4313(7)&1.613(3)&1.3670(4)&1.484(1)\\
    2.776&1.4977(8) &1.796(3)&-&-&-&- \\
	\hline
  \end{tabular}
\end{center}
  \caption{Values for the ratio ${\cal R}_{\rm MOM}(p_1,p_2)$ defined in 
Eq.~\eqref{eq:univ_ratio_MOM}.  For non-exceptional momenta, the quantity
${\cal R}_{\rm MOM}^{\rm non-ex}(p_1 = p_A, p_2 = p_B)$ is shown.  For exceptional 
momenta, the average of ${\cal R}_{\rm MOM}^{\rm non-ex}(p_1 = p_2 = p_A)$ and
${\cal R}_{\rm MOM}^{\rm non-ex}(p_1 = p_2 = p_B)$ is shown.  The first column
shows the value of $(p_1 a)^2 = (p_2 a)^2 = (p a)^2$.
Results from 12, 20 and 21 configurations have been averaged
to give the values for $\beta=1.70, 1.75$ and 1.82, respectively.  The quark mass 
values and lattice sizes used for these results are given in Tab.~\ref{tab:ratio_pi}.  
The significant variation among the results for a given value of $\beta$ indicate 
large $O\left((pa)^2\right)$ errors.}
\label{tab:npr}
\end{table}

The ratios presented in Tabs.~\ref{tab:ratio_pi} and \ref{tab:npr} and plotted in
Fig.~\ref{fig:npr} at a given value of $\beta$ are all expected to equal the common 
renormalization factor $Z_{{\rm tw} \to m_f}$.  However, as is evident from these 
tables and figure this expectation is realized at only the 20\% level, suggesting the 
presence of significant $O\left((pa)^2\right)$ errors and implying a similar uncertainty in extracting 
a consistent value for the important quantity $Z_{{\rm tw} \to m_f}$.  In fact, the
behavior of these results is consistent with an $O\left((pa)^2\right)$ origin for these discrepancies.
The larger dependence on momentum of the non-exceptional ratio 
${\cal R}^{\rm non-ex}_{\rm MOM}(p_1,p_2)$ than seen in 
${\cal R}^{\rm ex}_{\rm MOM}(p_1,p_2)$ and its larger deviation from the more 
consistent quantities ${\cal R}^{\rm ex}_{\rm MOM}(p_1,p_2)$ and  ${\cal R}_{\pi}$
is reasonable since the non-exceptional kinematics were originally introduced to
ensure that large momenta flow everywhere in the corresponding Green's 
function~\cite{Aoki:2007xm}.  The better agreement between the quantities
${\cal R}^{\rm ex}_{\rm MOM}(p_1,p_2)$ and  ${\cal R}_{\pi}$ and the smaller
momentum dependence of ${\cal R}^{\rm ex}_{\rm MOM}(p_1,p_2)$ is 
also consistent with the smaller internal momenta expected in these
Green's functions with exceptional kinematics.  Finally the decreasing differences
between these three quantities as $\beta$ increases from 1.70 to 1.82 with
the corresponding decrease in $a$ is also consistent with these violations
of universality arising from finite lattice spacing errors.

We therefore adopt the hypothesis that the discrepancies between these
different determinations of  $Z_{{\rm tw} \to m_f}$ arise from finite lattice
spacing effects and that the most reliable value for  $Z_{{\rm tw} \to m_f}$
will be obtained at smallest momentum.  Hence, we use the ratio
${\cal R}_{\pi}$ to provide values for $Z_{{\rm tw} \to m_f}$.  This choice has
the additional benefit that we have evaluated this ratio on the finite temperature
ensembles allowing us to use ${\cal R}_{\pi}$ to provide values of
$Z_{{\rm tw} \to m_f}$ for each of our values of $\beta$, avoiding extrapolation.
Note that the discrepancy between the finite and zero temperature results for
${\cal R}_{\pi}$ shown in Tab.~\ref{tab:ratio_pi} for the near-by $\beta$ values
$\beta = 1.700, 1.707$ and $\beta = 1.820, 1.829$ indicate remaining systematic
$a^2$ errors in our determination of  $Z_{{\rm tw} \to m_f}$ that are on the order 
of 5\%.

\subsection{Normalization conventions}

Using the methods described above, we can convert our results 
for the quark mass, chiral condensate, and Dirac spectrum into 
a single normalization scheme, allowing a meaningful comparison 
between the eigenvalues in the Dirac spectrum and the corresponding
quark mass.  We adopt the commonly-used $\overline{\rm MS}$ 
scheme, normalized at a scale $\mu=2$ GeV.  

We use the DWF results for the continuum, $\mu=2$ GeV, 
$\overline{\rm MS}$ quark masses determined in 
Ref.~\cite{Aoki:2010dy}, 
$m^{\overline{\rm MS}}_s(2\;{\rm GeV}) = (96.2 \pm 2.7)$MeV and 
$m^{\overline{\rm MS}}_{ud}(2\;{\rm GeV}) = (3.59\pm0.21)$MeV and
the accurate linear dependence of $m_\pi^2$ and $m_K^2$ on the quark 
masses in the region studied to convert a lattice light quark mass, 
$\widetilde{m}_l=m_f+m_{\rm res}$ corresponding to a pion mass 
$m_\pi(\widetilde{m}_l)$ into this same $\overline{\rm MS}$
scheme using the relation:
\begin{equation}
\label{eq:Zm}
m^{\overline{\rm MS}}_l(2{\rm GeV}) 
                 = (3.59 + 96.2) \mbox{MeV} 
               \frac{\Bigl(m_\pi(\widetilde{m}_l)\Bigr)^2}{2(m_K)^2},
\end{equation}
where $m_K = 495~\textrm{MeV}$ denotes the physical value of the Kaon mass.
The renormalization factor is then given by:
\begin{equation}
Z_{m_f \to \overline{\rm MS}} = \frac{99.79 ~\textrm{MeV}}{2\widetilde{m}} 
                             \left(\frac{m_\pi(\widetilde{m}_l)}{495 ~\textrm{MeV}}\right)^2
\label{eq:Z_mf_MSbar}
\end{equation}
for each of our ensembles.  Note the lattice quark mass, $\widetilde{m}$,
substituted in Eq.~\eqref{eq:Z_mf_MSbar} must be expressed in units of MeV to 
define a conventional, dimensionless value for $Z_{m_f \to \overline{\rm MS}}$. 
The resulting $Z_{m_f \to \overline{\rm MS}}$ factors for our seven ensembles 
are given in Tab.~\ref{tab:Z_MS_bar}.

\begin{table}[t]
\begin{center}
\begin{tabular}{ccc}
Label & $T$ (MeV)   & $Z_{m_f \to \overline{\rm MS}}(2 \mbox{GeV})$ \\ 
\hline
\runn{140} & 139  &  1.47(14) \\
\runn{150_48} & 149  &  1.49(10) \\
\runn{160} & 159  &  1.51(7) \\
\runn{170} & 168  &  1.53(6) \\
\runn{180} & 177  &  1.55(6) \\
\runn{190} & 186  &  1.57(7) \\
\runn{200} & 195  &  1.58(9) \\
\hline
\end{tabular}
\end{center}
\caption{Results for the factors $Z_{m_f \to \overline{\rm MS}}(2 \mbox{GeV})$ 
which convert a lattice quark mass, $\widetilde{m}$ into a mass normalized in 
the $\overline{\rm MS}$ conventions at $\mu=2$ GeV.}
\label{tab:Z_MS_bar}
\end{table}

The factors given in Tab.~\ref{tab:Z_MS_bar} will also be used to convert 
values of the chiral condensate $\overline{\psi}\psi$ (when constructed from 
the usual 
4-D surface, lattice operators) and Dirac spectrum (when normalized with the 
same conventions as $\overline{\psi}\psi$) into $\mu =2$ GeV, $\overline{\rm MS}$ 
values according to the relations:
\begin{eqnarray}
(\overline{\psi}\psi)^{\overline{\rm MS}}
             &=& \frac{(\overline{\psi}\psi)^{\rm lat}}{Z_{m_f \to \overline{\rm MS}}} \\
\rho(\lambda)^{\overline{\rm MS}}
             &=& \frac{\rho^{\rm lat}(\lambda/Z_{m_f \to \overline{\rm MS}})}
                 {Z_{m_f \to \overline{\rm MS}}}.
\end{eqnarray}

Of course, because the quark masses and lattices scales that we use are
interpolated and extrapolated from only three zero temperature ensembles, there
is significant uncertainty in our determination of the renormalization factors.
However, for the purposes of the present paper, we believe that these renormalization
factors in Tab.~\ref{tab:Z_MS_bar} have sufficient accuracy.

\subsection{Determining DWF Dirac eigenvalues and eigenvectors}

We directly diagonalize the five dimensional hermitian DWF Dirac operator $D_H=R_5\gamma_5D^{DWF}$
using the Kalkreuter-Simma (KS) version of the Ritz method 
\cite{Kalkreuter:1995mm}.
Details of this method have been described in \cite{GLiu:2003thesis} and \cite{Blum:2000kn}.

At each KS iteration, we use the conjugate gradient method to find the lowest $N_\mathrm{eig}$
eigenvalues of $D_H^2$ and corresponding eigenvectors one by one, by minimizing the Ritz functional,
\begin{equation}
  \mu(\Psi)=\frac{\langle\Psi|D_H^2|\Psi\rangle}{\langle\Psi|\Psi\rangle}.
  \label{eq:Ritz}
\end{equation}

We can then calculate the eigenvalues of $D_H$ by diagonalizing $D_H$ 
in the subspace spanned by the eigenvectors of $D_H^2$ previously obtained. 
The precision of the KS method is controlled by the maximum relative change of all the eigenvalues
between each KS iteration.

A spurious eigenmode problem may arise in the Jacobi diagonalization of $D_H$,
if only one of the paired eigenvectors is included in the subspace. The spurious
eigenmode's corresponding vector is the linear combination of two almost
degenerate eigenvectors with eigenvalues of opposite signs. We resolve this problem 
by applying $D_H$ to the problematic vector and find the proper linear combination
of the resulting vector and the original problematic vector which is the true eigenvector.

Using these methods we have computed the 100 eigenvalues with the smallest
magnitude of the DWF Dirac operator on the seven finite 
temperature ensembles 
in the temperature range 149 MeV $\le T \le 195$ MeV as well as the $\beta=1.75$, 
zero temperature ensemble discussed below.  Tab.~\ref{tab:EigenConfigs} identifies 
the configurations that were used in these calculations.

\begin{table}[t]
\begin{center}
\begin{tabular}{ccccccc}
Label & $T$ (MeV)   &$N_{\rm start}$  &$N_{\rm cfg}$ & ${\cal R}$ &${\cal R}\Lambda_0$ &$m_l+m_{\rm res}$ \\
\hline 
\runn{150_32} & 149&   300&   340 & 1.905& 0.00632& 0.00459 \\
\runn{150_48} & 149&   300&   340 & 1.980& 0.00606& 0.00469 \\
\runn{160}       & 159&   300&   408 & 1.725& 0.00828& 0.004321\\
\runn{170}       & 168&   300&   239 & 1.631& 0.01334& 0.00384 \\
\runn{180}       & 177&   300&   246 & 1.476& 0.02170& 0.00364 \\
\runn{190}       & 186&   300&   374 & 1.439& 0.03131& 0.00334 \\
\runn{200}       & 195&   302& 1140 & 1.365& 0.03837& 0.00311 \\
    \hline
\runn{1.75ml0.006}   &      0&  300& 252 & 1.568& 0.00489& 0.00488 \\
    \hline
\end{tabular}
\end{center}
\caption{List of the configurations used in the Dirac spectrum calculation as well
as the results for the average smallest normalized eigenvalue (${\cal R}\Lambda_0$).  
 Here $N_{\rm start}$
is the first configuration number on which the spectrum was computed, while 
$N_{\rm cfg}$ gives the total number of configurations on which the spectrum was 
determined.  In each case these configurations were separated by 5 time units.
(The sequence of trajectories used for run \#\runn{200} contained one anomaly:
samples 430 and 431 were separated by three instead of five time units.)}
\label{tab:EigenConfigs}
\end{table}

\subsection{Normalized spectral density}

The results for the Dirac spectrum at finite temperature obtained using these 
methods are presented and analyzed in Sec.~\ref{sec:ua1}, where the restoration 
of $U_A(1)$ symmetry is studied.  In this section we examine the Dirac spectrum 
obtained on the zero temperature ensemble labeled \# \runn{1.75ml0.006}, with
volume $16^4$ and $\beta = 1.75$. 

The discussion in the present section has three objectives.   First we explicitly
apply the normalization factors to convert the bare eigenvalues of the DWF Dirac
operator into the $\overline{\rm MS}$ scheme. 
The resulting spectral density is expressed in physical units and can easily be compared 
with both physical and simulated $\overline{\rm MS}$ values of the quark 
masses as well as with the QCD scale, $\Lambda_{QCD} \sim 300$ MeV.  Second, we convert 
the spectrum of the hermitian DWF Dirac operator, which includes the 
effects of the non-zero quark masses to the more conventional spectrum from 
which the mass has been removed, a step which depends critically on the 
normalization procedure and is sensitive to finite lattice spacing errors. Finally we 
examine the Banks-Casher relation between the resulting spectrum and the chiral 
condensate.

\begin{figure}[hbt]
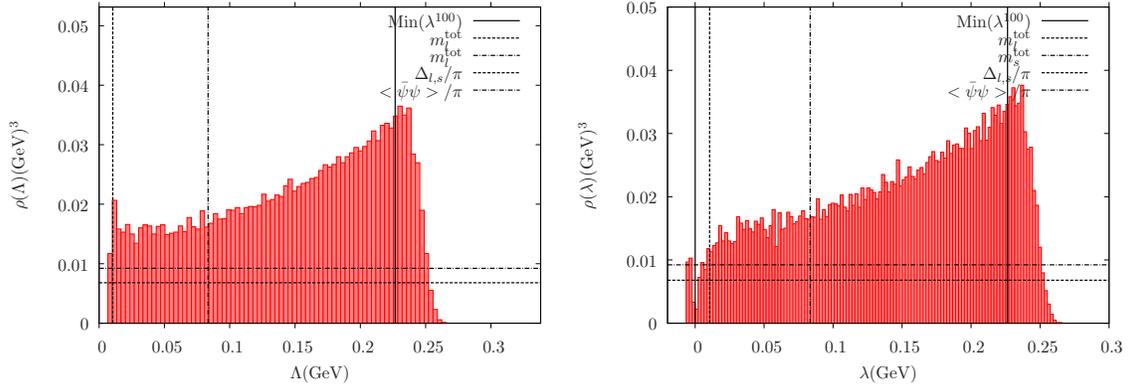

  \begin{center}
        \begin{minipage}[t]{0.5\linewidth}
                \centering
        \resizebox{\linewidth}{!}{\input{figs/000MeV_ml003_norm_wm.tex}}
        \end{minipage}%
        \begin{minipage}[t]{0.5\linewidth}
                \centering
        \resizebox{\linewidth}{!}{\input{figs/000MeV_ml003_norm_wom.tex}}
        \end{minipage}
  \end{center}
  \caption{Histogram of the spectrum of eigenvalues $\Lambda$ of the hermitian
DWF Dirac operator normalized in the $\overline{\rm MS}$ scheme at the scale 
$\mu=2$ GeV (left).  These eigenvalues are calculated on the zero-temperature
ensemble labeled \#\runn{1.75ml0.006}.  The right hand panel shows a histogram of the eigenvalues
$\lambda = \sqrt{\Lambda^2 - (m_f+m_{\rm res})^2}$ from which the quark mass 
has been removed.  In the this panel, the region $\lambda > 0$
shows those values for which $\Lambda^2 > (m_f + \mres)^2$, \textit{i.e.},
$\lambda$ is purely real, a condition that
should be obeyed in the continuum limit.  The region $\lambda < 0$ shows those eigenvalues with 
$\Lambda^2 < (m_f+m_{\rm res})^2$, \textit{i.e.}, $\lambda$ pure imaginary,
plotted on the negative part of the x-axis as $\lambda = -|\sqrt{\Lambda^2 - (m_f+m_{\rm res})^2}|$.
These unphysical values give a visible measure of the finite
lattice spacing distortions to the region of small $\lambda > 0$.}   \label{fig:0MeV}
\end{figure}

Fig. ~\ref{fig:0MeV} shows histograms of the Dirac eigenvalues measured
on 340 configurations from the zero-temperature, $16^4$ ensemble \#\runn{1.75ml0.006} in 
Tab.~\ref{tab:EigenConfigs}.  In the left-hand
panel of this figure, the histogram of eigenvalues $\Lambda$ is obtained by
converting the eigenvalues of the lattice DWF Dirac operator, as described above, to 
the $\overline{\rm MS}$ scheme with $\mu = 2$ GeV.  On each configuration 
the 100 eigenvalues of smallest magnitude have been determined. Figure~\ref{fig:0MeV} 
shows histograms of these 34,000 eigenvalues.  
The rightmost vertical line in both panels identifies the minimum value from the
set of the 100th largest eigenvalues on each of the 340 configurations.  For eigenvalues
less than this ``minmax'' value the histogram accurately represents the complete 
spectrum, undistorted by our cutoff of 100 eigenvalues per 
configuration.

Here, $\Lambda$ denotes an eigenvalue of the full hermitian DWF Dirac operator.   
These eigenvalues include the effect of the quark mass and in the continuum 
limit would have the form
\begin{equation}
\Lambda = \sqrt{\lambda^2 + \widetilde{m}^2}.
\label{eq:Lambda_def}
\end{equation}
The left-hand panel of Fig.~\ref{fig:0MeV} demonstrates the effect of using 
a consistent normalization scheme for the quark masses.   The two left-most
vertical lines in that plot correspond to the simulated light and strange quark 
masses, $\widetilde{m}_l$ and $\widetilde{m}_s$, in the same $\overline{\rm MS}$
normalization.  The expected coincidence between the peak in the $\Lambda$ distribution
at the smallest eigenvalues and the vertical line representing the light quark mass occurs
only after the relative 
normalization ${\cal R} = 1.570$ from Tab.~\ref{tab:EigenConfigs} between the 
DWF operator and the conventional input quark mass discussed above has been 
applied.

In the continuum theory the mass is conventionally removed from the Dirac operator 
before its eigenvalues are determined so that the usual eigenvalue distribution is 
given for the quantity $\lambda$ in Eq.~\eqref{eq:Lambda_def}.  In our case, the
transformation to this more usual eigenvalue distribution requires converting each
eigenvalue $\Lambda_n$ into a corresponding 
$\lambda_n=\sqrt{\Lambda_n^2-\widetilde{m}_l^2}$.  Unfortunately, this step is 
vulnerable to finite lattice spacing effects which allow an occasional value of
$\Lambda_n$ to be smaller than $\widetilde{m}_l$, leading to an unphysical,
imaginary result for $\lambda_n$.  This should become increasingly rare in
the limit $a \to 0$ of vanishing lattice spacing.  In this limit, the quantity 
$\widetilde{m}_l$ accurately corresponds to the light quark mass 
describing the long distance physics determined by our lattice theory.  Likewise,
the arguments given in Appendix~\ref{sec:spectral_normalization} imply
that in this limit, the spectral density $\rho(\Lambda)$ also approaches a
continuum limit which requires $\Lambda \ge \widetilde{m}_l$.  

However, in the calculation presented here the lattice spacing $a$ is relatively
large and deviations from the inequality $\Lambda \ge \widetilde{m}_l$ should
be expected.  In order to present the more conventional eigenvalue distribution
$\rho(\lambda)$ while at the same time displaying the imperfections arising
from finite $a$, we choose to plot the eigenvalue histograms in a hybrid form.
For each of the original eigenvalues $\Lambda$ we compute 
the derived eigenvalue $\lambda_n=\sqrt{\Lambda^2-\widetilde{m}_l^2}$.  If
$\lambda_n$ is real, it is included in the histogram in the normal way, along
the positive x-axis.  However, if $\lambda_n$ is imaginary it is displayed in
the same histogram along the negative x-axis in a bin corresponding to $-|\lambda|$.  

This has been done in the right-hand panel of Fig.~\ref{fig:0MeV}.  The 
histogram for $\lambda > 0$ is the conventional eigenvalue distribution, 
normalized in the $\mu = 2$ GeV, $\overline{\rm MS}$ scheme.  The histogram
bins for $\lambda < 0$ are unphysical and directly result from finite lattice spacing
artifacts.  By showing both on the same plot, we make it easy to
recognize the magnitude of the errors inherent in $\rho(\lambda), \lambda > 0$
introduced by lattice artifacts.  For example, it is likely that
a majority of the gap in $\rho(\lambda)$ for $\lambda$ positive but near zero
in the right-hand panel of Fig.~\ref{fig:0MeV} would be filled in
as $a \to 0$ by the imaginary values of $\lambda$ plotted as $-|\lambda| <0$,
and should not be attributed to the effects of finite volume.

An interesting test of these methods can be made by comparing the spectrum
shown  in the right-hand panel of Fig.~\ref{fig:0MeV} with the predictions of
the Banks-Casher formula which relates the eigenvalue density $\rho(\lambda)$
at $\lambda=0$ and the chiral condensate $\langle\overline{\psi}\psi\rangle$ when 
both are evaluated in the limit of infinite volume and vanishing quark mass,
\begin{equation}
\langle\overline{\psi}\psi\rangle = \pi\rho(0).
\label{eq:banks-casher1}
\end{equation}
The right and left-hand sides of Eq.~\eqref{eq:banks-casher1} can be compared by
examining the right-hand panel of Fig.~\ref{fig:0MeV} where we have superimposed
the quantity $\langle\overline{\psi}\psi\rangle/\pi$ as horizontal lines
on the histogram.  Two values for  $\langle\overline{\psi}\psi\rangle/\pi$ 
are shown.  The upper line corresponds to $\langle\overline{\psi_l}\psi_l\rangle/\pi$ with
finite light quark mass $m_l = 0.003$.  The lower horizontal line corresponds to 
the quantity $\Delta_{l,s}/\pi$ given by
\begin{equation}
\Delta_{l,s}  
          = \langle\overline{\psi}_l\psi_l\rangle
                   -\frac{m_l}{m_s}\langle\overline{\psi}_s\psi_s\rangle.
\label{eq:pbp_subtract}
\end{equation}

The subtraction is an attempt to remove a 
portion of the large, ultraviolet divergent contribution to 
$\langle\overline{\psi}\psi\rangle$, of the form $m/a^2$, expected for 
non-zero mass and finite $L_s$.
This subtracted quantity is a more realistic estimate 
of  $\langle\overline{\psi}\psi\rangle/\pi$ in the massless limit. To test the
Banks-Casher relation, we compare the value of
$\Delta_{l,s}/\pi$ with $\rho(\lambda)$ for small $\lambda$, as can be seen in
the right panel of Fig.~\ref{fig:0MeV}.  This shows a value for $\Delta_{l,s}/\pi$
about 30\% lower than $\rho(0)$, probably 
indicating that our $16^3$ lattice results are significantly distorted by finite volume 
effects.  

However, for the case of domain wall fermions there will be a residual mixing 
between the two fermion chiralities on the left and right walls when their separation, 
$L_s$, is finite.  For long-distance quantities, this just results in an additive 
renormalization of the quark masses by $\mres$.  However, as suggested by the results 
in \cite{Cheng:2009be}, the effects 
of residual chiral symmetry breaking on the dimension three operator 
$\overline{\psi}\psi$ may come from higher energies and be more perturbative 
than those contributing to $\mres$, and therefore may fall off exponentially with $L_s$
rather than as a power law.  If that is also the case for the present ensembles with 
$L_s\ge32$, the residual contribution to $\langle\overline{\psi}\psi\rangle$ is not very large and 
the subtraction in Eq.~\eqref{eq:pbp_subtract} may remove the dominant contributions to 
$\langle\overline{\psi}\psi\rangle$ from short-distance modes.  However, the use of the
DSDR action enhances the contribution of the 
exponential- relative to the power-suppressed residual chiral symmetry breaking, so neglecting
$\mres$ in Eq.~\eqref{eq:pbp_subtract} may not be as accurate on the DSDR ensembles 
as it would be on DWF ensembles where DSDR is not employed.
\section{Observables probing the chiral symmetries of QCD}
\label{sec:basics}

In this section we introduce some observables used in our
finite temperature calculations and discuss their connections to the
$\sua$ symmetry and the anomalous $\ua$ symmetry of QCD.

The most basic observable indicating chiral symmetry
restoration is the chiral condensate. In the chirally symmetric phase this quantity
should vanish in the chiral limit. The single flavor light and strange quark chiral
condensates are defined as
\begin{equation}
\langle\bar{\psi}_q \psi_q\rangle =  \frac{T}{V}
\frac{\partial \ln Z}{\partial m_q} = \frac{1}{N_\sigma^3N_\tau}\langle
\mathrm{Tr}M_q^{-1}\rangle\; ,\; q=l,\ s\,
\label{eq:pbp}
\end{equation}
where $M_q$ is the single-flavor Dirac matrix\footnote{For simplicity, we assign the quantity 
$\langle\pbp\rangle$ a positive sign corresponding to using the mass term $-m\pbp$ in
the Dirac Hamiltonian.}.
As discussed in the previous section,
the leading ultra-violet divergent part in the chiral condensate is of the form $\sim
m_q/a^2$. Thus, in order to eliminate this ultra-violet divergent contribution we
construct the subtracted chiral condensate, $\Delta_{l,s}$, as defined 
in Eq.~\eqref{eq:pbp_subtract}.

Chiral symmetry restoration can also be probed by studying various two-point
functions. For computational simplicity, we will focus on various
integrated two-point functions, {\it i.e.}, susceptibilities, instead of the
two-point correlations functions themselves.

The flavor non-singlet ($\delta$) and the flavor singlet ($\sigma$) two-point scalar
correlators are given by
\begin{eqnarray}
G_{\delta}(x) &=&
- {\rm tr} \langle \,M_l^{-1}(x,0) M_l^{-1}(0,x) \,\rangle  \; 
\qquad\mathrm{and} \label{eq:delta} \\
G_{\sigma}(x) &=& G_{\delta}(x)
+ \langle {\rm tr} M_l^{-1}(x,x) {\rm tr} M_l^{-1}(0,0)\rangle
-\langle {\rm tr} M_l^{-1}(x,x)\rangle\ \langle {\rm tr} M_l^{-1}(0,0)\rangle
\;,  \label{eq:tempcorsigma}
\end{eqnarray}
where the vacuum contribution to the $\sigma$ correlator has been explicitly subtracted.
By integrating these quantities over the four-volume one obtains the corresponding susceptibilities
\begin{eqnarray}
\chi_\delta
&=& \sum_{x} G_{\delta}(x)
= \chi_{\rm con}
\qquad\mathrm{and}
\label{eq:chi_delta} \\
\chi_\sigma
&=& \sum_{x} G_{\sigma}(x)
= \chi_{\rm con} + \chi_{\rm disc} \; ,
\label{eq:chi_sigma} 
\end{eqnarray}
where the quark-line disconnected and the quark-line connected parts of the
chiral susceptibilities\footnote{These quantities are referred to as chiral 
susceptibilities since they are related to the fluctuations of the quantity whose 
expectation value is the chiral condensate.} can be written respectively by 
\begin{eqnarray}
\chi_{\rm disc} &=&
{1 \over N_{\sigma}^3 N_{\tau}} \left\{
\langle\bigl( {\rm Tr} M_l^{-1}\bigr)^2  \rangle -
\langle {\rm Tr} M_l^{-1}\rangle^2 \right\}
 \qquad\mathrm{and} \label{eq:chi_dis} \\
\chi_{\rm con} &=&  -
{\rm tr} \sum_x \langle \,M_l^{-1}(x,0) M_l^{-1}(0,x) \,\rangle 
\equiv - {1 \over N_{\sigma}^3 N_{\tau}}  \langle {\rm Tr} M_l^{-2}\rangle\; .
\label{eq:chi_con}
\end{eqnarray}
The notation `$\rm tr$' indicates traces over spinor and color indices only, while
`$\rm Tr$' also includes a trace over the discrete points $x=(x_0, \vec{x})$ in the
four-volume.   Tables~\ref{tab:pbpLs96} and \ref{tab:pbp} summarize our results for the 
chiral condensates and disconnected chiral susceptibility, for the $L_s = 96$ and
the DSDR ensembles, respectively.  For both ensembles, the chiral condensates were obtained from a
stochastic approximation in which the trace in Eq.~\eqref{eq:pbp} 
is estimated by the average over the diagonal matrix elements of $M_l^{-1}$ evaluated on
ten Gaussian random sources at every fifth molecular dynamics time unit.  To compute the disconnected 
susceptibility, the term  $\langle \left({\rm Tr} M_l^{-1}\right)^2\rangle$ in Eq.
\ref{eq:chi_dis} is calculated by averaging on each configuration only the product of
matrix elements coming from different random sources.  This insures that the noise 
introduced by the Gaussian random vectors does not bias our estimate of $\chi_{\rm disc}$.  
(This strategy was also employed in computing the  
disconnected susceptibility, $\chi_{5,{\rm disc}}$, given later in Tab.~\ref{tab:pmd}).

\begin{table}[hbt]
\begin{center}
\begin{tabular}{ccccc}
$T$(MeV) & $\beta$ & $\left<\bar{\psi}_l \psi_l\right>/T^3$ & $\left<\bar{\psi}_s \psi_s\right>/T^3$ & $\chi_{\rm disc}/T^2$\\
\hline
137 & 1.965 & 15.1(2) & 37.6(1) & 20(2)\\
146 & 1.9875 & 13.2(1) & 35.99(7) & 26(4)\\
151 & 2.00 & 12.0(2) & 35.26 (9) & 24(4)\\
156 & 2.0125 & 10.3(2) & 33.92(12) & 30(5)\\
162 & 2.025 & 10.1(2) & 33.44(10) & 24(4)\\
167 & 2.0375 & 8.0(2) & 31.99(10) & 29(3)\\
173 & 2.05 & 7.4(2) & 31.48(10) & 20(3)\\
188 & 2.08 & 6.2(2) & 29.84(10) & 21(3)\\
198 & 2.10 & 5.2(2) & 28.68(10) & 16(3)\\
\hline
\end{tabular}
\caption{Chiral condensates and the disconnected light-quark chiral susceptibility for 
the $L_s = 96$ ensembles.}
\label{tab:pbpLs96}
\end{center}
\end{table}

\begin{table}[hbt]
\begin{center}
  \begin{tabular}{ccccccc}
   Label & T(MeV) &$\Tspace \Bspace \left<\pbp\right>_l/T^3$&
    $\left<\pbp\right>_s/T^3$&$\Delta_{l,s}/T^3$&$\chi^\textrm{bare}_{\rm disc}/T^2$ & $\chi^{\overline{\textrm{MS}}}_{\rm disc}/T^2$\\ \hline
\runn{140} & 139&9.23(14)&41.00(5)&10.30(14)&37(3) & 17.2(1.4) \\
\runn{150_32} & 149&6.26(12)&36.42(5)&7.74(12)&44(3) & 19.9(1.0)\\
\runn{150_48} & 149&8.39(10)&38.30(3)&7.06(10)&41(2)& 18.5(0.9)\\
\runn{160} & 159&5.25(17)&33.81(6)&4.83(17)&43(4)& 18.8(1.8)\\
\runn{170} & 168&4.03(18)&30.66(7)&2.78(18)&35(5)&14.9(2.1) \\
\runn{180} & 177&3.16(15)&27.88(6)&1.56(15)&25(4)&10.4(1.7) \\
\runn{190} & 186&2.44(9) &25.43(4)&0.71(9) &11(4) & 4.5(1.6)\\
\runn{200} & 195&2.07(9) &23.24(5)&0.34(9) &5(3) &2.0(1.2)  \\
	\hline
\end{tabular}
\caption{Chiral condensates and the disconnected light-quark chiral susceptibility for 
the DSDR ensembles.}
\label{tab:pbp}
\end{center}
\end{table}

Chiral symmetry restoration implies a massless $\sigma$ meson at the 
transition temperature. However, the
$\delta$ meson is expected to remain massive unless the $U(1)_A$ symmetry also 
becomes restored at that temperature. Thus, at the chiral transition
$\chi_\sigma$ will diverge, while $\chi_\delta$ remains finite. This implies (see
Eqs.\ \eqref{eq:chi_sigma} and \eqref{eq:chi_delta}) that the disconnected part of the chiral
susceptibility $\chi_{\rm disc}$ diverges at the chiral transition while the connected
part  $\chi_{\rm con}$ remains finite.  At the chiral
transition the diverging disconnected chiral susceptibility is expected to be related
to the $O(4)$ scaling properties of the chiral transition. This in turn suggests that
for non-zero light quark mass (or finite volume) the chiral crossover temperature
can be naturally identified by locating the maximum of the disconnected chiral susceptibility
as a function of the temperature.

We also introduce flavor non-singlet ($\pi$) and singlet ($\eta$) pseudo-scalar
two-point screening correlation functions,
\begin{eqnarray}
G_{\pi}(x) &=& {\rm tr} \langle \,\gamma_5 M_l^{-1}(x,0) \gamma_5
M_l^{-1}(0,x) \,\rangle
 \qquad\mathrm{and} \label{eq:cor_pi} \\
G_{\eta}(x) &=& G_{\pi}(x) -
\langle {\rm tr}\left[ \gamma_5 M_l^{-1}(x,x)\right] {\rm tr}\left[ \gamma_5 M_l^{-1}(0,0)\right]\rangle \; .
\label{eq:cor_eta}
\end{eqnarray}
Integrating these correlation functions over the four-volume we obtain the corresponding
pseudo-scalar susceptibilities
\begin{eqnarray}
\chi_\pi 
&=& \sum_{x} G_{\pi}(x)
\equiv \chi_{5,{\rm con}}
\qquad\mathrm{and} \label{eq:chi_pi}  \\
\chi_\eta
&=& \sum_{x} G_{\eta}(x)
\equiv \chi_{5,{\rm con}} - \chi_{5,{\rm disc}}.
\label{eq:chi_eta}
\end{eqnarray}
Table~\ref{tab:wspect} summarizes the details of our screening correlator measurements on the DSDR
ensembles.
 
\begin{table}[ht]
\begin{center}
\begin{tabular}{ccccc}
Label & T (MeV)& Trajectories & Step\\
\hline  
  \runn{140} & 139 & 200-2990 & 10\\
  \runn{150_48} & 149 & 300-7000 & 5\\
  \runn{160} & 159 & 300-3650 & 10\\
  \runn{170} & 168 & 300-3410 & 10\\
  \runn{180} & 177 & 300-1780 & 10\\
  \runn{190} & 186 & 300-4360 & 10\\
  \runn{200} & 195 & 302-2447 & 5\\
 &  & 2450-6000 & 5\\
	\hline
\end{tabular}
  \caption{Summary of screening correlator measurements.  All measurements are with a point source and 
  point sink with the source located at $(x,y,z,t) = (0,0,0,0)$.}
  \label{tab:wspect}
\end{center}
\end{table}

\begin{figure}[!t]
\begin{center}
\includegraphics[width=0.7\textwidth]{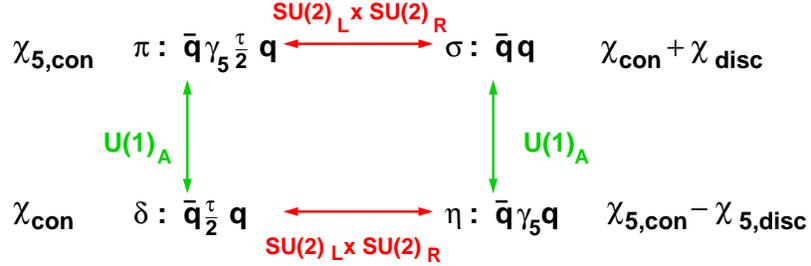}
\caption{Symmetry transformations relating scalar and pseudo-scalar
mesons in flavor singlet and non-singlet channels.}
\label{fig:trans}
\end{center}
\end{figure}

The scalar and pseudo-scalar correlation functions introduced above are related
through $SU(2)_L\times SU(2)_R$ flavor transformations, as illustrated by the
horizontal lines in  Fig.~\ref{fig:trans}. 
Hence, utilizing Eqs.\
\eqref{eq:chi_sigma}, \eqref{eq:chi_delta}, \eqref{eq:chi_pi} and \eqref{eq:chi_eta}, chiral symmetry
restoration is manifested through the following degeneracies among the
susceptibilities of the two-point correlation functions: 
\begin{eqnarray}
\chi_\pi = \chi_\sigma
\qquad &\Longrightarrow& \qquad \chi_\pi - \chi_\delta = \chi_{\rm disc}
\;, \qquad\mathrm{and} \label{eq:pi-delta_1}\\
\chi_\delta = \chi_\eta
\qquad &\Longrightarrow& \qquad \chi_\pi - \chi_\delta = \chi_{5,{\rm disc}}
\;. \label{eq:pi-delta_2} 
\end{eqnarray}

In the limit of two massless flavors, the anomalous $\ua$ symmetry
cannot be probed with a local expectation value such as the chiral condensate. In
this case it is necessary to use two-point correlation functions, as introduced
above \cite{Lee:1996zy,Evans:1996wf,Birse:1996dx}. Since the $\ua$ transformation does not
change the flavor quantum numbers, a restoration of $\ua$ symmetry
will be manifested by the equalities between the following susceptibilities, 
\begin{equation}
\chi_\pi = \chi_\delta
 \qquad \mathrm{and} \qquad
\chi_\sigma = \chi_\eta
\;.
\end{equation}
Thus, the susceptibility difference $\chi_{\pi} - \chi_{\delta}$
can be used to study restoration of $\ua$
symmetry at high temperatures. Note, while both the susceptibilities
$\chi_{\pi}$ and $\chi_\delta$ individually contain an additive ultra-violet
divergent term $\sim1/a^2$, their difference is free of this
divergence. Furthermore, in the chirally symmetric phase of QCD
one can use Eqs. \eqref{eq:pi-delta_1} and \eqref{eq:pi-delta_2} to obtain
\begin{equation}
\chi_{\pi} - \chi_{\delta} = \chi_{\rm disc} = \chi_{5,{\rm disc}}
\;, \qquad \mathrm{for} \quad T\ge T_c \;,\ m_l\rightarrow 0 \; .
\label{eq:Delta_rel}
\end{equation} 
Hence, in the chirally symmetric phase (in the chiral limit) the disconnected chiral
susceptibility itself can be used to probe the restoration of the $\ua$ symmetry. 

Further information about $\chi_{\pi} - \chi_\delta$ can be obtained by comparing
to the topological charge, $Q_{\rm top}$.  $Q_{\rm top}$ is defined as
\begin{equation}
Q_{\rm top} = \frac{g^2}{32 \pi^2} \int d^4 x F^a_{\mu \nu}(x) \tilde{F}^a_{\mu \nu}(x).
\label{eq:Qtop}
\end{equation}
On a smooth gauge configuration, if lattice artifacts are small, the topological charge 
and the integrated pseudo-scalar bilinear can be related:
\begin{equation}
Q_{\rm top}  =  m_l\int d^4 x \bar\psi_l(x)\gamma_5\psi_l(x).
\label{eq:Qtop_pbg5p}
\end{equation}
If this relation is squared, averaged over the gauge field and divided by the space-time
volume $V$ we obtain a relation between the topological susceptibility and the disconnected
pseudo-scalar susceptibility:
\begin{equation}
\chi_{\rm top} = \frac{\langle Q_{\rm top}^2 \rangle}{V}=m_l^2\,\chi_{5,{\rm disc}}.
\label{eq:top2chi_5disc}
\end{equation}
This equation can be obtained in the continuum theory by integrating the anomalous 
conservation law for the axial current over space-time, squaring the result, dividing by
the space-time volume and ignoring possible ambiguities in the operator product 
appearing in $Q_{\rm top}^2$.  If we assume $\sua$ symmetry and substitute 
Eq.~\eqref{eq:Delta_rel} into Eq.~\eqref{eq:top2chi_5disc} we can directly relate the 
measure of $\ua$ symmetry breaking $\chi_\pi-\chi_\delta$ and the topological 
susceptibility:
\begin{equation}
\chi_\pi-\chi_\delta = \frac{1}{m_l^2} \chi_{\rm top}.
\end{equation}

Finally, the eigenvalue spectrum of the Dirac operator is also intimately connected
with the chiral and anomalous axial symmetry.  The symmetry breaking quantities
$\left<\pbp\right>$ and $\chi_\pi-\chi_\delta$ can both be expressed in terms of the eigenvalue
spectrum of the Dirac operator in the following way:
\begin{eqnarray}
\langle \bar{\psi}_l \psi_l\rangle &=& \int_0^\infty \df\lambda\, 
\frac{2m_l\,\rho(\lambda)}{m_l^2+\lambda^2}
\; , \label{eq:pbp_rho} \\ 
\chi_\pi-\chi_\delta &=& \int_0^\infty \df\lambda\,
\frac{4m_l^2\,\rho(\lambda)}{\left(m_l^2+\lambda^2\right)^2}
\; . \label{eq:Delta_rho}
\end{eqnarray}
Equation\ \eqref{eq:pbp_rho} is the basis of the Banks-Casher relation \cite{Banks:1979yr} 
which connects the chiral condensate to the density of zero eigenvalues
$\lim_{m_l\to0}\langle\bar{\psi}_l\psi_l\rangle=\pi\rho(0)$.   While in the chirally broken 
phase a non-zero value of the chiral condensate demands $\rho(0)\ne0$, in the
chirally symmetric phase a vanishing chiral condensate leads to $\rho(0)=0$.  
However, Eq.~\eqref{eq:Delta_rho} shows that a non-zero
anomalous symmetry breaking difference $\chi_\pi-\chi_\delta$ in the limit of massless 
quarks requires complex behavior for $\rho(\lambda)$ as $\lambda$ approaches zero~\cite{Chandrasekharan:1995gt}. 
This required behavior is very different, for example, from that found in the case of a free 
field at finite temperature.  
For the free field case there is a gap in the spectrum between zero and the Matsubara 
frequency $\pi T$: $\rho(\lambda) = 0$ for $0 \le \lambda < \pi T$.  This question is 
studied in detail in Section~\ref{sec:ua1} .

\section{\boldmath $SU(2)_L\times SU(2)_R$ Restoration}
\label{sec:chiral}
We now turn to a discussion of $\sua$ chiral symmetry restoration.
We will first discuss the chiral transition using
conventional observables such as the chiral condensate and the related
chiral susceptibility. We then will turn to a discussion of several
hadronic susceptibilities.

\begin{figure}[t]
\begin{center}
\includegraphics[width=0.5\textwidth,angle=-90]{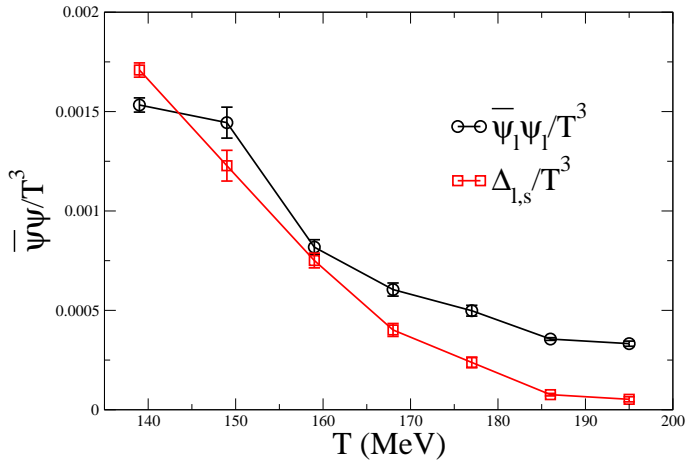}
\caption{The light quark chiral condensate, as well as the subtracted chiral condensate
plotted as a function of temperature. As discussed in the text, the values plotted for $T=139$ and 149 
MeV were computed using $L_s=48$ while those at higher temperatures used $L_s=32$.}
\label{fig:condensate}
\end{center}
\end{figure}

In Fig.~\ref{fig:condensate} we show results for the light quark chiral condensate 
calculated on the $16^3\times 8$
ensembles in the temperature range $139~{\rm MeV} \le T \le 195~{\rm MeV}$.
In this figure, we also show the subtracted chiral condensate
$\Delta_{l,s}$ introduced in Eq.~\eqref{eq:pbp_subtract}.  
The values plotted at the lower two
temperatures, $T=139$ and 149 MeV were obtained using $L_s=48$ while
the values at the five higher temperatures use $L_s=32$.  As discussed
in Sec.~\ref{sec:details}, the ultraviolet divergent piece of the chiral
condensate, $m_l/a^2$ is sensitive to the bare light quark mass.
This results in the irregular behavior for the light quark chiral condensate
seen in Fig.~\ref{fig:condensate} and the different values for this quantity
for ensembles \#\runn{150_32} and \#\runn{150_48} given in Tab.~\ref{tab:pbp}.
As also should be expected, this short distance contribution to
$\left<\pbp\right>$ is substantially reduced in the subtracted quantity $\Delta_{l,s}$, 
which agrees between $L_s=32$ and 48 at $T=149$ MeV at the 10\% 
level.

\begin{figure}[t]
\begin{center}
\includegraphics[width=0.43\textwidth,angle=-90]{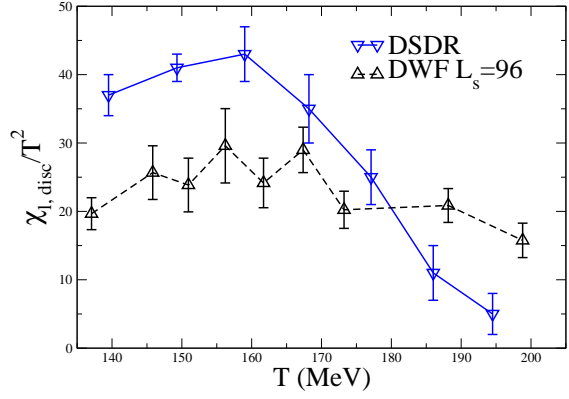}
\hspace{0.03\textwidth}
\includegraphics[width=0.43\textwidth,angle=-90]{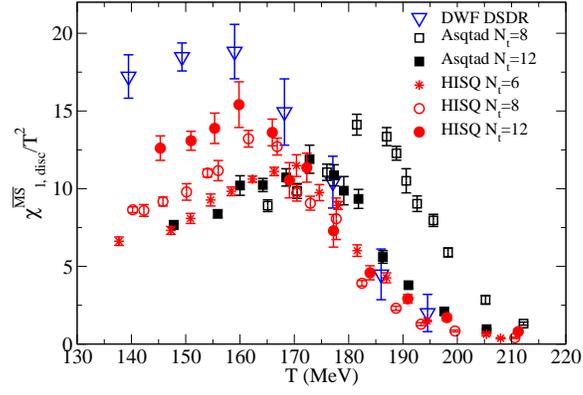}
\caption{In the upper panel, the unrenormalized, disconnected chiral susceptibility for 
DWF DSDR $L_s = 32, 48$ is compared with the DWF results with $L_s = 96$.  
In the lower panel, the renormalized chiral susceptibilities, converted to the 
$\overline{\textrm{MS}}$ scheme are compared between the DWF DSDR calculation 
and the HISQ and asqtad results from the HotQCD Collaboration, corresponding to
a pseudo-Goldstone pion mass of 161 and 179 MeV, respectively.}
\label{fig:chi_dis}
\end{center}
\end{figure}

As described in Sec.~\ref{sec:basics} we can use the fluctuations found in our calculation 
of the expectation values of $\pbp$ and $\pbgp$ to construct the disconnected part of the 
chiral susceptibility.  The upper panel of Fig.~\ref{fig:chi_dis} shows our results for
the disconnected chiral susceptibility from both the $L_s=96$ and the $L_s=32$ 
and 48 results calculated with the DSDR gauge action.  The discrepancy between the
two results for $T \le 170$ MeV can be explained by the different values of the light
quark mass used in the two calculations.  The $L_s=96$ calculation was performed 
with the quark mass fixed in lattice units and the resulting zero-temperature pion mass 
decreasing from approximately 275 MeV to 225 MeV as the temperature decreases 
from the highest to the lowest value.  In contrast, the DSDR calculation was performed 
at a fixed 200 MeV pion mass. Since the disconnected chiral susceptibility is expected 
to increase as the pion mass decreases for  $T \leq T_c$,  a larger value should be
expected from the DSDR calculation in this temperature range.  For temperatures
above the transition, the chiral condensate and to some degree its fluctuations
are suppressed by a decreasing physical quark mass, causing the DSDR values
for  $\chi_{\rm disc}$ to fall below those of the  $L_s=96$ ensemble.

In the lower panel of Fig.~\ref{fig:chi_dis} we compare the DSDR, DWF results with
those obtained previously using the asqtad and HISQ  staggered fermions by the HotQCD 
collaboration~\cite{Bazavov:2011nk}.  In order to make a comparison between different
 fermion actions, one must convert the unrenormalized results for the disconnected
chiral susceptibility into a common renormalization scheme, \textit{e.g.}
the $\overline{\rm MS}$ scheme that was discussed in Sec. ~\ref{sec:dirac}.
The renormalized chiral susceptibility is given by:
\begin{equation}
\chi_{\rm disc}^{\overline{MS}} = \left(\frac{1}{Z_{m_f \to \overline{\rm MS}}(\mu^2)}\right)^2~\chi_{\rm disc}^{bare},
\end{equation}
where an expression for $Z_{m_f\to\overline{\rm MS}}(\mu^2)$ is given in 
Eq.~\eqref{eq:Z_mf_MSbar}.  The values of $Z_{m_f\to\overline{\rm MS}}(\mu^2)$ are 
tabulated for the DWF+DSDR action with $\mu = 2$ GeV in 
Tab. \ref{tab:Z_MS_bar}.  Details for converting the
staggered results to the $\overline{\rm MS}$ scheme are
discussed in Appendix \ref{sec:renormalization}.

The difference between the DWF and staggered results shown in the lower panel
of Fig.~\ref{fig:chi_dis} may arise from more than one source.   While the staggered 
results are obtained with nominally lighter pion masses (the $N_t=12$ HISQ and 
asqtad results have $m_\pi =161$ and 179 MeV respectively)
this is the mass of the lightest Goldstone pion and taste breaking leads to a range of
masses for the other 15 taste-split pions, some of which are considerably larger. 
In contrast the DWF calculation has three degenerate 200 MeV pions.
However, the staggered calculations are performed at much larger physical volumes
than the DWF work reported here, with linear dimensions twice the size of those
in the DWF calculation.   In fact, a finite volume scaling study of an O(4) symmetric 
quark-meson model of the phase 
transition~\cite{Braun:2010vd} suggests that the height of  the peak in the chiral susceptibility 
associated with the transition should become smaller as the volume is increased,
which provides a second possible explanation of the discrepancy between the DWF
and staggered results found in Fig.~\ref{fig:chi_dis}.

To obtain the connected part
of the various susceptibilities we have calculated hadronic correlation
functions in different quantum number channels (for a more detailed 
discussion see Sec.~\ref{sec:basics}).   The sink position of these 
two-point correlation functions is then integrated over the full space-time 
volume to obtain the corresponding susceptibility.   For example, the 
integral over the scalar point-point correlation function gives the 
connected part of the chiral susceptibility $\chi_{l,{\rm con}} \equiv \chi_\delta$, 
with $\chi_\delta$ introduced in Eq.~\eqref{eq:chi_delta}. 

We find that susceptibilities calculated from 
connected correlation functions do not show significant temperature 
dependence.  This is quite similar to what has been found in calculations 
performed with staggered fermions.   While dramatic temperature 
dependence is expected in the connected susceptibilities, for example 
in $\chi_\pi$ associated with the small pion mass below $T_c$, these
quantities are likely dominated by the  $1/a^2$ divergence associated with 
the coincidence of the source and sink points when the correlation 
function is integrated over space-time.

\begin{figure}[htb]
\begin{center}
\includegraphics[width=0.5\textwidth]{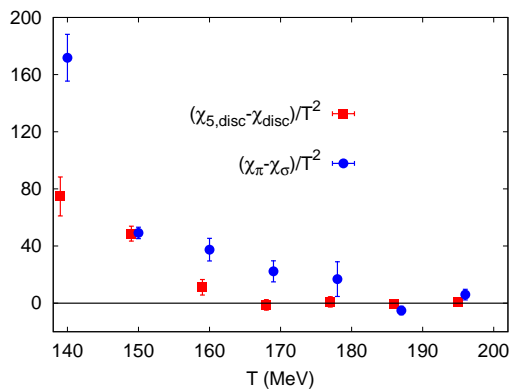}
\caption{
\label{fig:sigma}
The $\sua$-breaking differences between the disconnected pseudo-scalar 
and  disconnected scalar susceptibilities and between the flavor-triplet 
pseudo-scalar and flavor singlet scalar susceptibilities.}
\end{center}
\end{figure}

In the chiral limit the restoration of chiral flavor symmetry can also  
be seen in the vanishing of the susceptibilities differences $\chi_\pi - \chi_\sigma$
and $\chi_{\rm disc}-\chi_{5, {\rm disc}}$ as shown 
in Eq.~\eqref{eq:Delta_rel}.  We show these two measures of chiral symmetry 
breaking in Fig.~\ref{fig:sigma} where one sees a decrease
with increasing temperature that is even more rapid than that found in 
Fig.~\ref{fig:condensate} for the subtracted chiral order parameter 
$\Delta_{l,s}$.

The two differences $\chi_\pi-\chi_\sigma$ and $\chi_{\rm disc}-\chi_{5,{\rm disc}}$
provide information on chiral symmetry restoration that is consistent with the observed
peak in the disconnected chiral susceptibility.
All three observables suggest that the transition to the chirally symmetric,
high temperature phase occurs at a temperature of about 
$T\sim (160-170)$~MeV.
We should stress, however, that this result has been obtained at a single value of 
the lattice cut-off and from simulations performed in a rather small physical volume, 
$N_L/N_T = V^{1/3}T = 2$.  In an  $O(4)$  scaling study of a model of the transition,
Braun {\it et al.}~\cite{Braun:2010vd} find that the pseudo-critical transition 
temperature shifts to larger values when the volume is increased.  As
mentioned above, these finite volume effects also are expected to account for 
the larger height of the susceptibility peak found when comparing our DWF 
calculations to the larger-volume staggered results.

\section{Anomalous \boldmath{$U(1)_A$} breaking above \boldmath{$T_c$}}
\label{sec:ua1}

In this section we examine the strength of anomalous axial symmetry breaking as a function of temperature 
and attempt to determine its origin.  For temperatures below $T_c$ the non-vanishing light-quark chiral 
condensate, $\langle \overline{\psi}_l\psi_l\rangle$ which breaks the non-anomalous $SU(2)_L \times 
SU(2)_R$ chiral symmetry also breaks the anomalous symmetry.  This large vacuum $\ua$ asymmetry obscures 
other possible sources of anomalous symmetry breaking so that the effects of the axial anomaly are rather 
subtle, appearing, for example in the splitting between the mass of the SU(3) flavor singlet $\eta'$ meson 
and the SU(3) flavor octet of pseudo-Goldstone bosons.  However, as the temperature is increased above 
$T_c$ this vacuum symmetry breaking disappears (as discussed in Section~\ref{sec:chiral}) so that the 
remaining $\ua$ symmetry breaking must come from the axial anomaly present in the underlying quantum field 
theory.

At high temperatures the anomalous symmetry breaking can be described using a semi-classical expansion 
known as the dilute instanton gas approximation (DIGA).   In the DIGA, the Euclidean finite temperature 
path integral is described as an integral over quantum fluctuations about a series of classical Yang-Mills 
background fields constructed from a superposition of widely separated instanton and anti-instanton 
classical solutions.  Here the (anti-)instanton size will be on the order of or smaller than $1/T$ and the 
one-loop quantum corrections imply an instanton-anti-instanton density $\propto m_l^{N_f} 
\exp\{-8\pi^2/g(T)^2\}$~\cite{'tHooft:1976fv}.  The integer $N_f$ is the number of light flavors, which 
have a small common mass $m_l$, and $g(T)$ is the running Yang-Mills coupling constant evaluated at the 
momentum scale $\mu \sim T$.   The non-zero topological charge density, $(g^2/32\pi^2)F^{\mu\nu}
(x)\widetilde{F}^{\mu\nu}(x)$ in the DIGA can be directly related to the anomalous breaking of $\ua$ 
symmetry through the familiar anomaly equation:
\begin{equation}
\partial_\mu \sum_{i=1}^{N_f} \overline{\psi}_i\gamma^5\gamma^\mu\psi_i
  = 2m_l\sum_{i=1}^{N_f} \overline{\psi}_i\gamma^5\psi_i + N_f \frac{g^2}{16 \pi^2}   F^{\mu\nu}\widetilde{F}^{\mu\nu}.
\label{eq:axial_anomaly}
\end{equation}

The detailed mechanism of anomalous symmetry breaking which realizes the consequences of 
Eq.~\eqref{eq:axial_anomaly} is well understood as the effects of infra-red singularities associated with 
the $N_f$ fermion near-zero modes that are located at each of the instantons and anti-instantons in this 
semi-classical description.  For example, in Eq.~\eqref{eq:Delta_rho} the $\ua$-asymmetric difference 
between the isovector pseudo-scalar and scalar susceptibilities, $\Delta_{\pi-\delta}$ is 
expressed in terms of an integral over the Dirac eigenvalue density $\rho(\lambda)$, divided by an
infrared-singular denominator vanishing as $m_l$ and $\lambda$ approach zero.  The DIGA in the case of 
$N_f$ degenerate light flavors implies the existence of Dirac near-zero modes whose contribution to the 
eigenvalue spectrum should be well approximated by:
\begin{equation}
\rho(\lambda) \approx c(T) m^{N_f} \delta(\lambda).
\label{eq:dilute_gas_density}
\end{equation}
The use of the delta function $\delta(\lambda)$ neglects the small splitting from zero for these near-zero 
modes which results from the interactions between the widely separated instantons and anti-instantons in 
the ``dilute'' gas.  Although Eq.~\eqref{eq:Delta_rho} contains two powers of the fermion mass and naively 
vanishes in the chiral limit, this infrared divergent denominator $(\lambda^2+m^2)^2$, when combined with 
the eigenvalue density in Eq.~\eqref{eq:dilute_gas_density}, implies a non-zero value for
$\Delta_{\pi-\delta} = c(T)$ for the case of two light flavors in the limit of vanishing quark mass.

While the DIGA is expected to be the correct description of QCD thermodynamics at high temperature, one 
might imagine a more complex mechanism for anomalous symmetry breaking when the temperature is lower and 
this semi-classical, perturbative treatment of widely separated instantons and anti-instantons is invalid.  
For example, at lower temperatures still above $T_c$ one might imagine a non-perturbative accumulation of small 
eigenvalues which leads to a density $\rho(\lambda,m) = m^{\nu_m}\lambda^{\nu_\lambda}$.  For $T>T_c$ the 
vanishing of the chiral condensate and the Banks-Casher relation requires $\nu_m+\nu_\lambda > 0$.  
However, examining Eq.~\eqref{eq:Delta_rho} we see that the $\ua$-breaking difference 
$\chi_\pi-\chi_\delta$ will remain finite in the limit of vanishing quark mass for the present case of two 
light flavors if $\nu_m + \nu_\lambda \le 1$.  Similar possible $\ua$-symmetry breaking behaviors have been 
discussed previously ~\cite{Chandrasekharan:1995gt, Cohen:1997hz, Chandrasekharan:1998yx}.

We will now examine our numerical results for anomalous symmetry breaking and their correlation with gauge-
field topology as well as the Dirac eigenvalue spectrum itself.  In particular, we will discuss the 
anomalous symmetry breaking differences in both connected and disconnected susceptilities as well as in the 
underlying Green's functions evaluated in position space.  We will also compare our results with the 
predictions of the high-temperature DIGA and search for possible new mechanisms for $\ua$ symmetry breaking 
at temperatures closer to $T_c$.

\subsection{Connected and disconnected susceptibilities}
\label{sec:suscept}

As discussed in Section~\ref{sec:basics}, an accessible observable to examine is the $\ua$
symmetry breaking difference $\chi_\pi-\chi_\delta$.   In that Section we also showed in 
Eq.~\eqref{eq:Delta_rel} that the difference $\chi_\pi-\chi_\delta$, the disconnected chiral 
susceptibility $\chi_{\rm disc}$, and the disconnected pseudo-scalar susceptibility $\chi_{5,{\rm disc}}$ 
all become equal in the chiral limit for $T \ge T_c$ as a direct consequence of  $\sua$ symmetry.  In 
addition, $\chi_\pi-\chi_\delta$ is directly related to the Dirac eigenvalue density through 
Eq.~\eqref{eq:Delta_rho}.

\begin{figure}[hbt]
\begin{center}
\includegraphics[width=0.5\textwidth]{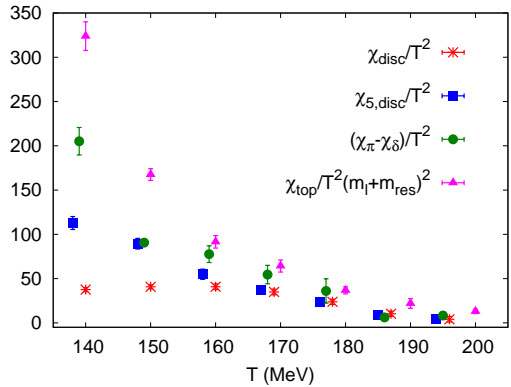}
\caption{The disconnected scalar (chiral) and pseudo-scalar susceptibilities plotted versus temperature as 
crosses and squares respectively.   The circles show the $\ua$-breaking difference $\chi_\pi-\chi_\delta$, 
which in the chiral limit will become equal to both disconnected susceptibilities above $T_c$.  Finally the 
triangles represent the topological susceptibility divided by the square of the total bare quark mass, 
$m_f+m_{\rm res}$, a combination which should equal the pseudo-scalar susceptibility at all temperatures, 
as in Eq.~\ref{eq:top2chi_5disc}.  The large discrepancy between $\chi_{\rm top}/(m_f+m_{\rm res})^2$ and 
$\chi_{5,{\rm disc}}$ is believed to arise from large lattice artifacts in the determination of $\chi_{\rm 
top}$ as discussed below and in Appendix~\ref{sec:g5-top_suscept}}
\label{fig:ua1delta}
\end{center}
\end{figure}

\begin{table}[hbt]
\centering
\begin{tabular}{ccccccc}
Label & $T$(MeV)& $\chi_\pi/T^2$ & $\chi_\delta/T^2$  & $(\pmd)/T^2$ & $\chi_{5,{\rm disc}}/T^2$ & $\chi_{\rm top}/T^2$\\ 
\hline
\runn{140} &   139 &  283(11)  & 78(6) & 205(16)   & 113(7)& 6.6(3) $\times 10^{-3}$\\ 
\runn{150_32} &  149 &  178(3) &  87(1) & 91(4) &  89(6)& 3.7(1) $\times 10^{-3}$\\ 
\runn{160} & 159 &  177(7) &  99(6) & 78(9)   & 55(6) & 1.7(1) $\times 10^{-3}$ \\ 
\runn{170} & 168 &  139(7) & 85(6) & 55(10)&  37(5) & 0.95(10) $\times 10^{-3}$\\ 
\runn{180} & 177 &  113(9)  & 77(6) & 36(14)&  24(4) & 0.49(5) $\times 10^{-3}$\\   
\runn{190} & 186 &  93(2) & 87(1) & 6(2)  & 9(3) & 0.24(6) $\times 10^{-3}$ \\ 
\runn{200} & 195 &  88(2) & 79(2) & 8(4)  & 5(4) & 0.13(3) $\times 10^{-3}$\\ 
\hline
\end{tabular}
\vspace{0.5cm}
\caption{Our results for the susceptibilities $\chi_\pi$, $\chi_\delta$, $\pmd$, $\chi_{5,{\rm disc}}$, and $\chi_{\rm top}$.}
\label{tab:pmd}
\end{table}

These three observables are plotted in Fig.~\ref{fig:ua1delta} and their numerical values for the DSDR 
ensembles are given in Tabs.~\ref{tab:pbp} and~\ref{tab:pmd}.  All three, $\chi_{\rm disc}$, $\chi_{5,{\rm 
disc}}$ and $\pmd$, agree within errors for $T\geqslant 168$ MeV suggesting both a restoration of vacuum 
$\sua$ symmetry and that our $\sim 10$ MeV quark mass and resulting 200 MeV pion introduce a sufficiently 
small explicit chiral symmetry breaking that its effects are not visible at our level of accuracy.  
Especially interesting is the fact that the $\ua$ breaking difference, $\chi_\pi-\chi_\delta$, is non-zero 
throughout the temperature range considered here. This suggests that $\ua$ remains explicitly broken even 
after chiral symmetry is restored.  Furthermore, since the symmetry breaking effects of the non-zero quark mass 
produce no visible discrepancies between $\chi_{\rm disc}$, $\chi_{5,{\rm disc}}$ and $\pmd$, it is 
reasonable to expect that the difference between $\chi_\pi$ and $\chi_\delta$ arises from the axial anomaly 
--- not the non-zero quark mass.

Also shown in Fig.~\ref{fig:ua1delta} is the combination $\chi_{\rm top}/(m_f+m_{\rm res})^2$ which is 
expected to be equal to the pseudo-scalar susceptibility  $\chi_{5,{\rm disc}}$, following 
Eq.~\ref{eq:top2chi_5disc}.  As can be seen in the figure this expectation is badly violated, with these 
two quantities differing by more than a factor of two at the lowest temperature.  As is discussed in 
greater detail in Appendix~\ref{sec:g5-top_suscept}, we have examined our results for these two quantities 
carefully and believe that our calculation of $\chi_{\rm top}$ is not reliable at the large lattice 
spacings and non-zero temperatures being explored here.  The quantity $\chi_{5,{\rm disc}}$ is determined 
directly from the Dirac propagator on the lattice and has a well-understood continuum limit.  In contrast, 
the topological susceptibility is obtained from an empirically justified procedure of gauge link smearing 
steps followed by the evaluation of an improved combination of links chosen to approximate the topological 
charge density $F\widetilde{F}$.  As shown in  Appendix~\ref{sec:g5-top_suscept}, these two quantities do 
not agree at non-zero temperature, despite the fact that there is good agreement at zero temperature, even 
at our coarsest lattice spacings. 

\subsection{Position-space corrrelators}

Additional understanding of this $\ua$ symmetry violation comes from examining the spatial correlators 
themselves. We begin by writing the iso-vector scalar and pseudo-scalar correlators (those for the $\delta$ 
and the $\pi$) in terms of their left- and right-handed components,
\begin{equation}
\begin{split}
G_{\pi/\delta}(x) &=   \big\langle \ubl\dr(x)\dbr\ul(0) + \ubr\dl(x)\dbl\ur(0) \big\rangle \\
                  &\pm \big\langle \ubl\dr(x)\dbl\ur(0) + \ubr\dl(x)\dbr\ul(0) \big\rangle.
\label{eq:c_sps}
\end{split}
\end{equation}
Here the left- and right-handed parts are defined as
\begin{align}
&& u_{L/R}(x) = \left(\frac{1\mp\gamma_5}{2}\right)u(x), && d_{L/R}(x) = \left(\frac{1\mp\gamma_5}{2}\right)d(x), && \\
&& \ub_{L/R}(x) = \ub(x)\left(\frac{1\pm\gamma_5}{2}\right), && \db_{L/R}(x) = \db(x)\left(\frac{1\pm\gamma_5}{2}\right) &&
\end{align}
In Eq.~\eqref{eq:c_sps}, the terms on the first line are invariant under $\ua$ rotations. These occur with the same sign for both the $\delta$ and the $\pi$ correlators. By contrast the terms on the second line, which occur with opposite signs for the two correlators, are not invariant under $\ua$ transformations and their expectation value should therefore vanish in a $\ua$-symmetric theory.

\begin{figure}[hbt]
\centering
\hspace{-0.12\textwidth}%
\includegraphics[width=0.45\textwidth]{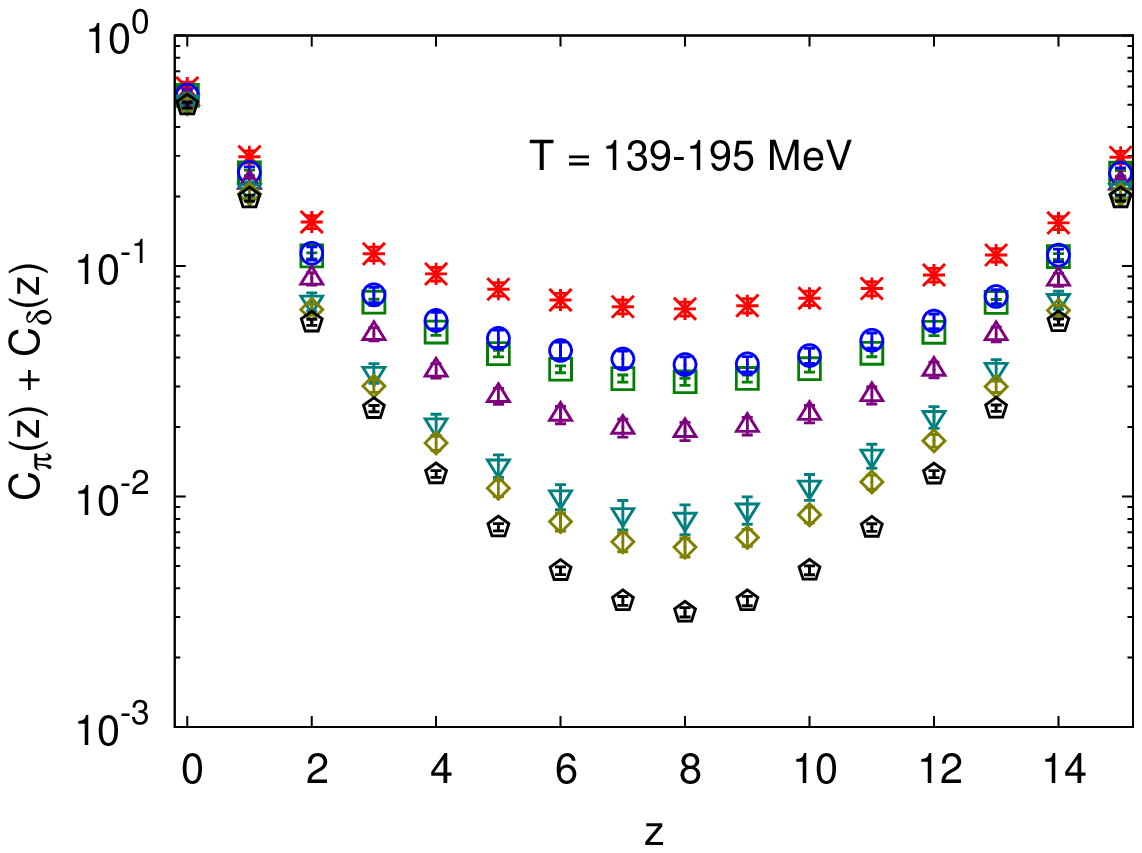}%
\hspace{0.05\textwidth}%
\includegraphics[width=0.45\textwidth]{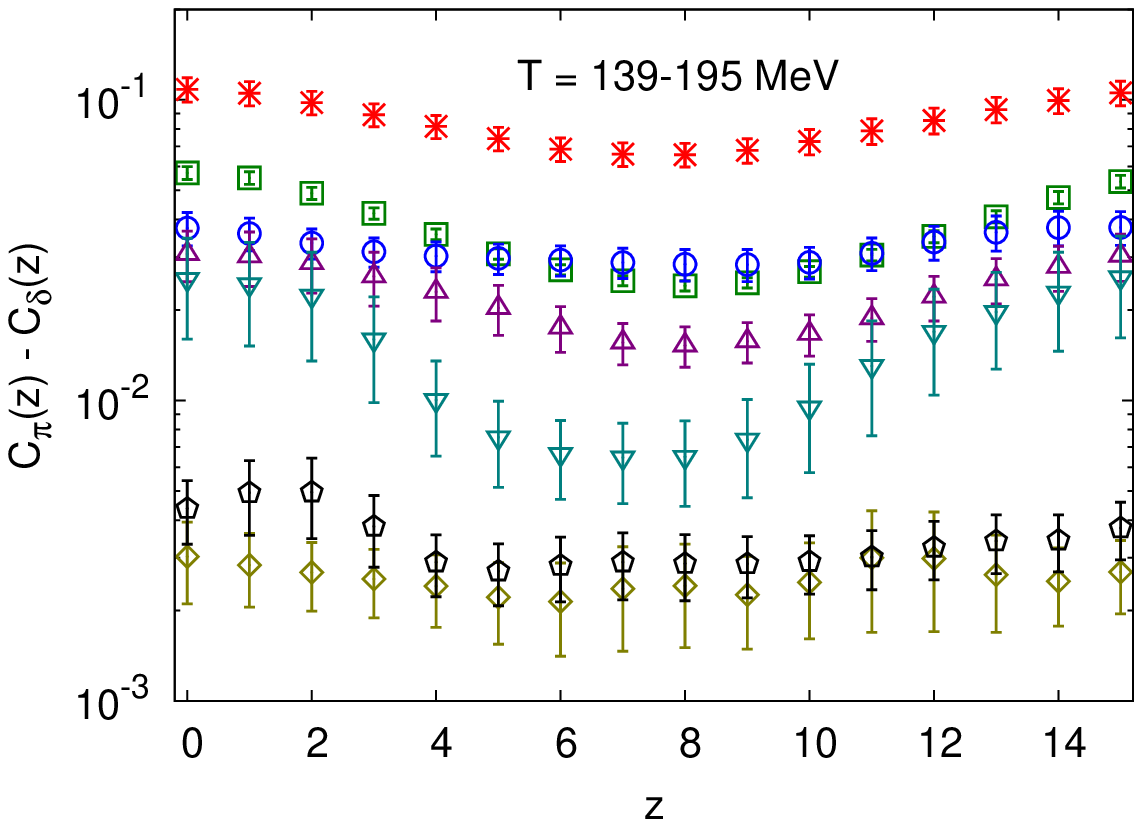}
\caption{\small (Left) The sum of the spatial $\pi$ and the $\delta$ correlators. The temperature increases from $T=139$ MeV to 195 MeV as one moves downward along the $y$-axis. (Right) The difference $C_\pi(z)-C_\delta(z)$. The temperatures are identified by the same symbols as in the sum.  The monotonic decreasing behavior seen with increasing temperature in the left panel is not seen for the highest temperatures in the right panel where the $T=195$ MeV data lies slightly above that for $T=186$.  However, this apparent diminished rate of decrease with increasing temperature may be an artifact of insufficient statistics since the statistical
errors on this signal, which, as discussed in Sec. \ref{subsec:topology}, arises from infrequent spikes in the data, may be underestimated.}
\label{fig:s_pm_ps}
\end{figure}

The invariant and non-invariant parts of these correlators may be isolated by taking the sum and difference respectively of the two correlators. These are shown in Fig.~\ref{fig:s_pm_ps} for all the temperatures.  Actually, what are plotted are the \emph{screening correlators} $C(z)$, which are related to the corresponding point-to-point correlators by
\begin{align}
& & C_H(z) = \sum_{x,y,\tau}G_H(x,y,z,\tau), & & H = \pi,\,\delta,\,\rho,\,\text{etc.} & &
\end{align}
We see that the difference $C_\pi(z)-C_\delta(z)$ is always nonzero.  For source-sink separations within a few lattice spacings of zero, this non-zero value is dwarfed by the much larger non-anomalous contribution to  $C_\pi(z)$ and $C_\delta(z)$ and this disparity grows with increasing temperature.  However, while its magnitude decreases as $T$ is increased, the difference is always comparable to the sum $C_\pi(z)+C_\delta(z)$ at the largest source-sink separations viz. $x\approx N_\sigma/2$.  This suggests a significant breaking of $\ua$ symmetry for this long-distance quantity, even with increasing temperature.  However, studies with a varying quark mass are required to
establish this as an effect of the anomaly.

\begin{figure}[hbt]
\centering
\includegraphics[width=0.45\textwidth]{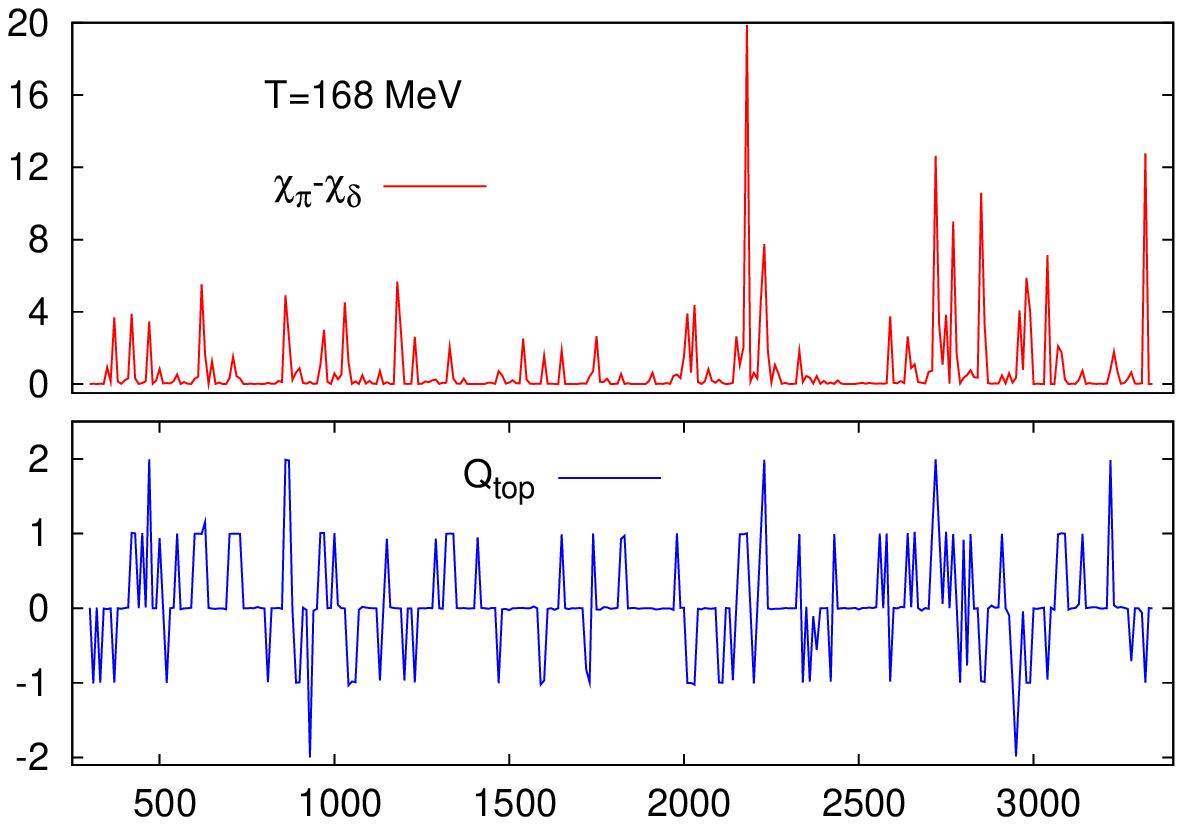}%
\hspace{0.05\textwidth}%
\includegraphics[width=0.45\textwidth]{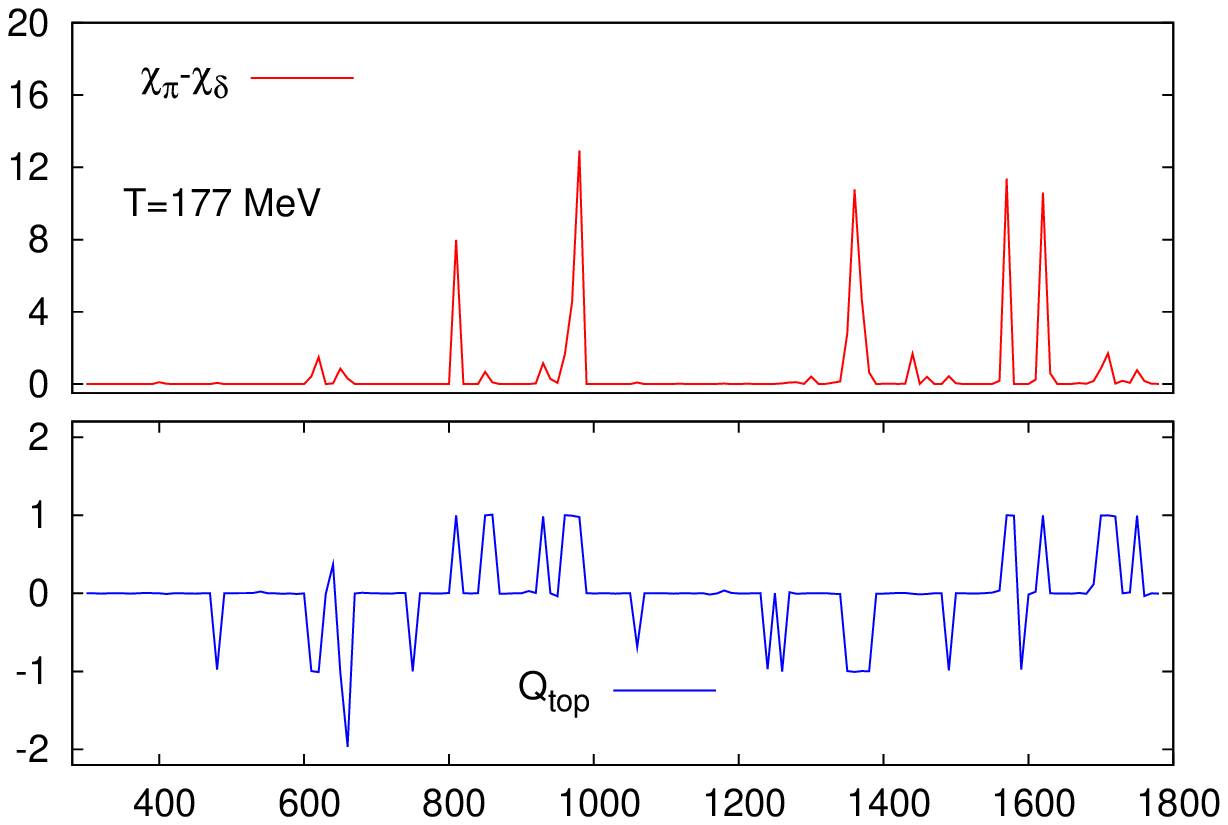}
\vspace{0.1\textheight}
\includegraphics[width=0.45\textwidth]{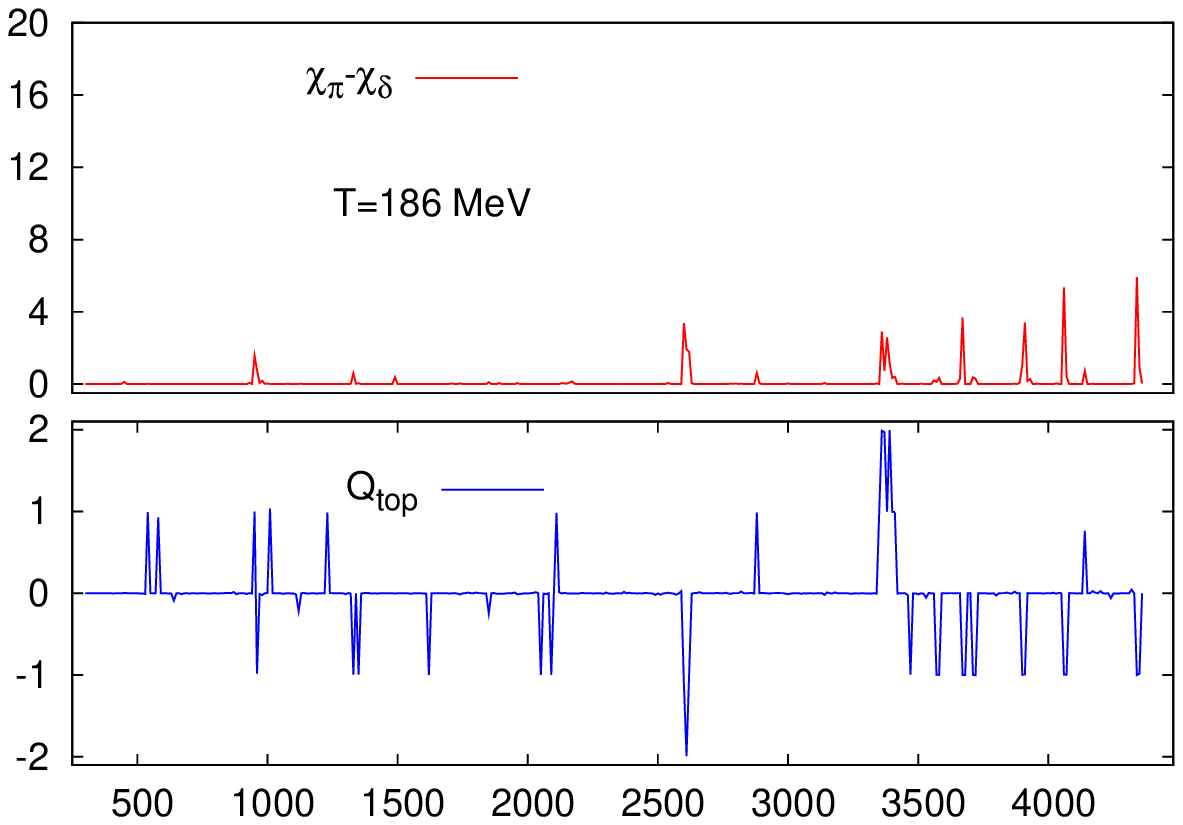}%
\hspace{0.05\textwidth}%
\includegraphics[width=0.45\textwidth]{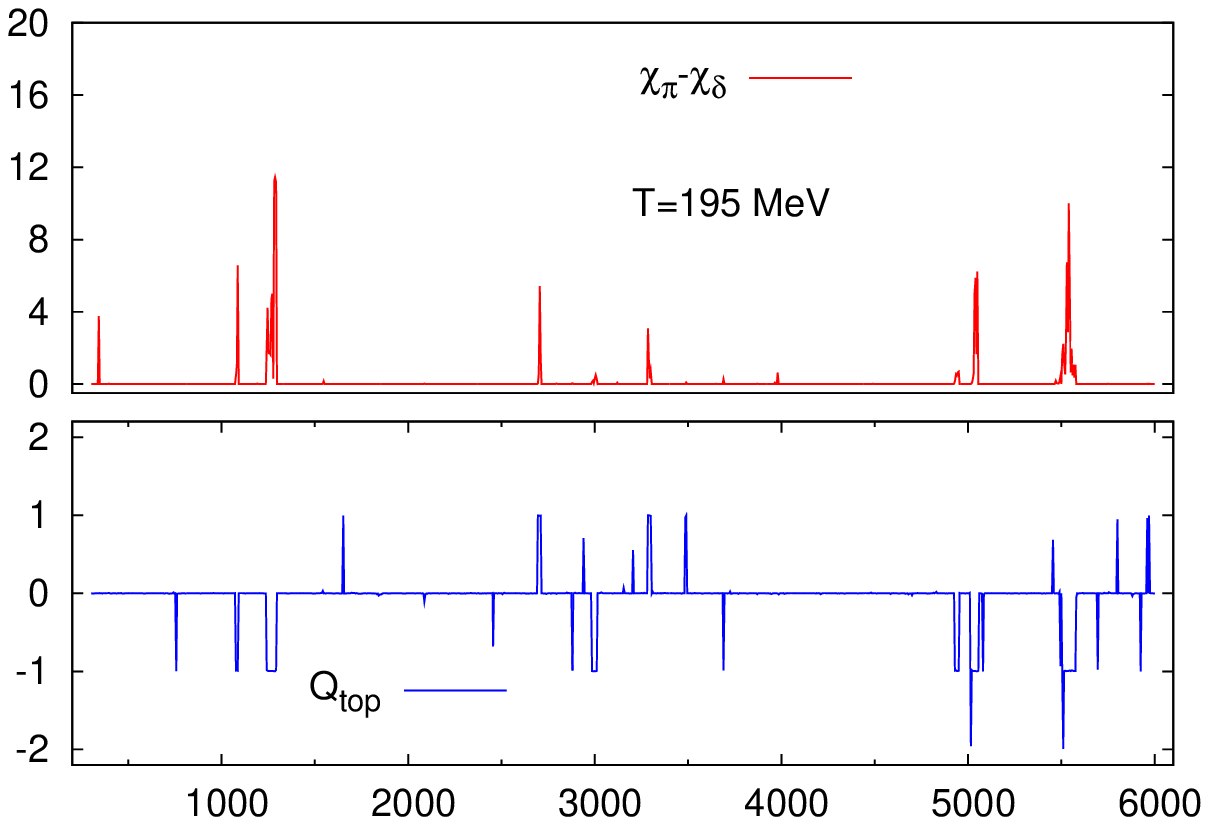}
\caption{\small The time histories for the topological charge (blue lines) and the integrated correlator $\chi_\pi-\chi_\delta$  (red lines) for $T=168$--195 MeV.  These time histories have been labeled with the quantities that result when those histories are time averaged.}
\label{fig:sps_time_hist}
\end{figure}

\subsection{Correlation with topology}
\label{subsec:topology}

The connection between the $\ua$-breaking difference $\chi_\pi-\chi_\delta$ and the topology of the gauge fields can be studied by comparing the Monte Carlo time histories for these two quantities. Figure~\ref{fig:sps_time_hist} contains plots of the time histories of the measurements whose average gives the connected susceptibility difference $\chi_\pi-\chi_\delta$ and the topological charge $\qtop$.   On our finite temperature gauge configurations, $\qtop$ is computed on each gauge configuration using the five loop 
improved (5Li) gauge field operator introduced in \cite{deForcrand:1997sq}.  $\qtop$ is measured 
after the gauge fields are smoothed by applying 60 APE smearing steps \cite{Albanese:1987ds} 
with smearing coefficient $\epsilon = 0.45$, so that $\qtop$ gives near-integer values.  We see that $\ua$ 
is not broken ``on average'' but rather only on specific configurations. These tend to
be the configurations with $\qtop\neq0$.

However, as discussed in Appendix~\ref{sec:g5-top_suscept}, the use of the 5Li method and cooled gauge fields to compute $\qtop$
is contaminated by significant lattice artifacts, particularly at stronger coupling.  This is reflected
by the less than perfect correlation between $\qtop$ and contributions to $\chi_\pi - \chi_\delta$
in Fig. \ref{fig:sps_time_hist}.  On a few configurations with $\qtop$ apparently non-zero there
is no evident contribution to $\chi_\pi - \chi_\delta$ while on some other configurations with $\qtop = 0$,
there is a non-zero contribution to $\chi_\pi - \chi_\delta$.

Despite the imperfections in $\qtop$, the correlation between  $\ua$-breaking and gauge field topology 
can still be qualitatively observed in our data.  This connection is similar to that predicted by the DIGA.  However, in that picture $\ua$-breaking is connected with the total number of instantons and anti-instantons, $N_I + N_{\overline{I}}$, not their difference, $N_I - N_{\overline{I}}$, which is determined by the gauge-field topology.   For example, we should expect to occasionally see a configuration containing a widely separated instanton and anti-instanton in which the resulting two near-zero modes produce a large spike in the time history of $\chi_\pi-\chi_\delta$ but which does not appear in the time history of the topology.  It is not obvious that there are examples of such a phenomena in Fig.~\ref{fig:sps_time_hist}.  Of course, our volume may be too small for multiple instantons/anti-instantons.  This is also suggested by the preponderance of three topological charges 0, $\pm 1$ and reflected in the direct determination of the density of Dirac near-zero modes presented in the following section.  Note, the fluctuations seen in the time histories of $\chi_\pi-\chi_\delta$ shown in Fig.~\ref{fig:sps_time_hist} arise in part from the method used to calculate this quantity and have only an indirect physical meaning.  At least a portion of these fluctuations arise from the occasional coincidence between the space-time location of the fixed point-source used in computing $\chi_\pi$ and $\chi_\delta$ and the random location of a localized near-zero mode, rather than from an increased number of near-zero modes.

\subsection{Dirac eigenvalue density}
\label{ssec:form_specdens}

Since the infra-red structure of QCD underlies the anomalous breaking of $\ua$ symmetry, we expect that much can be learned from explicitly examining the eigenvalue spectrum of the Dirac operator near zero eigenvalue.  
For earlier studies of the Dirac eigenvalue spectrum using staggered and overlap fermions see Refs.~\cite{Gockeler:2000jk, Damgaard:2000cx, Gavai:2001vx, Gavai:2008xe, Cossu:2010rc, Ohno:2011yr}.
Knowing the Dirac spectrum, we can directly examine the eigenvalue density $\rho(\lambda)$,
discussed in Section~\ref{sec:dirac}, looking for the behavior as $\lambda \to 0$ necessary to produce a $\ua$-breaking difference $\chi_\pi-\chi_\delta$ from Eq.~\eqref{eq:Delta_rho}.  We can compare our calculated density of eigenvalues  $\rho(\lambda)$ with what is expected in the case of a dilute instanton gas and look for possible new,  $\ua$-breaking behaviors as $T$ approaches $T_c$ from above.  In this subsection we will first present our numerical results and then discuss possible behaviors for  $\rho(\lambda,m)$ as the light quark mass $m_l$ and Dirac eigenvalue $\lambda$ approach zero.

\subsubsection{Numerical results for $\rho(\lambda)$}

\begin{figure}[h]
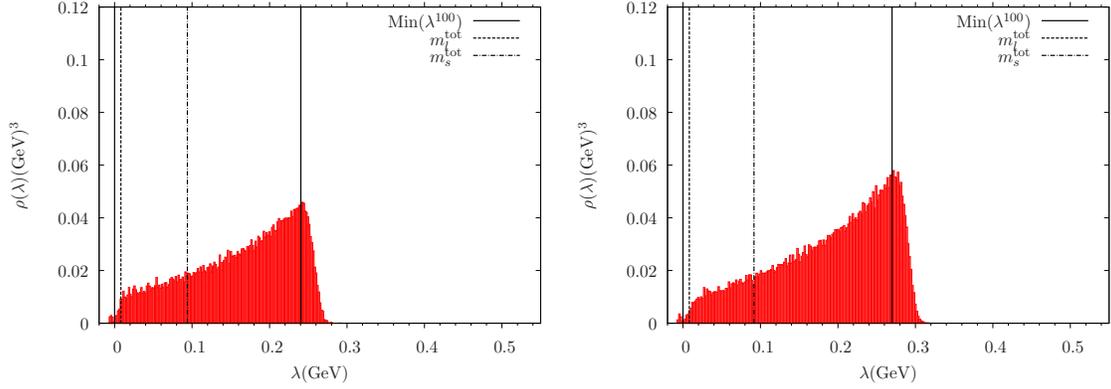

  \begin{center}
        \begin{minipage}[t]{0.5\linewidth}
                \centering
        \resizebox{\linewidth}{!}{\input{./figs/150MeV_1_norm.tex}}
        \end{minipage}%
        \begin{minipage}[t]{0.5\linewidth}
                \centering
        \resizebox{\linewidth}{!}{\input{./figs/160MeV_norm.tex}}
        \end{minipage}
  \end{center}
  \caption{Renormalized Dirac spectrum 149 MeV $L_s=32$ (left) and 159 MeV (right).}
  \label{fig:eig_150_160}
\end{figure}

\begin{figure}[h]
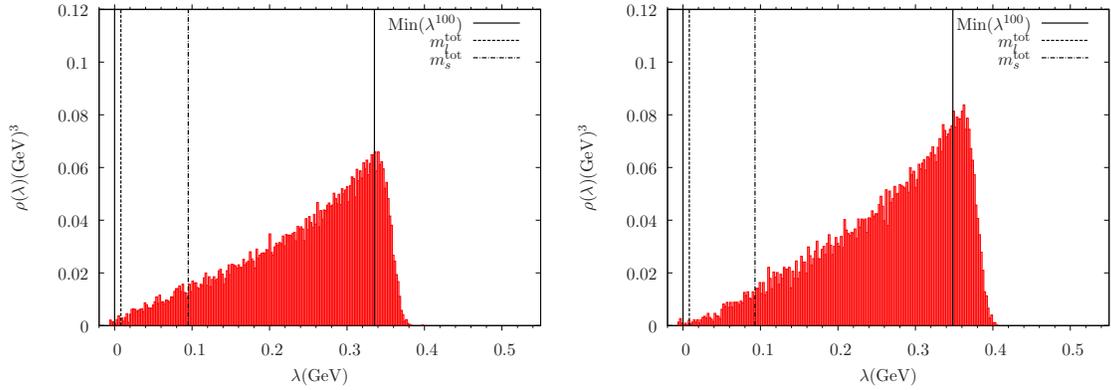

  \begin{center}
        \begin{minipage}[t]{0.5\linewidth}
                \centering
        \resizebox{\linewidth}{!}{\input{./figs/170MeV_norm.tex}}
        \end{minipage}%
        \begin{minipage}[t]{0.5\linewidth}
                \centering
        \resizebox{\linewidth}{!}{\input{./figs/180MeV_norm.tex}}
        \end{minipage}
  \end{center}
  \caption{Renormalized Dirac spectrum 168 MeV (left) and 177 MeV (right).}
  \label{fig:eig_170_180}
\end{figure}

\begin{figure}[h]
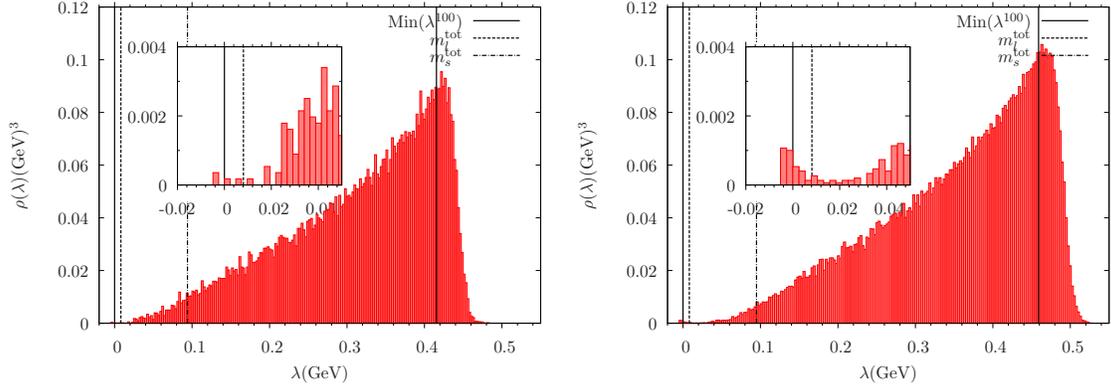

  \begin{center}
        \begin{minipage}[t]{0.5\linewidth}
                \centering
        \resizebox{\linewidth}{!}{\input{./figs/190MeV_norm.tex}}
        \end{minipage}%
        \begin{minipage}[t]{0.5\linewidth}
                \centering
        \resizebox{\linewidth}{!}{\input{./figs/200MeV_1_norm.tex}}
        \end{minipage}
  \end{center}
  \caption{Renormalized Dirac spectrum 186 MeV (left) and 195 MeV (right).}
  \label{fig:eig_190_200}
\end{figure}

In Figs.~\ref{fig:eig_150_160}, \ref{fig:eig_170_180} and \ref{fig:eig_190_200} we present our results for the $\rho(\lambda)$, with both $\rho$ and $\lambda$ normalized in the $\mu=2$ GeV, $\overline{\rm MS}$ scheme, determined from the 100 lowest eigenvalues calculated at each of six temperatures using the methods  
explained in Section~\ref{sec:dirac}.  The number of configurations used in each case varied from 239 to 1140 and is listed in Tab.~\ref{tab:EigenConfigs}.  Here we are presenting the lattice analogue of the usual Dirac eigenvalue $\lambda$ from which the quark mass has been removed, $\lambda = \sqrt{\Lambda^2-(m_f+m_{\rm res})^2}$.  As explained in Section~\ref{sec:dirac}, at finite lattice spacing this assumed mass dependence for the full Dirac eigenvalues $\Lambda$ is only approximate and in some cases the argument of the square root is negative.  In those cases the resulting $\lambda$ is placed on the histogram at the unphysical position $-|\lambda|$, allowing this type of $a^2$ error to be recognized.

At both $T=149$ and 159 MeV, the spectrum appears to be approaching a non-zero intercept as 
$\lambda$ approaches zero until $\lambda \sim 10$ MeV, when the eigenvalue density decreases rapidly toward zero.  As is suggested by the behavior of the chiral condensate in Fig.~\ref{fig:condensate} and the disconnected chiral susceptibility in Fig.~\ref{fig:chi_dis}, both the 149 and 159 MeV temperatures lie close to the crossover temperature and well within the transition region, broadened by the effects of finite size and finite quark mass.  Thus, it appears difficult to determine the character of either $\sua$ or $\ua$ symmetry restoration at these temperatures without examining larger volumes and smaller quark masses.

For  the temperatures $T=168$ and 177 MeV the small $\lambda$ behavior has qualitatively changed.  The 
pronounced shoulder near $\lambda = 10$ MeV has disappeared and instead the spectral density is approaching 
zero in a more linear fashion.  Looking carefully at the region $\lambda\approx 0$ for $T=168$ MeV, one 
sees what appears to be essentially linear behavior as $\lambda \to 0$.  At $T=177$ MeV similar behavior 
can be seen, although because of our limited statistics, $\rho(\lambda)$ could vanish with a higher-than-
linear power.   For $T=186$ MeV the behavior has changed again, with very few eigenvalues found below 20 
MeV.  At $T=195$ MeV, where larger statistics better populate this interesting region, $\rho(\lambda)$ 
decreases to a minimum near 20 MeV and then increases to a peak near $\lambda=0$.

This behavior at  $T=195$ MeV is consistent with that expected from the DIGA.  However, integrating over 
this small peak for $\lambda \le 20$ MeV and including those eigenvalues plotted to the left of zero, we 
find an average number of near-zero modes of 0.06/MeV.   With such a low density of near zero modes, we 
expect that the spectral broadening arising from the simultaneous presence of instantons and 
anti-instantons will be unimportant.  Thus, it appears likely that the spread of eigenvalues about zero seen for  
$T=195$ MeV is the result of finite lattice spacing.  This conclusion is consistent with the approximately 
equal number of eigenvalues $\Lambda$ slightly above $m_l+m_{\rm res}$ (giving $\lambda >0$) and the number 
slightly below (giving $\lambda$ imaginary and plotted as $-|\lambda|$ to the left of zero.  If this is 
correct, then we should expect that at  $T=195$ MeV and for a volume of spatial size $L\approx 2$ fm, 
$\rho(\lambda)$ will accurately approach a delta function, $\delta(\lambda)$ as $a \to 0$.

In summary, our study of the Dirac eigenvalue spectrum has provided limited but interesting results.
For our $\approx 10$ MeV quark mass and 2 fm spatial box, the transition region appears sufficiently broad 
that the spectral density found at $T=149$ and 159 MeV is strongly influenced by finite volume effects.  At 
$T=168$ and 177 MeV interesting, possibly non-perturbative behavior is seen in the
low-lying eigenvalue spectrum, $\rho(\lambda) \sim \lambda^\alpha$ with $\alpha \sim 1-2$, very different from the behavior of the free Dirac spectrum at finite temperature.  Determining whether this behavior can support the breaking of $\ua$ symmetry will require exploration with larger volumes and smaller masses.  Finally, near zero modes are clearly evident at the highest $T=186$ and 195 MeV temperatures, consistent with a very dilute instanton gas of density $\approx$ 0.01/fm$^4$.

\subsubsection{Possible behaviors for $\rho(\lambda,m)$}

Given the range of behaviors seen above for the function $\rho(\lambda)$ for $T$ above the transition region, $T \ge 168$ MeV, it may be useful to discuss the consequences of possible  functional forms of $\rho(\lambda,m)$ for the chiral condensate, the susceptibilities $\chi_\pi$, $\chi_\delta$, their difference,  $\chi_\pi - \chi_\delta$, and the disconnected chiral susceptibility $\chi_{\rm disc}$.   In addition to the Banks-Casher relation given in Eq.~\eqref{eq:pbp_rho}, and Eq.~\eqref{eq:Delta_rho} for the difference $\chi_\pi - \chi_\delta$, we can also relate $\chi_\pi$ to the eigenvalue density $\rho(\lambda)$ by inserting an eigenmode expansion in the expression for $\chi_\pi$ and obtain:
\begin{equation}
\chi_\pi  = \int_0^\infty\dL\; \density\frac{2}{m^2+\lambda^2} = \frac{\lpbpr}{m}.
\label{eq:pi_rho}
\end{equation}

Finally the full chiral susceptibility $\chi_\sigma=\chi_{\rm con}+\chi_{\rm disc}$ is given by
\begin{align}
\frac{\partial}{\partial m}\lpbpr 
     &= \int_0^\infty\dL\; \density\frac{\partial}{\partial m}\left[\frac{2m}{m^2+\lambda^2}\right]  
    \label{eq:chi_rho1} \\
     &+ \int_0^\infty\dL\;         \frac{\partial}{\partial m}\left[\density\right]\frac{2m}{m^2+\lambda^2},\notag\\
     &\equiv \chi_{\rm con}+\chi_{\rm disc}.
\label{eq:chi_rho}
\end{align}

We will now use these equations to determine the behavior of  $\pbp$, $\chi_\pi$, $\chi_\delta$ and $\chi_{\rm disc}$ in the limit $m\to 0$ for three different assumed behaviors of $\rho(\lambda,m)$.  The first is the behavior predicted by the DIGA, $\rho(\lambda,m)=C_0m^2\delta(\lambda)$.  Next we consider the hypothesis that above $T_c$ the density of eigenvalues is an analytic function of the quark mass and eigenvalue.  To linear order, this gives two possible terms for $T \ge T_c$ since the constant term $\rho(0,0)$ must vanish:
\begin{equation}
\density = C_1\lambda + C_2 m + O(\lambda m) + \dots
\label{eq:spec_taylor}
\end{equation}
Table ~\ref{tab:behaviors} lists the behavior for each of these four quantities that results from each Ansatz.

\begin{table}[htb]
\centering
\begin{tabular}{c|ccccc}
  Ansatz                         & $\lpbpr$         & $\chi_\pi$       & $\chi_\delta$  
                                                                                                                     & $\chi_\pi-\chi_\delta$
                                                                                                                                  & $\chi_{\rm disc}$ \\ \hline \hline
$m^2 \delta(\lambda)$  & $m$              & $1$          & $-1$
                                                                                                                     & $2$
                                                                                                                                  & $2$   \\
$\lambda$                     &  $-2m\ln(m)$ & $-2\ln(m)$ & $-2\ln(m)$
                                                                                                                     & $2$
                                                                                                                                  & $0$          \\ 
$m$                          &  $\pi m$            & $ \pi$       &$0$ 
                                                                                                                     & $\pi$
                                                                                                                                  &  $\pi$          
\end{tabular} 
\vspace{0.5cm}
\caption{Limiting behavior of various thermodynamic quantities as $m\to 0$ for three possible forms of $\rho(\lambda,m)$ for small $m$ and $\lambda$.  Note that the results in the right-hand columns have
the correct multiplicative coefficients, given the ans\"atze for $\rho(\lambda, m)$ in the leftmost column.}
\label{tab:behaviors}
\end{table}

The ansatz $\density\propto\lambda$ yields a finite $\pmd$ in the chiral limit. However the mechanism by which it does so is somewhat unusual. The chiral condensate of this theory vanishes as $m\ln m$ in the chiral limit. The logarithm shows up as a divergence in the susceptibilities $\chi_\pi$ and $\chi_\delta$. However it cancels out in the difference, leading to a finite $\pmd$. Lastly, since there is no $m$ dependence in the spectral density, the disconnected chiral susceptibility vanishes  according to Eq.~\eqref{eq:chi_rho} and $\chi_\pi-\chi_\delta \neq \chi_{\rm disc}$.  As we have already seen in Eq.~\eqref{eq:Delta_rel}, the failure of this equality would imply the breaking of $\sua$ symmetry for $T>T_c$.  

By contrast, the ansatz $\density\propto m$ does not give rise to logarithmic divergences. The chiral condensate vanishes linearly in the quark mass, the susceptibilities $\chi_\pi$ and $\pmd$ both remain finite and furthermore $\pmd=\chi_{\rm disc}$ as well. Interestingly however, the susceptibility $\chi_\delta$ vanishes in the chiral limit. The equality $\pmd=\chi_{\rm disc}$ is therefore just the equality $\chi_\pi=\chi_{\rm disc}$.

The contrasting possibilities shown in Tab.~\ref {tab:behaviors} suggest that future studies of these susceptibilities in the limit of small quark mass will also reveal which of these behaviors for $\rho(\lambda,m)$ is present and the underlying mechanism of $\ua$ symmetry breaking as a function of temperature for $T\ge T_c$.

\section{Conclusion}
\label{sec:conclusions}

The finite temperature properties of QCD are immediately accessible
to standard, Euclidean-space calculations in lattice QCD.  In fact,
lattice QCD has provided valuable,  {\it ab initio} information and 
insights into QCD thermodynamics since its inception.   However, 
the need to work in the large-volume, thermodynamic limit makes 
this a challenging application for lattice methods.  The needed 
large physical volumes are achieved by working at relatively large 
lattice spacing, making QCD thermodynamics calculations especially 
vulnerable to finite lattice spacing errors and restricting the range of 
lattice spacings available to carry out a reliable continuum limit.  As a 
result, it is important to examine the thermodynamic properties of 
QCD using a variety of lattice actions, as the effects of lattice
discretization errors are likely to vary between different choices of
lattice action. 

An appealing fermion action to use when studying the QCD chiral phase
transition is the domain wall action which accurately respects the 
chiral symmetry whose vacuum breaking and restoration drives 
this transition.   Unfortunately, the large lattice spacings which are 
needed for thermodynamics studies are a special problem for the 
domain wall formulation where the rough gauge fields characteristic 
of large lattice spacing induce sizable explicit chiral symmetry breaking 
unless the size of the fifth dimension is 
made very large.  As a result, earlier studies of QCD thermodynamics 
using domain wall fermions~\cite{Chen:2000zu,Cheng:2009be}  have 
been compromised by the resulting large residual chiral symmetry 
breaking effects.  Because the residual chiral symmetry breaking increases
at the larger lattice spacing associated with lower temperatures, these
effects can potentially distort the observed temperature dependence
seen in the transition region.

In the calculation reported here, we have succeeded in controlling
these effects.  First we have shown results from a brute force approach
using a very large fifth-dimensional extent of $L_s=96$.  Second, we
have employed the carefully tuned DSDR gauge action where the short
distance structure has been chosen to suppress the gauge field 
dislocations which induce explicit chiral symmetry breaking.  As a result, 
we are able to report a systematic study of the transition region on a 
line of constant physics with a pion mass of 200 MeV.  This has been 
achieved using the DSDR gauge action, $L_s=32$ or 48 and a small 
input bare quark mass which varies from positive to negative as the 
temperature is decreased below 159 MeV.

Using this chirally symmetric lattice fermion formulation  we have been 
able to confirm the expected chiral behavior of the QCD phase transition
seen using staggered fermions.  Specifically, in a lattice formulation 
with three degenerate light pions of fixed physical mass possessing the 
$SU(2)_L \times SU(2)_R$ chiral symmetry found in Nature, we see a crossover 
behavior going from the low temperature region, $T \le 159$ MeV, with 
vacuum chiral symmetry breaking to a chirally symmetric phase at higher 
temperature, $T \ge$ 168 MeV in which the large, low-temperature chiral 
condensate has dramatically decreased and the spatial Green's functions 
and screening lengths show good chiral symmetry.  

We have explored this phenomena microscopically by examining the 
spectrum of the fermion Dirac operator, normalized using standard 
$\overline{\mbox{MS}}$ conventions.   We find the expected non-zero 
eigenvalue density for small eigenvalues at low temperature required by 
vacuum chiral symmetry breaking and the Banks-Casher relation.   As 
the temperature increases, this density at small eigenvalue decreases 
dramatically until $T=$ 186 and 195 MeV where we find a striking absence 
of small eigenvalues.  In fact, except for a small density near zero, which 
may be attributed to semi-classical instanton effects, one might identify 
a gap in the spectrum below 20 MeV at these two highest temperatures.  
In the important region closer to $T_c$, $159~\textrm{MeV} < T < 177$ MeV, 
the behavior of the eigenvalue spectrum remains uncertain.  
While one might assign linear behavior, $\rho(\lambda) \propto \lambda$, 
at small $\lambda$ to the $T = 168$ MeV spectrum shown in 
Fig.~\ref{fig:eig_170_180}, the picture could also change dramatically 
with increased volume.

Of particular interest in the current study is the degree to which the
anomalous $U_A(1)$ symmetry is found to be broken
at high temperature.  For temperatures below the chiral transition,
both the anomalous and non-anomalous axial symmetries are broken
by the vacuum, making the effects of the axial anomaly difficult to
see.  (Only the relatively heavy $\eta'$ meson stands out at low energy
as a consequence of the axial anomaly.)  However, above the QCD
phase transition, the three non-anomalous axial symmetries are 
explicitly realized in a Wigner mode and the effects of the axial 
anomaly on the potential $U_A(1)$  symmetry can be easily explored.  
We find rapidly decreasing $\ua$-breaking susceptibilities and 
susceptibility differences with increasing temperature.  At our highest 
temperatures of 186 and 195 MeV, $\ua$ symmetry is largely realized 
with the small remaining asymmetries appearing to arise from relatively 
rare gauge field configurations carrying non-trivial topology.  The dearth 
of small Dirac eigenvalues at high temperatures mentioned above supports 
this picture of effective  $U_A(1)$ symmetry restoration.

It should be emphasized that the calculations reported here have been
carried out on a relative small, $16^3\times 8$ physical volume.  This 
aspect ratio of spatial to temporal size of 2 is much smaller than that
in the typical staggered fermion calculation and introduces important
uncertainties in our results.  While the disconnected chiral susceptibility 
as a function of temperature shown in Fig.~\ref{fig:chi_dis} shows 
interesting deviations from the results in earlier staggered work,
we expect that at least part of this difference is caused by our
small lattice volume.   Fortunately, while calculations on larger spatial 
volumes are difficult when using the five-dimensional DWF
formulation, the scale of computer resources now becoming
available for these calculations will allow an increase in lattice
volume from the present $16^3$ to $32^3$ and $48^3$.  Thus, over
the next one to two years, the methods introduced and demonstrated
here can be used to study appropriately large volumes allowing both
a careful comparison with earlier staggered fermion results and 
important exploration of those symmetry and spectral properties which
are best examined with a chiral fermion formulation.

\section*{Acknowledgments}
\addcontentsline{toc}{section}{Acknowledgements}

This work has been supported in part by contracts DE-AC02-98CH10886,
DE-AC52-07NA27344, \ DE-FC06-ER41446, \ DE-FG02-92ER40699, \
DE-FG02-91ER-40628, DE- FG02-91ER-40661, DE-FG02-04ER-41298, 
DE-KA-14-01-02 with the U.S. Department of Energy, Lawrence
Livermore National Laboratory under Contract DE-AC52-07NA27344, NSF grants 
PHY0903571, PHY08-57333, PHY07-57035, PHY07-57333 and PHY07-03296.  
NC, ZL, RM and HY were 
supported in part by U.S. DOE grant DE-FG02-92ER40699.  The numerical 
calculations have been performed on the QCDOC computers of the RIKEN BNL 
Reseach Center, the BlueGene/L and BlueGene/P computers at Lawrence Livermore
National Laboratory (LLNL) and the New York Center for Computational
Sciences (NYCCS) at Brookhaven National Laboratory.  We thank the LLNL 
Multiprogrammatic and Institutional Computing program for
time on the LLNL BlueGene/L and BlueGene/P supercomputer.

\appendix
\section{Normalization of DWF Dirac spectrum}
\label{sec:spectral_normalization}

In this appendix we repeat the arguments of Giusti and L\"uscher~\cite{Giusti:2008vb} 
to demonstrate that Dirac eigenvalue density $\rho(\lambda)$ has a scheme-dependent 
continuum limit which transforms under a change of conventions as shown in 
Eq.~\eqref{eq:spectral_renorm}.  Using these methods we then determine how such a 
``physical'' spectral density, $\rho(\lambda)$, can be determined from the 
eigenvalue distribution found for the DWF Dirac operator.

Following Giusti and L\"uscher we consider a single flavor of fermion with field 
variables $q(x)$ and $\overline{q}(x)$ which in a continuum formulation would 
have the Euclidean action density $\overline{q}(x)(\gamma^\nu D_\nu +m)q(x)$.  
This single fermion flavor is then replicated, creating $k$ doublet fields 
$q^j(x)$ and $\overline{q}^j(x)$, $1 \le j \le k$.  Finally a twisted mass term 
is added to the continuum action giving
\begin{equation}
{\cal L}(x) = \sum_{j=1}^k \overline{q}^j(x)
    \left(\gamma^\nu D_\nu + m + i\mu\gamma^5\tau^3\right)q^j(x),
\end{equation}
where $\tau^3$ is one of the standard Pauli matrices $\tau^i$ acting on the 
implicit doublet degrees of freedom of $q^j(x)$.  

This generalized action is then used to define the Green's function 
\begin{equation}
\sigma_3(\mu) = -\prod_{n=1}^6 \left\langle P^+_{1,2}(x_1) P^-_{2,3}(x_2)
                 P^+_{3,4}(x_3) P^-_{4,5}(x_4)P^+_{5,6}(x_5) P^-_{6,1}(x_6)
                   \right\rangle,
\label{eq:sigma_3}
\end{equation}
where $P^\pm_{ll'} = (P^1_{ll'} \pm P^2_{ll'})/2$ and the operators 
$P^i_{ll'}$ are defined by
\begin{equation}
P^i_{ll'} = \overline{q}\,^l(x)\tau^i q^{l'}(x).
\end{equation}
The Green's function given in Eq.~\eqref{eq:sigma_3} can be defined for
the case of six doublets, $k=6$ and can easily be generalized to define
$\sigma_{k/2}(\mu)$.  The structure of Eq.~\eqref{eq:sigma_3} insures
that the fermions flow in a single loop constructed from the product
of six fermion propagators which can be evaluated directly in QCD
perturbation theory.  The brackets $\langle\ldots\rangle$ in 
Eq.~\eqref{eq:sigma_3} describe the gauge average appropriate to the
original theory.  Thus, no fermion determinant should be 
introduced for any of the $k$ fermion fields
appearing in these Green's functions.

By design, the Green's function defined in Eq.~\eqref{eq:sigma_3} also can
be written as a path integral over the gauge degrees of freedom of a
product of fermion propagators, evaluated in each gauge background:
\begin{equation}
\sigma_3(\mu) = \left\langle{\rm tr}\left\{
                     \frac{1}{\Bigl((\gamma^5 D)^2 + \mu^2\Bigr)^3}
                               \right\}\right\rangle,
\label{eq:sigma_3_ga}
\end{equation}
where $\gamma^5 D = \gamma^5\gamma^\nu D_\nu + \gamma^5 m$ is the hermitian
 Euclidean Dirac operator and the $\gamma^5$
matrices which appear in the vertex operators $P^\pm_{ll'}$ have been
combined into the operators appearing in the propagators resulting in
the simple trace of products shown in Eq.~\eqref{eq:sigma_3_ga}.  

Finally the connection between $\sigma_3(\mu)$ and the eigenvalue density 
$\rho(\lambda)$ can established if, for each gauge configuration in the
average appearing in Eq.~\eqref{eq:sigma_3_ga}, we evaluate the trace of
products of Dirac propagators in the basis of eigenstates of the hermitian
Dirac operator $\gamma^5 D$:
\begin{eqnarray}
\sigma_3(\mu) &=& \left\langle\sum_n
                     \frac{1}{\Bigl(\lambda_n^2 + \mu^2\Bigr)^3}
                               \right\rangle \\
                &=& \int_{-\infty}^\infty d\lambda \rho(\lambda)
                           \frac{1}{\Bigl(\lambda^2 + \mu^2\Bigr)^3},
\label{eq:G_L}
\end{eqnarray}
where the $\lambda_n$ are the eigenvalues of $\gamma^5 D$ on each gauge
configuration over which the average is being performed.  In the final 
step we have made the usual replacement
\begin{equation}
\sum_n f(\lambda_n) = \int_{-\infty}^\infty d\lambda
                         \left(\sum_n \delta(\lambda - \lambda_n)\right)f(\lambda)
\end{equation}
for an arbitrary function $f(\lambda)$ and adopted the usual definition
\begin{equation}
\rho(\lambda) = \left\langle \sum_n \delta(\lambda - \lambda_n) \right\rangle.
\end{equation}

The transform given in Eq.~\eqref{eq:G_L} determining $\sigma_3(\mu)$ in 
terms of $\rho(\lambda)$ can be inverted, allowing $\rho(\lambda)$ to 
be defined from the Green's function $\sigma_3(\mu)$.  Since the operators 
$P^\pm_{ll'}$ and the related twisted mass term $\overline{q}\gamma^5 \tau^3 q$ 
can be given a meaning in the continuum limit, $\sigma_3(\mu)$ and 
hence $\rho(\lambda)$ can be defined in the continuum limit as well.  
If we work with a second regularization scheme, the corresponding 
mass operators ${P_{ll'}'}^i$ will have long distance matrix elements 
related to those of the first scheme by
\begin{equation}
P_{ll'}^{\prime i} = \frac{1}{Z_{m \to m'}} P_{ll'}^i.
\end{equation}
We can exploit this equation to relate the corresponding Green's 
functions $\sigma'_3(\mu)$ and $\sigma_3(\mu)$:
\begin{equation}
\sigma'_3(\mu') = \frac{1}{(Z_{m \to m'})^6}\sigma_3(\mu'/Z_{m \to m'})
\end{equation}
which in turn implies that $\rho'(\lambda')$ and $\rho(\lambda)$
are related by Eq.~\eqref{eq:spectral_renorm}.

We can now easily generalize this approach to the case of domain wall 
fermions.  We need only identify three operators which are the DWF analogue
of the $P_{ll'}^i$ defined above.  Since the product of the usual DWF Dirac
operator $D^{\rm DWF}$ with $\gamma^5$ and the reflection operator $R_5$
defined in Sec.~\ref{sec:dirac} is hermitian, we define:
\begin{equation}
P_{ll'}^{{\rm DWF},i}(x) = \sum_{s=0}^{L_s-1}
               \overline{\Psi}_l(x,s)\gamma^5\tau^i\Psi_{l'}(x,L_s-1-s).
\end{equation}
where, as above, we have introduced $k$ doublet five-dimensional fields
$\Psi_l(x)$, $1 \le l \le k$ in precise analogy with the generic treatment 
of Giusti and L\"uscher above.   As above we can use $P_{ll'}^{{\rm DWF},i}(x)$ 
to define a corresponding Green's function $\sigma^{\rm DWF}_3(\mu)$ 
which, as above, is directly related to the spectrum of DWF Dirac 
eigenvalues which we can obtain by numerically diagonalizing 
$D^{\rm DWF} \gamma^5 R_5$.  Again, as above, we can relate this spectrum 
to the Dirac spectrum found in a different lattice regularization or in a 
continuum scheme if we determine the needed normalization factor 
$Z_{\rm tw}$ connecting the operators $P_{ll'}^{{\rm DWF},i}(x)$ and those 
for the second scheme.

\section{Renormalization of staggered chiral susceptibilities}
\label{sec:renormalization}

In order to compare the chiral susceptibility between the DWF and staggered actions, we must also
calculate the renormalization factors for the HISQ and Asqtad actions used in \cite{Bazavov:2011nk}.  The
ensembles used in that work lie on slightly different lines of constant physics, given
by $m_\pi r_0$ = 0.381 and $m_\pi r_0$ = 0.425 for the HISQ and Asqtad actions, respectively.
This corresponds to $m_\pi = 161$ MeV and $m_\pi = 179$ MeV if one converts to
physical units using $r_0 = 0.468$ fm, the value for the Sommer parameter determined from staggered
calculations.
Using the $\overline{\textrm{MS}}$ masses $m_l = 3.2(2)$ MeV and $m_s = 88(5)$ MeV  at
$\mu = 2$ GeV determined in
 \cite{Bazavov:2009bb}, we can calculate the renormalization factors necessary to
convert to $\overline{\textrm{MS}}$ scheme:
\begin{equation}
Z_m = \frac{91.2 \textrm{MeV}}{2\widetilde{m}} 
                             \left(\frac{m_\pi}{495 \textrm{MeV}}\right)^2.
\label{eq:Z_m_stag}
\end{equation}
The renormalized, one-flavor susceptibility is then given by:
\begin{equation}
\label{eqn:susceptibility_renormalization}
\chi^\textrm{renorm}_{1f}/T^2 = \frac{1}{4}\left(\frac{1}{Z_{m_f \to \overline{\rm MS}}(\mu^2)}\right)^2 \chi^\textrm{bare}_{2f}/T^2,
\end{equation}
where $\chi^\textrm{bare}_{2f}$ is the bare two-flavor susceptibility tabulated in \cite{Bazavov:2011nk}, and
the factor of $1/4$ in Eqn. \eqref{eqn:susceptibility_renormalization} converts to the
one-flavor normalization used in this work.

\section{RHMC ensemble generation algorithms}
\label{sec:RHMC_evolution}

Here we give a brief description of the specific evolution algorithms used to 
generate the DSDR gauge field ensembles used in this paper.  Recall that  these
ensembles are generated using the Iwasaki gauge action, the DSDR action 
formed from the ratio of twisted-mass Wilson determinants given in 
Eq.~\eqref{eqn:DSDR} and the ratio of the DWF determinants for two flavors 
of light quarks with mass $m_l$ and one strange quark flavor with mass 
$m_s$ divided by three corresponding DWF Paul-Villars deteminants with 
mass $m_f=1$.  These DWF determinants are constructed from the following
ingredients.

A quotient fermion action is derived from the following fermion determinant
\begin{equation}
  \det\left(\frac{M^\dag(m)M(m)}{M^\dag(1)M(1)}\right)=
  \int \mathcal{D}\phi^\dag \mathcal{D}\phi \;
  \exp\left(-\phi^\dag M(1)\frac{1}{M^\dag(m)M(m)}M^\dag(1)\phi\right),
\end{equation}
where $M$ is the five-dimensional DWF Dirac operator.  The Hasenbusch 
factorization~\cite{Hasenbusch:2001ne} rewrites the above quotient action 
as a product of quotient actions by introducing $k$ intermediate masses
\begin{align}
  &\det\left(\frac{M^\dag(m)M(m)}{M^\dag(1)M(1)}\right)=
  \prod^{k+1}_{i=1}\det\left(
  \frac{M^\dag(m_{i-1})M(m_{i-1})}{M^\dag(m_i)M(m_i)}
  \right) \label{hasenbusch_1}\\
  =&\prod_{i=1}^{k+1}\int \mathcal{D}\phi_i^\dag \mathcal{D}\phi_i\;
  \exp\left(
  -\phi_i^\dag
  M(m_i)\frac{1}{M^\dag(m_{i-1})M(m_{i-1})}M^\dag(m_i)\phi_i
  \right), \label{hasenbusch_2}
\end{align}
where $m=m_0<m_1<\cdots<m_{k+1}=1$. 

In the following the symbol $S_\textrm{Q}(m_1, m_2)$ is used to
represent the quotient fermion action
\begin{equation}
  S_\textrm{Q}(m_1,m_2)
  =\phi^\dag M(m_2)\frac{1}{M^\dag(m_1)M(m_1)}M^\dag(m_2)\phi,
\label{eq:quotient_action}
\end{equation}
where Q means ``quotient''. Note that each quotient action has a
different pseudofermion field $\phi$. This fact is not represented in
Eq.~\eqref{eq:quotient_action}.

The quotient action discussed above accounts for two degenerate sea
quarks. This is used to simulate the two light quarks in the hybrid
Monte Carlo algorithm. For simulating the strange quark, the rational
approximation needs to be used:
\begin{eqnarray}
  \det\left(\frac{M^\dag(m)M(m)}{M^\dag(1)M(1)}\right)^{1/2}\label{rat_action} && \\
 && \hskip -1.65in  =\int \mathcal{D}\phi^\dag \mathcal{D}\phi\;
  \exp\left(
  -\phi^\dag
  \left(M^\dag(1)M(1)\right)^{1/4}
  \frac{1}{\left(M^\dag(m)M(m)\right)^{1/2}}
  \left(M^\dag(1)M(1)\right)^{1/4}\phi
  \right), \nonumber
\end{eqnarray}
where rational approximations to $x^{1/4}$ and $x^{-1/2}$ are used to
evaluate the non-integer powers of these matrices. In the following, the
symbol $S_\textrm{R}(m_1,m_2)$ is used to represent this rational
action
\begin{equation}
  S_\textrm{R}(m_1,m_2)=
  \phi^\dag
  \left(M^\dag(m_2)M(m_2)\right)^{1/4}
  \frac{1}{\left(M^\dag(m_1)M(m_1)\right)^{1/2}}
  \left(M^\dag(m_2)M(m_2)\right)^{1/4}\phi,
\end{equation}
where fractional powers such as $x^{1/4}$ and $x^{-1/2}$ are understood
to be shorthand notations for their corresponding rational
approximations.  The ``R'' in $S_\textrm{R}$ means ``rational''.

The final Hamiltonian used in the RHMC evolution contains the following parts:
\begin{equation}\label{thermo_hamiltonian}
  H=T(p)+S_\textrm{G}+S_\textrm{DSDR}+S_\textrm{R}(m_s, 1)+
  S_\textrm{Q}(m_l,1),
\end{equation}
Here $S_\textrm{G}$ and $S_\textrm{DSDR}$ represent the gauge action 
and the DSDR action, while $T(p)$ is the kinetic term.  We split 
$S_\textrm{Q}(m_l,1)$ into a few quotient actions using the Hasenbusch 
factorization as in Eqs.~\eqref{hasenbusch_1} and \eqref{hasenbusch_2}. 
A single quotient action can also be replaced by two rational actions given in
Eq.~\eqref{rat_action} using the ``Nroots'' acceleration method.

When evolving the above action, we use multiple levels of nested
integrators to separate different parts of the action. At each level
we use an Omelyan QPQPQ or force gradient QPQPQ integrator. A general
multi-level Sexton-Weingarten integration scheme can be written as
follows
\begin{align}
  H=T'_0=&T'_1+S_1 \\
  T'_i=&T'_{i+1}+S_{i+1}\;\;\; i=1,2,\cdots,N-1,
\end{align}
where $T'_i$,  $i=0, 1, N-1$ is the Hamiltonian to be integrated at level $i$.  
The $i^{th}$-level Hamiltonian $T'_i$ is further split into $T'_{i+1}$ and $S_i$, 
which are the Q and P parts used by the Omelyan or force gradient
integrator.  The Hamiltonian $T'_N$ at the last level is the kinetic term $T(p)$. 
The above equations separate the entire action into $N$ levels.

The details of the RHMC algorithms used in this paper are listed in the following two 
tables. The column labeled level$(i)$  in these tables contains the
integer $n_i$ which specifies the number of $T'$ steps in the 
Sexton-Weingarten integration scheme for each level while $S_i$ specifies 
the part of the action in Eq.~\eqref{thermo_hamiltonian} included in each level.

\begin{table}[hbt]
  \centering
  \begin{tabular}{cccc}
    level($i$) & $S_i$ & integrator type & $n_i$  \\
    \hline
    1 & $S_\textrm{Q}(m_l, 0.01)+S_\textrm{Q}(0.01, m_s)$ &
    Omelyan QPQPQ & 1  \\
    2 & $S_\textrm{R}(m_s, 1)+S_\textrm{R}(m_s,
    1)+S_\textrm{R}(m_s, 1)$ & Omelyan QPQPQ & 4  \\
    3 & $S_\textrm{DSDR}$ & Omelyan QPQPQ & 6  \\
    4 & $S_\textrm{G}$ & Omelyan QPQPQ & 1 \\
    \hline
  \end{tabular}
  \caption{Scheme 1 with a total of $N=4$ levels of nested
    integrators. The quotient action $S_\textrm{Q}(m_l, 1)$ is split
    into $S_\textrm{Q}(m_l, 0.01) + S_\textrm{Q}(0.01, m_s) +
    S_\textrm{R}(m_s, 1)+S_\textrm{R}(m_s, 1)$. Note that two copies
    of the rational action $S_\textrm{R}(m_s, 1)$ are used to replace
    a single quotient action $S_\textrm{Q}(m_s, 1)$. Ensembles
    \runn{160} (159MeV), \runn{170} (168MeV), \runn{180} (177MeV) and
    \runn{190} (186MeV) were generated using this scheme, using top
    level step size 1/4. The light and strange quark masses $m_l$ and
    $m_s$ can be found in Tab.~\protect{\ref{tab:para}}.}
  \label{tab:evolution_1}
\end{table}

\begin{table}[hbt]
  \centering
  \begin{tabular}{cccc}
    level($i$) & $S_i$ & integrator type & $n_i$ \\
    \hline
    1 & $\sum_{i=1}^6 S_\textrm{Q}(m_{i-1}, m_i)+S_\textrm{R}(m_s, 1)$ &
    Omelyan/FG QPQPQ & 4  \\
    2 & $S_\textrm{DSDR}$ & Omelyan/FG QPQPQ & 1 \\
    3 & $S_\textrm{G}$ & Omelyan/FG QPQPQ & 1 \\
    \hline
  \end{tabular}
  \caption{Scheme 2 with a total of $N=3$ levels of nested
    integrators. Ensemble \runn{140} (139MeV), \runn{150_32} \&
    \runn{150_48} (149MeV) and \runn{200} (195MeV) were generated
    using this scheme. Ensemble \runn{140} \runn{150_32} and
    \runn{150_48} used the force gradient QPQPQ integrator
    \cite{Kennedy:2009fe} with top level step size 1/7, while
    \runn{200} used the Omelyan QPQPQ integrator with top level step
    size 1/8. Here $m_i$, $i=0,1,\cdots6$, represent different
    Hasenbusch masses, with $m_0=m_l$, $m_1=0.01$, $m_2=0.06$,
    $m_3=0.18$, $m_4=0.37$, $m_5=0.67$ and $m_6=1$.  The masses $m_l$
    and $m_s$ can be found in Tab.~\protect{\ref{tab:para}}.}
  \label{tab:evolution_2}
\end{table}

\section{Comparison of $\chi_{\rm top}$ and $\chi_{5,{\rm disc}}$}
\label{sec:g5-top_suscept}

In this appendix we investigate the large discrepancy between the topological susceptibility $\chi_{\rm top}$ and the pseudo-scalar susceptibility $m_{l,{\rm tot}} ^2\chi_{5,{\rm disc}}$
shown in Fig.~\ref{fig:ua1delta} and described in Sec.~\ref{sec:suscept}.  The relation between $\chi_{\rm top}$  and  $m_{l,{\rm tot}} ^2\chi_{5,{\rm disc}}$ given in Eq.~\ref{eq:top2chi_5disc} is often viewed as providing a good definition of $\chi_{\rm top}$ since the fermionic quantity has a better understood continuum limit~\cite{Giusti:2004qd,Luscher:2004fu,Giusti:2008vb,Luscher:2010ik}.  However, we compute $\chi_{\rm top}$ using a widely used method which usually gives consistent results so the discrepancy found here caused us to look carefully at our code and to seek further tests of our results for both $\chi_{\rm top}$  and  $\chi_{5,{\rm disc}}$.  

For both quantities our computational procedures appear to be robust.  We increased the number of random sources used to determine $\chi_{5,{\rm disc}}$ from ten to 100 and saw only the expected decease in statistical errors.  Independent code gave consistent results.  We increased the number of smearing steps performed before the determination of $\chi_{\rm top}$ from 60 to 150 and saw no systematic change in the result.  

We cannot make a meaningful comparison of the relationship given in Eq.~\ref{eq:Qtop_pbg5p} on individual configurations because at least the right side of this relation takes on its continuum meaning only after a gauge average is performed.  Because both sides are parity odd, a gauge average will give a non-zero result only if the equation is squared, leading us back to the relation we are trying to test.  However, more information can be obtained by examining other products of similar parity-odd operators.  Specifically we examine $ \chi_\mathrm{top}$ and the four additional quantities:
\begin{eqnarray}
  X_l &=& m^2_{l,\mathrm{tot}}\chi^5_{l,\mathrm{disc}}\\
  X_s &=& \frac{1}{V} m^2_{s,\mathrm{tot}}\Biggl\langle\Bigl(
                        \int d^4x \overline{\psi}_s(x)\gamma^5\psi_s(x) \Bigr)
                  \Bigl(\int d^4y \overline{\psi}_s(y)\gamma^5\psi_s(y)\Bigr)\Biggr\rangle \\
  X_{l,s}  &=&  \frac{1}{V} m_{l,\mathrm{tot}}m_{s,\mathrm{tot}}\Biggl\langle\Bigl(
                        \int d^4x \overline{\psi}_l(x)\gamma^5\psi_l(x) \Bigr)
                  \Bigl(\int d^4y \overline{\psi}_s(y)\gamma^5\psi_s(y)\Bigr)\Biggr\rangle \\
 X_{l,{\rm top}}  &=&  \frac{1}{V} m_{l,\mathrm{tot}}\Biggl\langle\Bigl(
                        \int d^4x \overline{\psi}_l(x)\gamma^5\psi_l(x) \Bigr)
                        \Bigl( Q_\mathrm{top}\Bigr)\Biggr\rangle,
  \label{eq:chi5all}
\end{eqnarray}
all five of which should agree.  The results are shown in Tab.~\ref{tab:chi5all}.

\begin{table}[H] 
  \centering   
   \[ 
  \begin{array}{cc|ccccc}
    \# & T(\mathrm{MeV})& X_l& X_s&X_{l,s}& \chi_\mathrm{top}&X_{l,{\rm top}}\\ 
    \hline 
    \runn{140}   &139 & 36(3)& 51(20)& 42(5)&107(5)& 37(3)\\ 
    \runn{150_32}&149 & 27(3)& 35(20)& 29(4)& 54(2)& 26(2)\\ 
    \runn{150_48}&149 & 31(2)& 44(19)& 33(4)& 57(2)& 30(2)\\ 
    \runn{160}   &159 & 16(2)&  6(12)& 15(3)& 27(2)& 15(2)\\ 
    \runn{170}   &168 &  9(2)&-11(12)&  6(2)& 15(2)&  9(2)\\ 
    \runn{180}   &177 &  5(1)& -1( 8)&  4(2)&7.6(9)&4.8(8)\\ 
    \runn{190}   &186 &1.7(7)& -3( 6)&  1(1)&  4(1)&2.0(8)\\ 
    \runn{200}   &195 &1.4(5)&  4( 7)&1.3(9)&2.2(5)&1.5(5)\\ 
    \hline 
    \runn{1.70ml0.006}&-&50(9)&67(22)&55(12)&49(7)&44(8)\\ 
    \runn{1.75ml0.006}&-&54(8)&33(56)&43(16)&62(6)&47(6)\\ 
    \runn{1.82ml0.007}&-&20(3)& 2(20)&16(53)&23(4)&21(4)\\ 
    \hline 
  \end{array}\] 
  \caption{Results for five different susceptibilities computed on both finite and zero
  temperature ensembles.   All the values are given in lattice units with a factor of  
  $10^{-6}$ removed.} 
  \label{tab:chi5all} 
\end{table} 

While the errors on the  strange quark susceptibilities $X_s$ are too large  to allow
a meaningful test, the light quark susceptibilities $X_l$  and the light-strange product $X_{l,s}$
agree within their $10\%$ to $20\%$ errors. This reaffirms the consistency of the results computed directly from the fermion fields and supports the view that the fermionic quantities, which are the basis of most of the results in this paper, are behaving as expected.  Note, this includes the interpretation of the total bare quark mass as the sum of the input plus the residual mass $m_f+m_{\rm res}$ since the ratio of $m_{\rm res}$ to $m_f$ various substantially among the rows in Tab.~\ref{tab:chi5all}.  However, those susceptibilities are much smaller than $\chi_\mathrm{top}$ at temperatures near or below the transition region (see also Fig.~\ref{fig:ua1delta}). This  discrepancy is not visible at higher temperatures or for the zero-temperature ensembles. 

The right-most column in Tab.~\ref{tab:chi5all} offers some insight into this discrepancy.   Comparing the $X_l$ and $X_{l,{\rm top}}$ columns shows agreement between the pure fermionic susceptiblity $X_l$ and the cross, fermion-topological susceptibility $X_{l,{\rm top}}$ within their $10\%$ to $20\%$ errors for all the ensembles.  This suggests the presence of unphysical fluctuations in the gauge field observable $Q_\mathrm{top}$ at lower temperatures.  These unphysical fluctuations are uncorrelated with the fermionic degrees of freedom and hence do not pollute the cross correlator $X_{l,{\rm top}}$.  However, they do add to the fluctuations in $Q_\mathrm{top}$, leading to an unphysical increase in $\chi_\mathrm{top}$.  At $T=140$ MeV these unphysical fluctuations appear to have the same size as those which are physical.

\bibliographystyle{apsrev}
\bibliography{main} 

\begin{thebibliography}{71}
\expandafter\ifx\csname natexlab\endcsname\relax\def\natexlab#1{#1}\fi
\expandafter\ifx\csname bibnamefont\endcsname\relax
  \def\bibnamefont#1{#1}\fi
\expandafter\ifx\csname bibfnamefont\endcsname\relax
  \def\bibfnamefont#1{#1}\fi
\expandafter\ifx\csname citenamefont\endcsname\relax
  \def\citenamefont#1{#1}\fi
\expandafter\ifx\csname url\endcsname\relax
  \def\url#1{\texttt{#1}}\fi
\expandafter\ifx\csname urlprefix\endcsname\relax\def\urlprefix{URL }\fi
\providecommand{\bibinfo}[2]{#2}
\providecommand{\eprint}[2][]{\url{#2}}

\bibitem[{\citenamefont{Adler}(1969)}]{Adler:1969gk}
\bibinfo{author}{\bibfnamefont{S.~L.} \bibnamefont{Adler}},
  \bibinfo{journal}{Phys. Rev.} \textbf{\bibinfo{volume}{177}},
  \bibinfo{pages}{2426} (\bibinfo{year}{1969}).

\bibitem[{\citenamefont{Bell and Jackiw}(1969)}]{Bell:1969ts}
\bibinfo{author}{\bibfnamefont{J.~S.} \bibnamefont{Bell}} \bibnamefont{and}
  \bibinfo{author}{\bibfnamefont{R.}~\bibnamefont{Jackiw}},
  \bibinfo{journal}{Nuovo Cim.} \textbf{\bibinfo{volume}{A60}},
  \bibinfo{pages}{47} (\bibinfo{year}{1969}).

\bibitem[{\citenamefont{'t~Hooft}(1976{\natexlab{a}})}]{'tHooft:1976up}
\bibinfo{author}{\bibfnamefont{G.}~\bibnamefont{'t~Hooft}},
  \bibinfo{journal}{Phys. Rev. Lett.} \textbf{\bibinfo{volume}{37}},
  \bibinfo{pages}{8} (\bibinfo{year}{1976}{\natexlab{a}}).

\bibitem[{\citenamefont{Kaplan}(1992)}]{Kaplan:1992bt}
\bibinfo{author}{\bibfnamefont{D.~B.} \bibnamefont{Kaplan}},
  \bibinfo{journal}{Phys. Lett.} \textbf{\bibinfo{volume}{B288}},
  \bibinfo{pages}{342} (\bibinfo{year}{1992}).

\bibitem[{\citenamefont{Furman and Shamir}(1995)}]{Furman:1994ky}
\bibinfo{author}{\bibfnamefont{V.}~\bibnamefont{Furman}} \bibnamefont{and}
  \bibinfo{author}{\bibfnamefont{Y.}~\bibnamefont{Shamir}},
  \bibinfo{journal}{Nucl.Phys.} \textbf{\bibinfo{volume}{B439}},
  \bibinfo{pages}{54} (\bibinfo{year}{1995}), \eprint{hep-lat/9405004}.

\bibitem[{\citenamefont{Sharpe}(2006)}]{Sharpe:2006re}
\bibinfo{author}{\bibfnamefont{S.~R.} \bibnamefont{Sharpe}},
  \bibinfo{journal}{PoS} \textbf{\bibinfo{volume}{LAT2006}},
  \bibinfo{pages}{022} (\bibinfo{year}{2006}).

\bibitem[{\citenamefont{Donald et~al.}(2011)\citenamefont{Donald, Davies,
  Follana, and Kronfeld}}]{Donald:2011if}
\bibinfo{author}{\bibfnamefont{G.~C.} \bibnamefont{Donald}},
  \bibinfo{author}{\bibfnamefont{C.~T.} \bibnamefont{Davies}},
  \bibinfo{author}{\bibfnamefont{E.}~\bibnamefont{Follana}}, \bibnamefont{and}
  \bibinfo{author}{\bibfnamefont{A.~S.} \bibnamefont{Kronfeld}}
  (\bibinfo{year}{2011}), \eprint{1106.2412 [hep-lat]}.

\bibitem[{\citenamefont{Adams}(2010)}]{Adams:2009eb}
\bibinfo{author}{\bibfnamefont{D.~H.} \bibnamefont{Adams}},
  \bibinfo{journal}{Phys. Rev. Lett.} \textbf{\bibinfo{volume}{104}},
  \bibinfo{pages}{141602} (\bibinfo{year}{2010}).

\bibitem[{\citenamefont{Gross et~al.}(1981)\citenamefont{Gross, Pisarski, and
  Yaffe}}]{Gross:1980br}
\bibinfo{author}{\bibfnamefont{D.~J.} \bibnamefont{Gross}},
  \bibinfo{author}{\bibfnamefont{R.~D.} \bibnamefont{Pisarski}},
  \bibnamefont{and} \bibinfo{author}{\bibfnamefont{L.~G.} \bibnamefont{Yaffe}},
  \bibinfo{journal}{Rev. Mod. Phys.} \textbf{\bibinfo{volume}{53}},
  \bibinfo{pages}{43} (\bibinfo{year}{1981}).

\bibitem[{\citenamefont{Pisarski and Wilczek}(1984)}]{Pisarski:1983ms}
\bibinfo{author}{\bibfnamefont{R.~D.} \bibnamefont{Pisarski}} \bibnamefont{and}
  \bibinfo{author}{\bibfnamefont{F.}~\bibnamefont{Wilczek}},
  \bibinfo{journal}{Phys. Rev.} \textbf{\bibinfo{volume}{D29}},
  \bibinfo{pages}{338} (\bibinfo{year}{1984}).

\bibitem[{\citenamefont{Butti et~al.}(2003)\citenamefont{Butti, Pelissetto, and
  Vicari}}]{Butti:2003nu}
\bibinfo{author}{\bibfnamefont{A.}~\bibnamefont{Butti}},
  \bibinfo{author}{\bibfnamefont{A.}~\bibnamefont{Pelissetto}},
  \bibnamefont{and} \bibinfo{author}{\bibfnamefont{E.}~\bibnamefont{Vicari}},
  \bibinfo{journal}{JHEP} \textbf{\bibinfo{volume}{0308}}, \bibinfo{pages}{029}
  (\bibinfo{year}{2003}).

\bibitem[{\citenamefont{Basile et~al.}(2006)\citenamefont{Basile, Pelissetto,
  and Vicari}}]{Basile:2005hw}
\bibinfo{author}{\bibfnamefont{F.}~\bibnamefont{Basile}},
  \bibinfo{author}{\bibfnamefont{A.}~\bibnamefont{Pelissetto}},
  \bibnamefont{and} \bibinfo{author}{\bibfnamefont{E.}~\bibnamefont{Vicari}},
  \bibinfo{journal}{PoS} \textbf{\bibinfo{volume}{LAT2005}},
  \bibinfo{pages}{199} (\bibinfo{year}{2006}).

\bibitem[{\citenamefont{Rapp and Wambach}(2000)}]{Rapp:1999ej}
\bibinfo{author}{\bibfnamefont{R.}~\bibnamefont{Rapp}} \bibnamefont{and}
  \bibinfo{author}{\bibfnamefont{J.}~\bibnamefont{Wambach}},
  \bibinfo{journal}{Adv. Nucl. Phys.} \textbf{\bibinfo{volume}{25}},
  \bibinfo{pages}{1} (\bibinfo{year}{2000}).

\bibitem[{\citenamefont{Shuryak}(1994)}]{Shuryak:1993ee}
\bibinfo{author}{\bibfnamefont{E.~V.} \bibnamefont{Shuryak}},
  \bibinfo{journal}{Comments Nucl. Part. Phys.} \textbf{\bibinfo{volume}{21}},
  \bibinfo{pages}{235} (\bibinfo{year}{1994}).

\bibitem[{\citenamefont{Huang and Wang}(1996)}]{Huang:1995fc}
\bibinfo{author}{\bibfnamefont{Z.}~\bibnamefont{Huang}} \bibnamefont{and}
  \bibinfo{author}{\bibfnamefont{X.-N.} \bibnamefont{Wang}},
  \bibinfo{journal}{Phys. Rev.} \textbf{\bibinfo{volume}{D53}},
  \bibinfo{pages}{5034} (\bibinfo{year}{1996}).

\bibitem[{\citenamefont{Kapusta et~al.}(1996)\citenamefont{Kapusta, Kharzeev,
  and McLerran}}]{Kapusta:1995ww}
\bibinfo{author}{\bibfnamefont{J.~I.} \bibnamefont{Kapusta}},
  \bibinfo{author}{\bibfnamefont{D.}~\bibnamefont{Kharzeev}}, \bibnamefont{and}
  \bibinfo{author}{\bibfnamefont{L.~D.} \bibnamefont{McLerran}},
  \bibinfo{journal}{Phys. Rev.} \textbf{\bibinfo{volume}{D53}},
  \bibinfo{pages}{5028} (\bibinfo{year}{1996}).

\bibitem[{\citenamefont{Csorgo et~al.}(2010)\citenamefont{Csorgo, Vertesi, and
  Sziklai}}]{Csorgo:2009pa}
\bibinfo{author}{\bibfnamefont{T.}~\bibnamefont{Csorgo}},
  \bibinfo{author}{\bibfnamefont{R.}~\bibnamefont{Vertesi}}, \bibnamefont{and}
  \bibinfo{author}{\bibfnamefont{J.}~\bibnamefont{Sziklai}},
  \bibinfo{journal}{Phys. Rev. Lett.} \textbf{\bibinfo{volume}{105}},
  \bibinfo{pages}{182301} (\bibinfo{year}{2010}).

\bibitem[{\citenamefont{Petreczky}(2012)}]{Petreczky:2012rq}
\bibinfo{author}{\bibfnamefont{P.}~\bibnamefont{Petreczky}}
  (\bibinfo{year}{2012}), \eprint{1203.5320}.

\bibitem[{\citenamefont{Mukherjee}(2011)}]{Mukherjee:2011td}
\bibinfo{author}{\bibfnamefont{S.}~\bibnamefont{Mukherjee}},
  \bibinfo{journal}{J.Phys.G} \textbf{\bibinfo{volume}{G38}},
  \bibinfo{pages}{124022} (\bibinfo{year}{2011}), \eprint{1107.0765}.

\bibitem[{\citenamefont{Bernard et~al.}(1997)\citenamefont{Bernard, Blum,
  Detar, Gottlieb, Heller et~al.}}]{Bernard:1996iz}
\bibinfo{author}{\bibfnamefont{C.~W.} \bibnamefont{Bernard}},
  \bibinfo{author}{\bibfnamefont{T.}~\bibnamefont{Blum}},
  \bibinfo{author}{\bibfnamefont{C.~E.} \bibnamefont{Detar}},
  \bibinfo{author}{\bibfnamefont{S.~A.} \bibnamefont{Gottlieb}},
  \bibinfo{author}{\bibfnamefont{U.~M.} \bibnamefont{Heller}},
  \bibnamefont{et~al.}, \bibinfo{journal}{Phys. Rev. Lett.}
  \textbf{\bibinfo{volume}{78}}, \bibinfo{pages}{598} (\bibinfo{year}{1997}).

\bibitem[{\citenamefont{Chandrasekharan
  et~al.}(1999)\citenamefont{Chandrasekharan, Chen, Christ, Lee, Mawhinney
  et~al.}}]{Chandrasekharan:1998yx}
\bibinfo{author}{\bibfnamefont{S.}~\bibnamefont{Chandrasekharan}},
  \bibinfo{author}{\bibfnamefont{D.}~\bibnamefont{Chen}},
  \bibinfo{author}{\bibfnamefont{N.~H.} \bibnamefont{Christ}},
  \bibinfo{author}{\bibfnamefont{W.-J.} \bibnamefont{Lee}},
  \bibinfo{author}{\bibfnamefont{R.}~\bibnamefont{Mawhinney}},
  \bibnamefont{et~al.}, \bibinfo{journal}{Phys. Rev. Lett.}
  \textbf{\bibinfo{volume}{82}}, \bibinfo{pages}{2463} (\bibinfo{year}{1999}).

\bibitem[{\citenamefont{Kogut et~al.}(1998)\citenamefont{Kogut, Lagae, and
  Sinclair}}]{Kogut:1998rh}
\bibinfo{author}{\bibfnamefont{J.}~\bibnamefont{Kogut}},
  \bibinfo{author}{\bibfnamefont{J.}~\bibnamefont{Lagae}}, \bibnamefont{and}
  \bibinfo{author}{\bibfnamefont{D.}~\bibnamefont{Sinclair}},
  \bibinfo{journal}{Phys. Rev.} \textbf{\bibinfo{volume}{D58}},
  \bibinfo{pages}{054504} (\bibinfo{year}{1998}).

\bibitem[{\citenamefont{Cheng et~al.}(2011)\citenamefont{Cheng, Datta, Francis,
  van~der Heide, Jung et~al.}}]{Cheng:2010fe}
\bibinfo{author}{\bibfnamefont{M.}~\bibnamefont{Cheng}},
  \bibinfo{author}{\bibfnamefont{S.}~\bibnamefont{Datta}},
  \bibinfo{author}{\bibfnamefont{A.}~\bibnamefont{Francis}},
  \bibinfo{author}{\bibfnamefont{J.}~\bibnamefont{van~der Heide}},
  \bibinfo{author}{\bibfnamefont{C.}~\bibnamefont{Jung}}, \bibnamefont{et~al.},
  \bibinfo{journal}{Eur. Phys. J.} \textbf{\bibinfo{volume}{C71}},
  \bibinfo{pages}{1564} (\bibinfo{year}{2011}).

\bibitem[{\citenamefont{Ohno et~al.}(2011)\citenamefont{Ohno, Heller, Karsch,
  and Mukherjee}}]{Ohno:2011yr}
\bibinfo{author}{\bibfnamefont{H.}~\bibnamefont{Ohno}},
  \bibinfo{author}{\bibfnamefont{U.}~\bibnamefont{Heller}},
  \bibinfo{author}{\bibfnamefont{F.}~\bibnamefont{Karsch}}, \bibnamefont{and}
  \bibinfo{author}{\bibfnamefont{S.}~\bibnamefont{Mukherjee}},
  \bibinfo{journal}{PoS} \textbf{\bibinfo{volume}{LATTICE2011}},
  \bibinfo{pages}{210} (\bibinfo{year}{2011}), \eprint{1111.1939}.

\bibitem[{\citenamefont{Chen et~al.}(2001)}]{Chen:2000zu}
\bibinfo{author}{\bibfnamefont{P.}~\bibnamefont{Chen}} \bibnamefont{et~al.},
  \bibinfo{journal}{Phys. Rev.} \textbf{\bibinfo{volume}{D64}},
  \bibinfo{pages}{014503} (\bibinfo{year}{2001}), \eprint{hep-lat/0006010}.

\bibitem[{\citenamefont{Cheng et~al.}(2010)\citenamefont{Cheng, Christ, Li,
  Mawhinney, Renfrew et~al.}}]{Cheng:2009be}
\bibinfo{author}{\bibfnamefont{M.}~\bibnamefont{Cheng}},
  \bibinfo{author}{\bibfnamefont{N.~H.} \bibnamefont{Christ}},
  \bibinfo{author}{\bibfnamefont{M.}~\bibnamefont{Li}},
  \bibinfo{author}{\bibfnamefont{R.~D.} \bibnamefont{Mawhinney}},
  \bibinfo{author}{\bibfnamefont{D.}~\bibnamefont{Renfrew}},
  \bibnamefont{et~al.}, \bibinfo{journal}{Phys. Rev.}
  \textbf{\bibinfo{volume}{D81}}, \bibinfo{pages}{054510}
  (\bibinfo{year}{2010}).

\bibitem[{\citenamefont{Borsanyi et~al.}(2012)\citenamefont{Borsanyi, Delgado,
  Durr, Fodor, Katz et~al.}}]{Borsanyi:2012xf}
\bibinfo{author}{\bibfnamefont{S.}~\bibnamefont{Borsanyi}},
  \bibinfo{author}{\bibfnamefont{Y.}~\bibnamefont{Delgado}},
  \bibinfo{author}{\bibfnamefont{S.}~\bibnamefont{Durr}},
  \bibinfo{author}{\bibfnamefont{Z.}~\bibnamefont{Fodor}},
  \bibinfo{author}{\bibfnamefont{S.}~\bibnamefont{Katz}}, \bibnamefont{et~al.}
  (\bibinfo{year}{2012}), \eprint{1204.4089}.

\bibitem[{\citenamefont{Giusti and Luscher}(2009)}]{Giusti:2008vb}
\bibinfo{author}{\bibfnamefont{L.}~\bibnamefont{Giusti}} \bibnamefont{and}
  \bibinfo{author}{\bibfnamefont{M.}~\bibnamefont{Luscher}},
  \bibinfo{journal}{JHEP} \textbf{\bibinfo{volume}{0903}}, \bibinfo{pages}{013}
  (\bibinfo{year}{2009}).

\bibitem[{\citenamefont{de~Forcrand et~al.}(1997)\citenamefont{de~Forcrand,
  Garcia~Perez, and Stamatescu}}]{deForcrand:1997sq}
\bibinfo{author}{\bibfnamefont{P.}~\bibnamefont{de~Forcrand}},
  \bibinfo{author}{\bibfnamefont{M.}~\bibnamefont{Garcia~Perez}},
  \bibnamefont{and} \bibinfo{author}{\bibfnamefont{I.-O.}
  \bibnamefont{Stamatescu}}, \bibinfo{journal}{Nucl.Phys.}
  \textbf{\bibinfo{volume}{B499}}, \bibinfo{pages}{409} (\bibinfo{year}{1997}),
  \eprint{hep-lat/9701012}.

\bibitem[{\citenamefont{Antonio et~al.}(2008)}]{Antonio:2008zz}
\bibinfo{author}{\bibfnamefont{D.~J.} \bibnamefont{Antonio}}
  \bibnamefont{et~al.} (\bibinfo{collaboration}{RBC Collaboration, UKQCD
  Collaboration}), \bibinfo{journal}{Phys.Rev.} \textbf{\bibinfo{volume}{D77}},
  \bibinfo{pages}{014509} (\bibinfo{year}{2008}), \eprint{0705.2340}.

\bibitem[{\citenamefont{Vranas}(1999)}]{Vranas:1999rz}
\bibinfo{author}{\bibfnamefont{P.~M.} \bibnamefont{Vranas}}
  (\bibinfo{year}{1999}), \eprint{hep-lat/0001006}.

\bibitem[{\citenamefont{Vranas}(2006)}]{Vranas:2006zk}
\bibinfo{author}{\bibfnamefont{P.~M.} \bibnamefont{Vranas}},
  \bibinfo{journal}{Phys.Rev.} \textbf{\bibinfo{volume}{D74}},
  \bibinfo{pages}{034512} (\bibinfo{year}{2006}), \eprint{hep-lat/0606014}.

\bibitem[{\citenamefont{Fukaya et~al.}(2006)}]{Fukaya:2006vs}
\bibinfo{author}{\bibfnamefont{H.}~\bibnamefont{Fukaya}} \bibnamefont{et~al.}
  (\bibinfo{collaboration}{JLQCD Collaboration}), \bibinfo{journal}{Phys.Rev.}
  \textbf{\bibinfo{volume}{D74}}, \bibinfo{pages}{094505}
  (\bibinfo{year}{2006}), \eprint{hep-lat/0607020}.

\bibitem[{\citenamefont{Iwasaki}(1983)}]{Iwasaki:1983ck}
\bibinfo{author}{\bibfnamefont{Y.}~\bibnamefont{Iwasaki}},
  \bibinfo{journal}{UTHEP-118}  (\bibinfo{year}{1983}).

\bibitem[{\citenamefont{Antonio et~al.}(2007)}]{Antonio:2006px}
\bibinfo{author}{\bibfnamefont{D.}~\bibnamefont{Antonio}} \bibnamefont{et~al.}
  (\bibinfo{collaboration}{RBC and UKQCD Collaborations}),
  \bibinfo{journal}{Phys.Rev.} \textbf{\bibinfo{volume}{D75}},
  \bibinfo{pages}{114501} (\bibinfo{year}{2007}), \eprint{hep-lat/0612005}.

\bibitem[{\citenamefont{Allton et~al.}(2007)}]{Allton:2007hx}
\bibinfo{author}{\bibfnamefont{C.}~\bibnamefont{Allton}} \bibnamefont{et~al.}
  (\bibinfo{collaboration}{RBC and UKQCD Collaborations}),
  \bibinfo{journal}{Phys.Rev.} \textbf{\bibinfo{volume}{D76}},
  \bibinfo{pages}{014504} (\bibinfo{year}{2007}), \eprint{hep-lat/0701013}.

\bibitem[{\citenamefont{Allton et~al.}(2008)}]{Allton:2008pn}
\bibinfo{author}{\bibfnamefont{C.}~\bibnamefont{Allton}} \bibnamefont{et~al.}
  (\bibinfo{collaboration}{RBC-UKQCD Collaboration}),
  \bibinfo{journal}{Phys.Rev.} \textbf{\bibinfo{volume}{D78}},
  \bibinfo{pages}{114509} (\bibinfo{year}{2008}), \eprint{0804.0473}.

\bibitem[{\citenamefont{Aoki et~al.}(2011)}]{Aoki:2010dy}
\bibinfo{author}{\bibfnamefont{Y.}~\bibnamefont{Aoki}} \bibnamefont{et~al.}
  (\bibinfo{collaboration}{RBC Collaboration, UKQCD Collaboration}),
  \bibinfo{journal}{Phys. Rev.} \textbf{\bibinfo{volume}{D83}},
  \bibinfo{pages}{074508} (\bibinfo{year}{2011}).

\bibitem[{\citenamefont{Renfrew et~al.}(2008)\citenamefont{Renfrew, Blum,
  Christ, Mawhinney, and Vranas}}]{Renfrew:2009wu}
\bibinfo{author}{\bibfnamefont{D.}~\bibnamefont{Renfrew}},
  \bibinfo{author}{\bibfnamefont{T.}~\bibnamefont{Blum}},
  \bibinfo{author}{\bibfnamefont{N.}~\bibnamefont{Christ}},
  \bibinfo{author}{\bibfnamefont{R.}~\bibnamefont{Mawhinney}},
  \bibnamefont{and} \bibinfo{author}{\bibfnamefont{P.}~\bibnamefont{Vranas}},
  \bibinfo{journal}{PoS} \textbf{\bibinfo{volume}{LATTICE2008}},
  \bibinfo{pages}{048} (\bibinfo{year}{2008}), \eprint{0902.2587}.

\bibitem[{\citenamefont{Blum et~al.}(2012)\citenamefont{Blum, Boyle, Christ,
  Garron, Goode et~al.}}]{Blum:2011ng}
\bibinfo{author}{\bibfnamefont{T.}~\bibnamefont{Blum}},
  \bibinfo{author}{\bibfnamefont{P.}~\bibnamefont{Boyle}},
  \bibinfo{author}{\bibfnamefont{N.}~\bibnamefont{Christ}},
  \bibinfo{author}{\bibfnamefont{N.}~\bibnamefont{Garron}},
  \bibinfo{author}{\bibfnamefont{E.}~\bibnamefont{Goode}},
  \bibnamefont{et~al.}, \bibinfo{journal}{Phys.Rev.Lett.}
  \textbf{\bibinfo{volume}{108}}, \bibinfo{pages}{141601}
  (\bibinfo{year}{2012}), \bibinfo{note}{5 pages, 1 figure},
  \eprint{1111.1699}.

\bibitem[{\citenamefont{Kelly}(2012)}]{Kelly:2012uy}
\bibinfo{author}{\bibfnamefont{C.}~\bibnamefont{Kelly}} (\bibinfo{year}{2012}),
  \eprint{1201.0706}.

\bibitem[{\citenamefont{Golterman and Shamir}(2003)}]{Golterman:2003qe}
\bibinfo{author}{\bibfnamefont{M.}~\bibnamefont{Golterman}} \bibnamefont{and}
  \bibinfo{author}{\bibfnamefont{Y.}~\bibnamefont{Shamir}},
  \bibinfo{journal}{Phys.Rev.} \textbf{\bibinfo{volume}{D68}},
  \bibinfo{pages}{074501} (\bibinfo{year}{2003}), \eprint{hep-lat/0306002}.

\bibitem[{\citenamefont{Golterman
  et~al.}(2005{\natexlab{a}})\citenamefont{Golterman, Shamir, and
  Svetitsky}}]{Golterman:2004cy}
\bibinfo{author}{\bibfnamefont{M.}~\bibnamefont{Golterman}},
  \bibinfo{author}{\bibfnamefont{Y.}~\bibnamefont{Shamir}}, \bibnamefont{and}
  \bibinfo{author}{\bibfnamefont{B.}~\bibnamefont{Svetitsky}},
  \bibinfo{journal}{Phys.Rev.} \textbf{\bibinfo{volume}{D71}},
  \bibinfo{pages}{071502} (\bibinfo{year}{2005}{\natexlab{a}}),
  \eprint{hep-lat/0407021}.

\bibitem[{\citenamefont{Golterman
  et~al.}(2005{\natexlab{b}})\citenamefont{Golterman, Shamir, and
  Svetitsky}}]{Golterman:2005fe}
\bibinfo{author}{\bibfnamefont{M.}~\bibnamefont{Golterman}},
  \bibinfo{author}{\bibfnamefont{Y.}~\bibnamefont{Shamir}}, \bibnamefont{and}
  \bibinfo{author}{\bibfnamefont{B.}~\bibnamefont{Svetitsky}},
  \bibinfo{journal}{Phys.Rev.} \textbf{\bibinfo{volume}{D72}},
  \bibinfo{pages}{034501} (\bibinfo{year}{2005}{\natexlab{b}}),
  \eprint{hep-lat/0503037}.

\bibitem[{\citenamefont{Blum et~al.}(2004)\citenamefont{Blum, Chen, Christ,
  Cristian, Dawson et~al.}}]{Blum:2000kn}
\bibinfo{author}{\bibfnamefont{T.}~\bibnamefont{Blum}},
  \bibinfo{author}{\bibfnamefont{P.}~\bibnamefont{Chen}},
  \bibinfo{author}{\bibfnamefont{N.~H.} \bibnamefont{Christ}},
  \bibinfo{author}{\bibfnamefont{C.}~\bibnamefont{Cristian}},
  \bibinfo{author}{\bibfnamefont{C.}~\bibnamefont{Dawson}},
  \bibnamefont{et~al.}, \bibinfo{journal}{Phys. Rev.}
  \textbf{\bibinfo{volume}{D69}}, \bibinfo{pages}{074502}
  (\bibinfo{year}{2004}).

\bibitem[{\citenamefont{Martinelli et~al.}(1995)\citenamefont{Martinelli,
  Pittori, Sachrajda, Testa, and Vladikas}}]{Martinelli:1995ty}
\bibinfo{author}{\bibfnamefont{G.}~\bibnamefont{Martinelli}},
  \bibinfo{author}{\bibfnamefont{C.}~\bibnamefont{Pittori}},
  \bibinfo{author}{\bibfnamefont{C.~T.} \bibnamefont{Sachrajda}},
  \bibinfo{author}{\bibfnamefont{M.}~\bibnamefont{Testa}}, \bibnamefont{and}
  \bibinfo{author}{\bibfnamefont{A.}~\bibnamefont{Vladikas}},
  \bibinfo{journal}{Nucl. Phys.} \textbf{\bibinfo{volume}{B445}},
  \bibinfo{pages}{81} (\bibinfo{year}{1995}), \eprint{hep-lat/9411010}.

\bibitem[{\citenamefont{Aoki et~al.}(2008)}]{Aoki:2007xm}
\bibinfo{author}{\bibfnamefont{Y.}~\bibnamefont{Aoki}} \bibnamefont{et~al.},
  \bibinfo{journal}{Phys. Rev.} \textbf{\bibinfo{volume}{D78}},
  \bibinfo{pages}{054510} (\bibinfo{year}{2008}), \eprint{0712.1061}.

\bibitem[{\citenamefont{Gockeler et~al.}(1999)\citenamefont{Gockeler, Horsley,
  Oelrich, Perlt, Petters et~al.}}]{Gockeler:1998ye}
\bibinfo{author}{\bibfnamefont{M.}~\bibnamefont{Gockeler}},
  \bibinfo{author}{\bibfnamefont{R.}~\bibnamefont{Horsley}},
  \bibinfo{author}{\bibfnamefont{H.}~\bibnamefont{Oelrich}},
  \bibinfo{author}{\bibfnamefont{H.}~\bibnamefont{Perlt}},
  \bibinfo{author}{\bibfnamefont{D.}~\bibnamefont{Petters}},
  \bibnamefont{et~al.}, \bibinfo{journal}{Nucl.Phys.}
  \textbf{\bibinfo{volume}{B544}}, \bibinfo{pages}{699} (\bibinfo{year}{1999}),
  \eprint{hep-lat/9807044}.

\bibitem[{\citenamefont{Kalkreuter and Simma}(1996)}]{Kalkreuter:1995mm}
\bibinfo{author}{\bibfnamefont{T.}~\bibnamefont{Kalkreuter}} \bibnamefont{and}
  \bibinfo{author}{\bibfnamefont{H.}~\bibnamefont{Simma}},
  \bibinfo{journal}{Comput. Phys. Commun.} \textbf{\bibinfo{volume}{93}},
  \bibinfo{pages}{33} (\bibinfo{year}{1996}).

\bibitem[{\citenamefont{Liu}(2003)}]{GLiu:2003thesis}
\bibinfo{author}{\bibfnamefont{G.}~\bibnamefont{Liu}} (\bibinfo{year}{2003}),
  \bibinfo{note}{\uppercase{P}h.D. thesis, unpublished}.

\bibitem[{\citenamefont{Lee and Hatsuda}(1996)}]{Lee:1996zy}
\bibinfo{author}{\bibfnamefont{S.~H.} \bibnamefont{Lee}} \bibnamefont{and}
  \bibinfo{author}{\bibfnamefont{T.}~\bibnamefont{Hatsuda}},
  \bibinfo{journal}{Phys. Rev.} \textbf{\bibinfo{volume}{D54}},
  \bibinfo{pages}{1871} (\bibinfo{year}{1996}).

\bibitem[{\citenamefont{Evans et~al.}(1996)\citenamefont{Evans, Hsu, and
  Schwetz}}]{Evans:1996wf}
\bibinfo{author}{\bibfnamefont{N.~J.} \bibnamefont{Evans}},
  \bibinfo{author}{\bibfnamefont{S.~D.} \bibnamefont{Hsu}}, \bibnamefont{and}
  \bibinfo{author}{\bibfnamefont{M.}~\bibnamefont{Schwetz}},
  \bibinfo{journal}{Phys. Lett.} \textbf{\bibinfo{volume}{B375}},
  \bibinfo{pages}{262} (\bibinfo{year}{1996}).

\bibitem[{\citenamefont{Birse et~al.}(1996)\citenamefont{Birse, Cohen, and
  McGovern}}]{Birse:1996dx}
\bibinfo{author}{\bibfnamefont{M.~C.} \bibnamefont{Birse}},
  \bibinfo{author}{\bibfnamefont{T.~D.} \bibnamefont{Cohen}}, \bibnamefont{and}
  \bibinfo{author}{\bibfnamefont{J.~A.} \bibnamefont{McGovern}},
  \bibinfo{journal}{Phys. Lett.} \textbf{\bibinfo{volume}{B388}},
  \bibinfo{pages}{137} (\bibinfo{year}{1996}).

\bibitem[{\citenamefont{Banks and Casher}(1980)}]{Banks:1979yr}
\bibinfo{author}{\bibfnamefont{T.}~\bibnamefont{Banks}} \bibnamefont{and}
  \bibinfo{author}{\bibfnamefont{A.}~\bibnamefont{Casher}},
  \bibinfo{journal}{Nucl. Phys.} \textbf{\bibinfo{volume}{B169}},
  \bibinfo{pages}{103} (\bibinfo{year}{1980}).

\bibitem[{\citenamefont{Chandrasekharan and
  Christ}(1996)}]{Chandrasekharan:1995gt}
\bibinfo{author}{\bibfnamefont{S.}~\bibnamefont{Chandrasekharan}}
  \bibnamefont{and} \bibinfo{author}{\bibfnamefont{N.~H.}
  \bibnamefont{Christ}}, \bibinfo{journal}{Nucl.Phys.Proc.Suppl.}
  \textbf{\bibinfo{volume}{47}}, \bibinfo{pages}{527} (\bibinfo{year}{1996}),
  \eprint{hep-lat/9509095}.

\bibitem[{\citenamefont{Bazavov et~al.}(2012)\citenamefont{Bazavov,
  Bhattacharya, Cheng, DeTar, Ding et~al.}}]{Bazavov:2011nk}
\bibinfo{author}{\bibfnamefont{A.}~\bibnamefont{Bazavov}},
  \bibinfo{author}{\bibfnamefont{T.}~\bibnamefont{Bhattacharya}},
  \bibinfo{author}{\bibfnamefont{M.}~\bibnamefont{Cheng}},
  \bibinfo{author}{\bibfnamefont{C.}~\bibnamefont{DeTar}},
  \bibinfo{author}{\bibfnamefont{H.}~\bibnamefont{Ding}}, \bibnamefont{et~al.},
  \bibinfo{journal}{Phys.Rev.} \textbf{\bibinfo{volume}{D85}},
  \bibinfo{pages}{054503} (\bibinfo{year}{2012}), \bibinfo{note}{published
  version, typos corrected, minor revisions in section I and VII, conclusions
  unchanged}, \eprint{1111.1710}.

\bibitem[{\citenamefont{Braun et~al.}(2011)\citenamefont{Braun, Klein, and
  Piasecki}}]{Braun:2010vd}
\bibinfo{author}{\bibfnamefont{J.}~\bibnamefont{Braun}},
  \bibinfo{author}{\bibfnamefont{B.}~\bibnamefont{Klein}}, \bibnamefont{and}
  \bibinfo{author}{\bibfnamefont{P.}~\bibnamefont{Piasecki}},
  \bibinfo{journal}{Eur. Phys. J.} \textbf{\bibinfo{volume}{C71}},
  \bibinfo{pages}{1576} (\bibinfo{year}{2011}).

\bibitem[{\citenamefont{'t~Hooft}(1976{\natexlab{b}})}]{'tHooft:1976fv}
\bibinfo{author}{\bibfnamefont{G.}~\bibnamefont{'t~Hooft}},
  \bibinfo{journal}{Phys.Rev.} \textbf{\bibinfo{volume}{D14}},
  \bibinfo{pages}{3432} (\bibinfo{year}{1976}{\natexlab{b}}).

\bibitem[{\citenamefont{Cohen}(1997)}]{Cohen:1997hz}
\bibinfo{author}{\bibfnamefont{T.~D.} \bibnamefont{Cohen}}, pp.
  \bibinfo{pages}{100--114} (\bibinfo{year}{1997}), \eprint{nucl-th/9801061}.

\bibitem[{\citenamefont{Albanese et~al.}(1987)}]{Albanese:1987ds}
\bibinfo{author}{\bibfnamefont{M.}~\bibnamefont{Albanese}} \bibnamefont{et~al.}
  (\bibinfo{collaboration}{APE Collaboration}), \bibinfo{journal}{Phys.Lett.}
  \textbf{\bibinfo{volume}{B192}}, \bibinfo{pages}{163} (\bibinfo{year}{1987}).

\bibitem[{\citenamefont{Gockeler et~al.}(2001)\citenamefont{Gockeler, Hehl,
  Rakow, Schafer, Soldner et~al.}}]{Gockeler:2000jk}
\bibinfo{author}{\bibfnamefont{M.}~\bibnamefont{Gockeler}},
  \bibinfo{author}{\bibfnamefont{H.}~\bibnamefont{Hehl}},
  \bibinfo{author}{\bibfnamefont{P.~E.} \bibnamefont{Rakow}},
  \bibinfo{author}{\bibfnamefont{A.}~\bibnamefont{Schafer}},
  \bibinfo{author}{\bibfnamefont{W.}~\bibnamefont{Soldner}},
  \bibnamefont{et~al.}, \bibinfo{journal}{Nucl.Phys.Proc.Suppl.}
  \textbf{\bibinfo{volume}{94}}, \bibinfo{pages}{402} (\bibinfo{year}{2001}),
  \eprint{hep-lat/0010049}.

\bibitem[{\citenamefont{Damgaard et~al.}(2000)\citenamefont{Damgaard, Heller,
  Niclasen, and Rummukainen}}]{Damgaard:2000cx}
\bibinfo{author}{\bibfnamefont{P.}~\bibnamefont{Damgaard}},
  \bibinfo{author}{\bibfnamefont{U.~M.} \bibnamefont{Heller}},
  \bibinfo{author}{\bibfnamefont{R.}~\bibnamefont{Niclasen}}, \bibnamefont{and}
  \bibinfo{author}{\bibfnamefont{K.}~\bibnamefont{Rummukainen}},
  \bibinfo{journal}{Nucl.Phys.} \textbf{\bibinfo{volume}{B583}},
  \bibinfo{pages}{347} (\bibinfo{year}{2000}), \eprint{hep-lat/0003021}.

\bibitem[{\citenamefont{Gavai et~al.}(2002)\citenamefont{Gavai, Gupta, and
  Lacaze}}]{Gavai:2001vx}
\bibinfo{author}{\bibfnamefont{R.~V.} \bibnamefont{Gavai}},
  \bibinfo{author}{\bibfnamefont{S.}~\bibnamefont{Gupta}}, \bibnamefont{and}
  \bibinfo{author}{\bibfnamefont{R.}~\bibnamefont{Lacaze}},
  \bibinfo{journal}{Phys.Rev.} \textbf{\bibinfo{volume}{D65}},
  \bibinfo{pages}{094504} (\bibinfo{year}{2002}), \eprint{hep-lat/0107022}.

\bibitem[{\citenamefont{Gavai et~al.}(2008)\citenamefont{Gavai, Gupta, and
  Lacaze}}]{Gavai:2008xe}
\bibinfo{author}{\bibfnamefont{R.}~\bibnamefont{Gavai}},
  \bibinfo{author}{\bibfnamefont{S.}~\bibnamefont{Gupta}}, \bibnamefont{and}
  \bibinfo{author}{\bibfnamefont{R.}~\bibnamefont{Lacaze}},
  \bibinfo{journal}{Phys.Rev.} \textbf{\bibinfo{volume}{D77}},
  \bibinfo{pages}{114506} (\bibinfo{year}{2008}), \eprint{0803.0182}.

\bibitem[{\citenamefont{Cossu et~al.}(2010)}]{Cossu:2010rc}
\bibinfo{author}{\bibfnamefont{G.}~\bibnamefont{Cossu}} \bibnamefont{et~al.}
  (\bibinfo{collaboration}{JLQCD Collaboration}), \bibinfo{journal}{PoS}
  \textbf{\bibinfo{volume}{LATTICE2010}}, \bibinfo{pages}{174}
  (\bibinfo{year}{2010}), \eprint{1011.0257}.

\bibitem[{\citenamefont{Bazavov et~al.}(2010)\citenamefont{Bazavov, Toussaint,
  Bernard, Laiho, DeTar et~al.}}]{Bazavov:2009bb}
\bibinfo{author}{\bibfnamefont{A.}~\bibnamefont{Bazavov}},
  \bibinfo{author}{\bibfnamefont{D.}~\bibnamefont{Toussaint}},
  \bibinfo{author}{\bibfnamefont{C.}~\bibnamefont{Bernard}},
  \bibinfo{author}{\bibfnamefont{J.}~\bibnamefont{Laiho}},
  \bibinfo{author}{\bibfnamefont{C.}~\bibnamefont{DeTar}},
  \bibnamefont{et~al.}, \bibinfo{journal}{Rev.Mod.Phys.}
  \textbf{\bibinfo{volume}{82}}, \bibinfo{pages}{1349} (\bibinfo{year}{2010}),
  \eprint{0903.3598}.

\bibitem[{\citenamefont{Hasenbusch}(2001)}]{Hasenbusch:2001ne}
\bibinfo{author}{\bibfnamefont{M.}~\bibnamefont{Hasenbusch}},
  \bibinfo{journal}{Phys. Lett.} \textbf{\bibinfo{volume}{B519}},
  \bibinfo{pages}{177} (\bibinfo{year}{2001}), \eprint{hep-lat/0107019}.

\bibitem[{\citenamefont{Kennedy et~al.}(2009)\citenamefont{Kennedy, Clark, and
  Silva}}]{Kennedy:2009fe}
\bibinfo{author}{\bibfnamefont{A.}~\bibnamefont{Kennedy}},
  \bibinfo{author}{\bibfnamefont{M.}~\bibnamefont{Clark}}, \bibnamefont{and}
  \bibinfo{author}{\bibfnamefont{P.}~\bibnamefont{Silva}},
  \bibinfo{journal}{PoS} \textbf{\bibinfo{volume}{LAT2009}},
  \bibinfo{pages}{021} (\bibinfo{year}{2009}), \eprint{0910.2950}.

\bibitem[{\citenamefont{Giusti et~al.}(2004)\citenamefont{Giusti, Rossi, and
  Testa}}]{Giusti:2004qd}
\bibinfo{author}{\bibfnamefont{L.}~\bibnamefont{Giusti}},
  \bibinfo{author}{\bibfnamefont{G.}~\bibnamefont{Rossi}}, \bibnamefont{and}
  \bibinfo{author}{\bibfnamefont{M.}~\bibnamefont{Testa}},
  \bibinfo{journal}{Phys.Lett.} \textbf{\bibinfo{volume}{B587}},
  \bibinfo{pages}{157} (\bibinfo{year}{2004}), \eprint{hep-lat/0402027}.

\bibitem[{\citenamefont{Luscher}(2004)}]{Luscher:2004fu}
\bibinfo{author}{\bibfnamefont{M.}~\bibnamefont{Luscher}},
  \bibinfo{journal}{Phys.Lett.} \textbf{\bibinfo{volume}{B593}},
  \bibinfo{pages}{296} (\bibinfo{year}{2004}), \eprint{hep-th/0404034}.

\bibitem[{\citenamefont{Luscher and Palombi}(2010)}]{Luscher:2010ik}
\bibinfo{author}{\bibfnamefont{M.}~\bibnamefont{Luscher}} \bibnamefont{and}
  \bibinfo{author}{\bibfnamefont{F.}~\bibnamefont{Palombi}},
  \bibinfo{journal}{JHEP} \textbf{\bibinfo{volume}{1009}}, \bibinfo{pages}{110}
  (\bibinfo{year}{2010}), \eprint{1008.0732}.

\end{thebibliography}

\end{document}